\documentclass{article}

\usepackage[margin=1in]{geometry}
\usepackage[numbers,sort&compress]{natbib}
\usepackage{url}

\usepackage{hyperref}
\hypersetup{
    colorlinks=true,
}

\usepackage{zref-clever}
\zcsetup{
    cap=true,
    nameinlink=false
}
\zcRefTypeSetup{prob}{Name-sg=Problem,name-sg=problem}

\usepackage{tepper}

\usepackage[table,dvipsnames]{xcolor}
\usepackage{adjustbox}
\usepackage{enumitem}

\usepackage{tabularray}
\UseTblrLibrary{booktabs}
\UseTblrLibrary{siunitx}
\usepackage{listings}
\definecolor{mygreen}{rgb}{0,0.6,0}
\definecolor{mygray}{rgb}{0.5,0.5,0.5}
\definecolor{mymauve}{rgb}{0.58,0,0.82}
\usepackage{caption}
\DeclareCaptionType[fileext=loc,name=Code]{code}
\zcRefTypeSetup{code}{Name-sg=Code,name-sg=code} 

\usepackage{tcolorbox}
\tcbset{
    colback=white,colframe=black,
    arc=0mm, 
    left=0pt,
    right=0mm,
    top=0mm,
    bottom=0mm,
    boxrule=0.1mm
}

\begin{document}
\title{ASH: Asymmetric Scalar Hashing With Learned Dimensionality Reduction for High-Fidelity Vector Quantization}

\author{
  Mariano Tepper\\
  Elastic\\
  \texttt{mariano.tepper@elastic.co}
  \and
  Theodore Willke\\
  IBM WatsonX\\
  \texttt{ted.willke@ibm.com}
}

\date{}

\maketitle

\begin{abstract}
    For a long time, additive quantizers, such as product quantization, have been considered the gold standard in terms of accuracy and efficiency. Recently, scalar quantization has re-emerged from the depths of history with a new wave of data-agnostic techniques. Inscribed in this general framework, we turn our attention to data-driven methods, showing that new highs in recall and speed can be achieved by reducing the number of dimensions while increasing the bitrate per dimension. Critically, this dimensionality reduction needs to be learned from data to be successful. We present ASH (Asymmetric Scalar Hashing), a data-driven encoder-decoder framework that applies dimensionality reduction to database vectors via a learned orthonormal projection, followed by scalar quantization, while keeping queries in their original form. This asymmetric design enables higher accuracy than the best additive and scalar quantizers at iso-compression, while admitting highly efficient similarity computations via SIMD operations. ASH has short learning and encoding times, making it attractive for real-world deployment. Extensive experiments on a variety of datasets demonstrate that ASH achieves state-of-the-art ANN recall and speeds across all compression regimes.
\end{abstract}

\section{Introduction}

Approximate nearest neighbors (ANN), defined as the optimization problem of finding the vector in a given set that is most similar to a given query vector, has a long history and has recently become a critical component of AI infrastructure (e.g., in retrieval-augmented generation (RAG) \cite{lewis_retrieval-augmented_2020}).
The ANN literature evolves quickly, trying to keep up with ever-increasing requirements: more vectors with higher dimensionality, higher speeds, and lower deployment costs, all while maintaining high accuracy. In particular, the literature is rich in parsimonious vector representations equipped with computationally efficient similarity computations.

In the 2000s, the dominating paradigm for ANN used a symmetric vector representation, where the similarity between the query $\vect{q} \in \Real^D$ and the indexed vector $\vect{x} \in \Real^D$ is approximated as
\begin{equation}
    \langle \vect{q}, \vect{x} \rangle \approx \langle \operatorname{quant}(\vect{q}), \operatorname{quant}(\vect{x}) \rangle .
    \label{eq:symmetric_similarity}
\end{equation}
Here, both $\vect{q}$ and $\vect{x}$ are encoded using the same quantizer, i.e., they are quantized \emph{symmetrically}.
Hashing \cite{indyk_approximate_1998} is perhaps the most iconic representative of this approach. There, the input vectors are represented as low-dimensional binary vectors $\operatorname{quant}(\vect{x}) \in \{ 0, 1 \}^{d}$, i.e., $D > d$. This representation enables computing similarities by counting the number of ones of $\operatorname{quant}(\vect{q}) \wedge \operatorname{quant}(\vect{x})$, where $\wedge$ is the bitwise logical conjunction operator; in an era when compute was still scarce, the efficiency of this operation was critical to the success of hashing. Different forms of hashing have been proposed over the years, encompassing both data-agnostic \cite[e.g.,][]{andoni_near-optimal_2008,datar_locality-sensitive_2004,charikar_similarity_2002,dasgupta_neural_2017,jafari_survey_2021} and data-driven \cite[e.g.,][]{weiss_spectral_2008,gong_iterative_2011,kalantidis_locally_2014,dai_stochastic_2017,carreira-perpinan_hashing_2015,luo_survey_2023} approaches.

In the 2010s, the seminal product quantization (PQ) \cite{jegou_product_2011} introduced asymmetric similarity computations,
\begin{equation}
    \langle \vect{q}, \vect{x} \rangle \approx \langle \vect{q}, \operatorname{quant}(\vect{x}) \rangle ,
    \label{eq:asymmetric_similarity}
\end{equation}
where the query retains its original form and only the indexed vector is quantized. Additionally, the PQ quantizer takes an additive form that leads to a simple and fast algorithm for learning its parameters from the indexed vectors. The combination of these two innovations proved to be very successful, yielding much higher accuracy than hashing at iso-compression. This success had two main consequences: (1) a plethora of additive quantizers were proposed in the subsequent years \cite[e.g.,][]{babenko_additive_2014,babenko_tree_2015,martinez_revisiting_2016,ge_optimized_2014,kalantidis_locally_2014,guo_accelerating_2020}, and (2) the belief that other forms of quantization cannot compete with additive quantization was established in the literature \cite[e.g.,][]{babenko_tree_2015,martinez_revisiting_2016}. Additionally, very fast approaches to compute PQ similarities have subsequently been introduced \cite{andre_cache_2015,andre_accelerated_2017,andre_quicker_2021}, further cementing the popularity and empirical success of PQ.

In recent years, new research \cite{gao_rabitq_2024,aguerrebere_similarity_2023,tepper_leanvec_2024,gao_practical_2025,zandieh_turboquant_2025} has shown that when extreme levels of compression are not needed, other forms of quantization, such as scalar quantization or hashing, can outperform additive quantization when used in an asymmetric fashion. Additionally, scalar quantization and hashing lead to faster similarity computations than additive quantizers. These exciting lines of research have focused mainly on the data-agnostic regime, where the quantizer is chosen independently of the data to be quantized.

In this work, we show that data-driven asymmetric scalar quantization combines the best of both worlds: it comes with speedy similarity computations while further boosting the accuracy of the approximation.
To this effect, we present \textbf{ASH} (Asymmetric Scalar Hashing), a new quantization method where the parameters are learned from the indexed vectors. Critically, ASH relies on combining dimensionality reduction with scalar quantization to achieve \textbf{higher accuracy} than the best additive and scalar quantizers \textbf{at iso-compression}. Our use of scalar quantization leads to \textbf{highly efficient computations} of the asymmetric similarity in \zcref{eq:asymmetric_similarity}. Furthermore, ASH has short learning and encoding times, comparable to those of state-of-the-art data-agnostic quantizers.
Our full list of contributions is:
\begin{itemize}[topsep=0pt,parsep=0pt,partopsep=0pt,leftmargin=12pt]
    \item We present ASH, an encoder-decoder framework to generate asymmetric scalar-quantized vector representations (\zcref{sec:ash}). ASH enables fast similarity computations with efficient SIMD and cache-friendly implementations for the sequential and random access patterns that arise when traversing inverted and graph indices, respectively. Critically, given a fixed payload per vector, we show that applying dimensionality reduction while increasing the bitrate per dimension gives access to new configurations that are faster and more accurate.
    \item We present an algorithm (\zcref{sec:optimization}) to efficiently learn ASH representations from data by minimizing the reconstruction error. The proposed learning technique clearly outperforms sensible data-agnostic alternatives.
    \item We include a detailed comparative analysis with PQ, LOPQ, EDEN, RaBitQ, TurboQuant, and LeanVec in terms of accuracy and efficiency (\zcref{sec:comparisons}).
    \item We show with extensive experimental results (\zcref{sec:experiments}) that ASH achieves state-of-the-art ANN accuracy and speed at iso-footprint in all compression regimes (high, medium, and low compression).
\end{itemize}
For reproducibility, we have published the implementation of these contributions at \url{https://github.com/marianotepper/ash} and \url{https://github.com/tlwillke/ivf-ash-fastscan}.

\textbf{Notation.} We denote vectors/matrices by lowercase/uppercase bold letters, e.g., $\vect{v} \in \Real^{n}$ and $\vect{A} \in \Real^{m \times n}$.
The unit $d$-dimensional hypersphere is defined as $S^{d-1} \defeq \left\{ \vect{x} \in \Real^d \,|\, \norm{\vect{x}}{2} = 1 \right\}$.
The Stiefel manifold (i.e., the set of row-orthonormal matrices) is defined as $\operatorname{St} (d, D) \defeq \{ \mat{U} \in \Real^{d \times D} \,|\, \mat{U} \transpose{\mat{U}} = \mat{I} \}$ and the special orthogonal group (i.e., the set of rotation matrices) is defined as $\operatorname{SO} (d) \defeq \{ \mat{U} \in \Real^{d \times d} \,|\, \transpose{\mat{U}} \mat{U} = \mat{U} \transpose{\mat{U}} = \mat{I} \}$.

\subsection*{Related work}

Binary hashing \cite{jafari_survey_2021,wang_survey_2018,luo_survey_2023} has been extensively studied. Locality-sensitive hashing (LSH) \cite{indyk_approximate_1998,datar_locality-sensitive_2004,charikar_similarity_2002,dasgupta_neural_2017} is one of the few ANN approaches with approximation certificates \cite[e.g.,][]{andoni_near-optimal_2008}. Additionally, when compute was still scarce, the efficiency of implementing similarities as bit-wise in-register operations was very appealing. Many approaches \cite[e.g.,][]{weiss_spectral_2008,gong_iterative_2011,kalantidis_locally_2014,dai_stochastic_2017,carreira-perpinan_hashing_2015,zhang_fast_2015,yu_circulant_2014} improve accuracy over data-agnostic methods by learning the linear quantizer from data. With the rise of neural networks, many architectures have been proposed to create a deep non-linear quantizer \cite{luo_survey_2023}.

Product Quantization (PQ) \cite{jegou_product_2011} is perhaps the most successful vector representation to date because of its attractive combination of accuracy, speed, footprint, and ease of implementation. Many derived methods appeared over the years, in general improving accuracy at the cost of efficiency \cite{ge_optimized_2014,kalantidis_locally_2014}. QuickerADC \cite{andre_quicker_2021} is an accelerated SIMD approach to compute PQ similarities in bulk, becoming the de facto standard. SCANN improves the quantization error in PQ by treating it as an anisotropic distribution \cite{guo_accelerating_2020}.
Additive Quantization (AQ) \cite{babenko_additive_2014,zhang_composite_2014,martinez_revisiting_2016,babenko_tree_2015} is a generalization of PQ with significantly improved accuracy.
Residual Quantization (RQ) \cite{juang_multiple_1982,chen_approximate_2010,liu_improved_2015}, although older than AQ, can be interpreted as a form of AQ with stage-wise training. While AQ and RQ provide much better representations than PQ, their training times skyrocket in higher dimensions (in the thousands with modern embedding models), rendering them prohibitive in real-world scenarios.

Using an asymmetric representation with binary or scalar hashing, although not new \cite{gordo_asymmetric_2011}, has recently become popular \cite{vargaftik_eden_2022,aguerrebere_similarity_2023,gao_rabitq_2024,gao_practical_2025,tepper_leanvec_2024,zandieh_turboquant_2025} and can increase search accuracy with respect to symmetric hashing and PQ. However, the technique in \cite{gordo_asymmetric_2011} is unable to improve the accuracy of PQ, only matching it. The approaches in \cite{vargaftik_eden_2022,gao_rabitq_2024,gao_practical_2025} do not learn the quantizer from data. The technique in \cite{tepper_leanvec_2024} performs dimensionality reduction without considering the quantizer, only applying it at the end as a heuristic.
Our work can be regarded as a spiritual successor that unleashes the full potential of asymmetric scalar hashing by learning it from data.

For completeness, we also mention the data structures that are commonly used to index these vector representations. The most relevant types of indices are trees \cite[e.g.,][]{beygelzimer_cover_2006,bentley_multidimensional_1975}, inverted indices \cite[e.g.,][]{muja_scalable_2014,johnson_billion-scale_2021}, and graphs \cite[e.g.,][]{malkov_efficient_2020,subramanya_diskann_2019}.

\section{Hashing with asymmetric similarity computations}
\label{sec:ash}

In this work, we seek to represent a vector $\vect{z}_i \in S^{D-1}$ using the lossy approximation
\begin{equation}
    \vect{z} \approx \hat{\vect{z}} = \norm{\vect{v}}{2}^{-1} \transpose{\mat{W}} \vect{v}
\end{equation}
for $\mat{W} \in \operatorname{St} (d, D)$ and $\vect{v} \in \set{V}_{b}^{d}$, where $\set{V}_{b}^{d}$ is the Cartesian power of
\begin{equation}
    \set{V}_b = \left\{ 2c - 2^b + 1 \,|\, c = 0,\dots, 2^b - 1 \right\} .
    \label{eq:quantized_values}
\end{equation}
The parameter $b$ indicates the bitrate per dimension.
For example, $\set{V}_1 = \{ -1, 1 \}$ for $b=1$ and $\set{V}_2 = \{ -3, -1, 1, 3 \}$ for $b=2$.
In general, for $\vect{v} \in \set{V}_{b}^{d}$, $\norm{\vect{v}}{2} \neq 1$. Thus, the normalization and the constraint $\mat{W} \in \operatorname{St} (d, D)$ ensure that $\hat{\vect{z}} \in S^{D-1}$.

We learn our representation from data $\{ \vect{z}_i \in S^{D-1} \}_{i=1}^{n}$ by solving the optimization problem
\begin{align}
    \min_{\substack{
        \left\{ \vect{v}_i \in \set{V}_{b}^{d} \right\}_{i=1}^n
        \\
        \mat{W} \in \operatorname{St} (d, D)
    }}
    \tfrac{1}{2} \sum_{i=1}^{n} \norm{ \vect{z}_i - \norm{\vect{v}_i}{2}^{-1} \transpose{\mat{W}} \vect{v}_i }{2}^2
    \label{prob:reconstruction_loss}
\end{align}
that minimizes the average reconstruction error.
The parameter $\mat{W}$ is learned from \zcref{prob:reconstruction_loss} during indexing, as described in \zcref{sec:optimization}, and stored as a global index parameter.
For the subproblem of finding $\vect{v}_i$ with $\mat{W}$ fixed, we have the following result.
\begin{proposition}
    For $\vect{z} \in S^{D-1}$, we have
    \begin{equation}
        \argmin_{\vect{v} \in \set{V}_{b}^{d}}
        \tfrac{1}{2} \norm{ \vect{z} - \norm{\vect{v}}{2}^{-1} \transpose{\mat{W}} \vect{v} }{2}^2
        = \operatorname{quant}_b \left( \mat{W} \vect{z} \right)
    \end{equation}
    where $\operatorname{quant}_b$ is defined by
    \begin{equation}
        \operatorname{quant}_b (\vect{u}) \defeq \argmax_{\vect{v} \in \set{V}_{b}^{d}} \operatorname{cosSim} \left( \vect{v}, \vect{u} \right) .
        \label{eq:quant}
    \end{equation}
\end{proposition}
The proof follows directly from developing the norm $\norm{ \vect{z} - \norm{\vect{v}}{2}^{-1} \transpose{\mat{W}} \vect{v} }{2}^2$ and dropping the constant terms, i.e.,
$
    \argmin_{\vect{v} \in \set{V}_{b}^{d}}
    \tfrac{1}{2} \norm{ \vect{z} - \norm{\vect{v}}{2}^{-1} \transpose{\mat{W}} \vect{v} }{2}^2
    =
    \argmax_{\vect{v} \in \set{V}_{b}^{d}}
    \norm{\vect{v}}{2}^{-1} \left\langle \mat{W} \vect{z} , \vect{v} \right\rangle 
$.
This result indicates that the approximation can be interpreted as an encoder-decoder architecture \cite{amara_nearest_2022},
\begin{equation}
    \vect{z} \approx f(g(\vect{z}; \mat{W}) ; \mat{W}) ,
    \label{eq:encoder_decoder}
\end{equation}
where the encoder $g$ reduces the dimensionality from $D$ to $d$ dimensions and performs quantization while the decoder $f$ expands it back. Here, the encoder $g : S^{D-1} \to \set{V}_{b}^{d}$ and the decoder $f : \set{V}_{b}^{d} \to S^{D-1}$ take the forms
\begin{align}
    g(\vect{z}; \mat{W}) &\defeq \operatorname{quant}_b \left( \mat{W} \vect{z} \right) ,
    \label{eq:g}
    \\
    f(\vect{z} ; \mat{W})
    &\defeq
    \norm{\vect{z}}{2}^{-1} \transpose{\mat{W}} \vect{z} .
    \label{eq:f}
\end{align}
By construction, the reconstruction error in our architecture is purely angular, in contrast to other quantizers that carry angular and norm errors~\cite{guo_accelerating_2020}.
Since our architecture works in $S^{D-1}$, we need to normalize the encoder inputs.
Let $\set{X} = \{\vect{x}_i \}_{i=1}^n$ be the set of vectors to quantize and let $\{ \vect{\mu}_c \}_{c=1}^C$ be a set of landmark vectors. For any vector $\vect{x}_i \in \set{X}$, we use $\tilde{\vect{x}}_i$ as the input to the encoder $g$, where
\begin{equation}
    \tilde{\vect{x}}_i \defeq (\vect{x}_i - \vect{\mu}_i^*) / \norm{\vect{x}_i - \vect{\mu}_i^*}{2} 
    \label{eq:x_normalized}
\end{equation}
and
\begin{equation}
    \vect{\mu}_i^* = \argmin_{\{ \vect{\mu}_c \}_{c=1}^C} \norm{\vect{x}_i - \vect{\mu}_{c}}{2} .
    \label{eq:c_star}
\end{equation}
The landmark set $\{ \vect{\mu}_c \}_{c=1}^C$ can be learned stage-wise using k-means clustering (the approach followed in this work) or learned jointly with the other parameters of the autoencoder as in VQ-VAE~\cite{van_den_oord_neural_2017}. In the case of an inverted index (IVF) \cite{johnson_billion-scale_2021}, the landmarks can be defined as the centroids of the IVF clusters. If $C=1$, $\vect{\mu}_1$ can be the dataset mean or a robust statistic, such as the median.

We point out that RaBitQ \cite{gao_rabitq_2024} and extended RaBitQ \cite{gao_practical_2025} are particular instantiations of our model with $D = d$, $C=1$, and $\mat{W} = \mat{R}$ for a random orthogonal matrix $\mat{R} \in \Real^{D \times D}$. In this configuration, RaBitQ and extended RaBitQ support $b=1$ and $b \geq 1$, respectively.

\subsection{Analysis of the ASH error}
\label{sec:ash_error}

Let $X$ be a random variable on the $D-1$-dimensional hypersphere.
We are interested in studying
\begin{equation}
    \epsilon \defeq \expectation{\norm{X - f(g(X ; \mat{W} ) ; \mat{W})}{2}^2} .
\end{equation}
The quantization error can be modeled by a $d$-dimensional random variable $E$,
\begin{equation}
    \epsilon = \expectation{\norm{X - \norm{\mat{W} X}{2}^{-1} \transpose{\mat{W}} \left( \mat{W} X + E \right)}{2}^2} .
\end{equation}
A few algebraic manipulations yield
\begin{equation}
    \epsilon = \expectation{
        2
        - 2 \norm{\mat{W} X}{2}
        + \frac{\norm{E}{2}^2}{\norm{\mat{W} X}{2}^2}
    } .
\end{equation}
Naturally, as the bitrate $b$ increases, $\norm{E}{2}$ decreases. Thus, at higher bitrates, the term involving $E$ vanishes as $\norm{\mat{W} X}{2}^2$ dominates. In this regime, the total expected error is mainly determined by the dimensionality reduction error $\norm{X}{2}^2 - 2 \norm{\mat{W} X}{2}$. Thus, at higher bitrates, reducing the dimensionality becomes the main source of error.
At lower bitrates, the tradeoff between quantization and dimensionality reduction error is more nuanced. Our experiments in \zcref{sec:experiments} show that interesting gains are obtained by reducing the number of dimensions and using a slightly higher bitrate.

\subsection{Efficient asymmetric dot products}

We now show how our representation leads to computationally efficient dot product calculations. While we focus on the dot product, our framework extends naturally to the Euclidean distance and cosine similarity, see \zcref{sec:other_metrics} in the appendix.
We start by manipulating the dot product $\langle \vect{q}, \vect{x}_i \rangle$ between a query $\vect{q}$ and a vector $\vect{x}_i \in \set{X}$ to work with our centered and normalized vector $\tilde{\vect{x}}_i$,
\begin{align}
    \langle \vect{q}, \vect{x}_i \rangle
    &=
    \langle \vect{q} - \vect{\mu}_i^* , \vect{x}_i - \vect{\mu}_i^* \rangle
    + \langle \vect{q} , \vect{\mu}_i^* \rangle
    + \langle \vect{x}_i , \vect{\mu}_i^* \rangle
    - \norm{ \vect{\mu}_i^* }{2}^2 \nonumber \\
    &=
    \norm{ \vect{x}_i - \vect{\mu}_i^* }{2} \left\langle \vect{q} - \vect{\mu}_i^* , \tilde{\vect{x}}_i \right\rangle
    + \langle \vect{q} , \vect{\mu}_i^* \rangle
    + \langle \vect{x}_i , \vect{\mu}_i^* \rangle
    - \norm{ \vect{\mu}_i^* }{2}^2 .
    \label{eq:spherical_approach}
\end{align}
The dot product $\left\langle \vect{q} - \vect{\mu}_i^* , \tilde{\vect{x}}_i \right\rangle$ is the main quantity of interest as the remaining terms can be pre-computed.
Applying the approximation in \zcref{eq:encoder_decoder} to $\tilde{\vect{x}}_i$, we get
\begin{align}
    \left\langle \vect{q} - \vect{\mu}_i^* , \tilde{\vect{x}}_i \right\rangle
    &\approx
    \norm{g(\tilde{\vect{x}}_i)}{2}^{-1}
    \left\langle \vect{q} - \vect{\mu}_i^* , \transpose{\mat{W}} g(\tilde{\vect{x}}_i) \right\rangle
    \nonumber \\
    &=
    \norm{g(\tilde{\vect{x}}_i)}{2}^{-1}
    \left( \left\langle \breve{\vect{q}} , g(\tilde{\vect{x}}_i) \right\rangle - \left\langle \mat{W} \vect{\mu}_i^* , g(\tilde{\vect{x}}_i) \right\rangle \right)
    \label{eq:dotprod_approx_efficient}
\end{align}
where
\begin{equation}
    \breve{\vect{q}} \defeq \mat{W} \vect{q} .
    \label{eq:query_low_dim}
\end{equation}
Let $\vect{v}_i = \operatorname{quant}_b (\mat{W} \tilde{\vect{x}}_i)$. Plugging \zcref{eq:dotprod_approx_efficient} into \zcref{eq:spherical_approach}, we get the desired approximation,
\begin{multline}
    \langle \vect{q}, \vect{x}_i \rangle
    \approx
    \underbrace{
        \norm{\vect{v}_i}{2}^{-1}
        \norm{ \vect{x}_i - \vect{\mu}_i^* }{2}
    }_\textsc{scale}
    \underbrace{
        \left\langle \breve{\vect{q}} , \vect{v}_i \right\rangle
    }_\textsc{dot-prod}
    +
    \underbrace{
        \langle \vect{q} , \vect{\mu}_i^* \rangle
    }_\textsc{query-compute}
    +
    \underbrace{
        \langle \vect{x}_i , \vect{\mu}_i^* \rangle
        - \norm{\vect{v}_i}{2}^{-1} \norm{ \vect{x}_i - \vect{\mu}_i^* }{2} \left\langle \mat{W} \vect{\mu}_i^* , \vect{v}_i \right\rangle
        - \norm{\vect{\mu}_i^*}{2}^2
    }_\textsc{offset}.
    \label{eq:spherical_approach_approximation}
\end{multline}
The importance of a simple linear decoder becomes clearer now as it enables the one-time application of $\mat{W}$ as an encoder for the query instead of applying $\transpose{\mat{W}}$ as a decoder once per dot product $\langle \vect{q}, \vect{x} \rangle$. This yields significant savings, as thousands of dot products are performed for each query in state-of-the-art vector search indices~\cite{subramanya_diskann_2019,malkov_efficient_2020,chen_spann_2021}. Moreover, the transformation reduces the dimensionality of the query, accelerating the computation of $\left\langle \breve{\vect{q}} , \vect{v}_i \right\rangle$. Computing similarities symmetrically is additionally required when building graph indices; we extend our formulation to this symmetric setting in \zcref{sec:symmetric_dot_product} in the appendix.

\subsection{Payload}

The encoder $g$ transforms the input vectors in $S^{D-1}$ to $\set{V}_{b}^{d}$. As $|\set{V}_{b}^{d}| = 2^{bd}$, ASH reduces the footprint by a factor of $32 D / (b d)$ for vectors encoded in single-precision format (float32). 

For each vector $\vect{x}_i$, we store the $d$-dimensional vector $g( \tilde{\vect{x}}_i )$ as a bit string of length $bd$. Along with it, we store the values of the terms \textsc{scale} and \textsc{offset} in \zcref{eq:spherical_approach_approximation} in a 16-bit format (FP16 or BF16). The integer $c_i^* \in [0, C)$ is encoded using $\lceil \log_2 C \rceil$ bits.
Thus, to encode a vector using $B$ total bits at a bitrate of $b$ bits, we set $d = \lfloor (B - 2 \cdot 16 - \lceil \log_2 C \rceil) / b \rfloor$. We include a summary in \zcref{tab:payload}.

\begin{table*}[t]
    \caption{The payload of an ASH vector representation ($c_i^*$ and $\vect{\mu}_i^*$ are defined in \zcref{eq:c_star}). To obtain a $B$-bit representation, we subtract the header payload to obtain the target dimensionality, $d = \lfloor (B - 2 \cdot 16 - \lceil \log_2 C \rceil) / b \rfloor$.}
    \label{tab:payload}

    \small
    \centering
    \begin{tblr}{
        colspec = {|c|c|c|c|},
        rowspec = {|c|c|c|c|},
        cell{1,2}{1} = {red!15},
        cell{1,2}{2} = {red!15},
        cell{1,2}{3} = {red!15},
        cell{1,2}{4} = {NavyBlue!15},
    }
        \SetCell[c=3]{c} Header
        &&&
        Body
        \\
        $\frac{\norm{\vect{x}_i - \vect{\mu}_i^*}{2}}{\norm{\operatorname{quant}_b ( \tilde{\vect{x}}_i )}{2}}$ &
        $\langle \vect{x}_i , \vect{\mu}_i^* \rangle - \frac{\norm{ \vect{x}_i - \vect{\mu}_i^* }{2}}{\norm{\operatorname{quant}_b ( \tilde{\vect{x}}_i )}{2}} \left\langle \mat{W} \vect{\mu}_i^* , \operatorname{quant}_b ( \tilde{\vect{x}}_i ) \right\rangle - \norm{\vect{\mu}_i^*}{2}^2$ &
        $c_i^*$ &
        $\operatorname{quant}_b ( \tilde{\vect{x}}_i )$
        \\
        $\Real$ &
        $\Real$ &
        $\Nat \cap [1 , C]$ &
        $\set{V}_{b}^{d}$
        \\
        16 bits &
        16 bits &
        $\lceil \log_2 C \rceil$ bits &
        $bd$ bits
        \\
    \end{tblr}
\end{table*}

\subsection{Efficiency}

Let us analyze the terms in \zcref{eq:spherical_approach_approximation}.
For each query $\vect{q} \in \Real^D$, we first compute $\breve{\vect{q}}$ and 
$\{ \langle \vect{q} , \vect{\mu}_c \rangle \}_{c=1}^C$
once per query.
Hence, computing the term \textsc{query-compute} only requires fetching from the latter sets.
The terms \textsc{scale} and \textsc{offset} are stored with the quantized vector.
The term \textsc{dot-prod} can be implemented as a fused multiply-add (FMA) \cite{intel_intrinsics_guide__mm512_fmadd_ps_2025}. Combined with special memory layouts \cite{afroozeh_fastlanes_2023,aguerrebere_locally-adaptive_2024}, these operations become extremely efficient.

Alternatively, for sequential scans, ASH can be implemented using a FastScan-style \cite{andre_quicker_2021} lookup table. The quantized code is packed into 4-bit groups: for $b = 1$, $b = 2$, $b = 4$, each group stores four, two and one binary coordinates, respectively. For a query $\vect{q}$, we first compute the projected query $\mat{W} \vect{q}$. Then, for each 4-bit group, we precompute a 16-entry table containing the contribution $\langle (\mat{W} \vect{q})^{(m)}, v^{(m)} \rangle$ for each possible codeword value in that group. Scoring an indexed vector reduces to reading its packed 4-bit codes and summing the corresponding table entries, followed by the scalar scale and offset correction in \zcref{eq:spherical_approach_approximation}.

\subsection{The 1-bit case}

The scenario where $b=1$ presents further optimization opportunities.
In this case, $\set{V}_1 = \{ -1, 1 \}$ and
$\operatorname{quant}_b (\vect{u}) \defeq \operatorname{sign} \left( \mat{W} \vect{u} \right)$,
where $\operatorname{sign} (x) = 1$ if $x \geq 0$, and $\operatorname{sign} (x) = -1$ otherwise.
Trivially, $\norm{\operatorname{quant}_1 (\cdot)}{2} = \sqrt{d}$.
With this specialization, further expanding \zcref{eq:dotprod_approx_efficient} yields
\begin{equation}
    \left\langle \vect{q} - \vect{\mu}_i^* , \tilde{\vect{x}}_i \right\rangle
    \approx
    d^{-1/2}
    \Big(
    2 
    \left\langle \breve{\vect{q}} , \operatorname{bin} \left( \mat{W} \tilde{\vect{x}}_i \right) \right\rangle
    - \left\langle \mat{W} \vect{\mu}_i^* , 2 \operatorname{bin} \left( \mat{W} \tilde{\vect{x}}_i \right) - \vect{1} \right\rangle 
    - \left\langle \breve{\vect{q}} , \vect{1} \right\rangle
    \Big),
    \label{eq:dotprod_approx_binary}
\end{equation}
where $\operatorname{bin} (\cdot) \in \{ 0, 1 \}^{d}$ is defined as $\operatorname{bin} (\vect{z}) \defeq (\operatorname{sign} (\vect{z}) + 1) / 2$. 
Finally, plugging \zcref{eq:dotprod_approx_binary} into \zcref{eq:spherical_approach}, we get
\begin{multline}
    \langle \vect{q}, \vect{x}_i \rangle
    \approx
    \underbrace{
        2 d^{-1/2}
        \norm{ \vect{x}_i - \vect{\mu}_i^* }{2}
    }_\textsc{scale}
    \underbrace{
        \left\langle \breve{\vect{q}} , \vect{b}_i \right\rangle
    }_\textsc{masked-add}
    -
    \underbrace{
        2 d^{-1/2} \norm{\vect{x}_i - \vect{\mu}_i^*}{2}
    }_\textsc{scale}
    \underbrace{
        \left\langle \breve{\vect{q}} , \vect{1} \right\rangle
        + \langle \vect{q} , \vect{\mu}_i^* \rangle
    }_\textsc{query-compute}
    +
    \\
    +
    \underbrace{
        \langle \vect{x}_i , \vect{\mu}_i^* \rangle
        - d^{-1/2} \norm{ \vect{x}_i - \vect{\mu}_i^* }{2} \left\langle \mat{W} \vect{\mu}_i^* , 2 \vect{b}_i - \vect{1} \right\rangle
        - \norm{\vect{\mu}_i^*}{2}^2
    }_\textsc{offset}.
    \label{eq:spherical_approach_approximation_binary}
\end{multline}
where $\vect{b}_i = \operatorname{bin} \left( \mat{W} \tilde{\vect{x}}_i \right)$.
Let us compare the terms in \zcref{eq:spherical_approach_approximation_binary} with those in \zcref{eq:spherical_approach_approximation}.
For each query $\vect{q} \in \Real^D$, we additionally compute the set
$\{ \langle \breve{\vect{q}}_c , \vect{1} \rangle \}_{c=1}^C$
once per query.
The term \textsc{scale} now appears twice.
The term \textsc{masked-add} can be implemented efficiently for $\breve{\vect{q}} \in \Real^d$ and $\vect{b} \in \{ 0, 1 \}^d$ as
\begin{equation}
    \left\langle \breve{\vect{q}} , \vect{b} \right\rangle
    =
    \sum_{j=1}^{d} \indicator{b_j = 1} \cdot \breve{q}_{j} ,
    \label{eq:masked-add}
\end{equation}
where $b_j$ is the $j$-th element of $\vect{b}$. This operation can be written as a masked-load instruction~\cite{intel_intrinsics_guide__mm512_maskz_loadu_ps_2025} followed by a horizontal add, see \zcref{tab:binary_dot_product_efficient}. Once this computation occurs, only a handful of scalar multiplications and additions remain in \zcref{eq:spherical_approach_approximation}. An example implementation using AVX-512 instructions is provided in \zcref{code:ash_dot_product}. We point out that there are computational schemes that are more efficient than \zcref{code:ash_dot_product} while being more involved; we include a simple example as an illustration.

\begin{code*}
\caption{Basic implementation of the ASH dot product approximation in \zcref{eq:spherical_approach_approximation} using AVX-512 for $b=2$. The lengths of the query and binary vectors are $d$ and $d / 16$, respectively (we assume that $d$ is a multiple of 16). The parameters \texttt{scale} and \texttt{offset} summarize the multiplicative and additive components of the terms $\textsc{scale}$, $\textsc{query-compute}$, and $\textsc{offset}$ in \zcref{eq:spherical_approach_approximation}.}
\label{code:ash_dot_product}

\begin{lstlisting}[language=C++]
float ash_dot_product_2bit(const float* query, const uint16_t* low_bits,
                           const uint16_t* high_bits, uint32_t d, float scale, float offset)
{
    __m512 vec_low  = _mm512_setzero_ps();
    __m512 vec_high = _mm512_setzero_ps();
    for (uint32_t i = 0, j = 0; j < d; i++, j += 16) {
        __mmask16 mask_low  = _load_mask16((__mmask16*) (low_bits  + i));
        __mmask16 mask_high = _load_mask16((__mmask16*) (high_bits + i));
        __m512 q_low  = _mm512_maskz_load_ps(mask_low,  (void const*) (query + j));
        __m512 q_high = _mm512_maskz_load_ps(mask_high, (void const*) (query + j));
        vec_low  = _mm512_add_ps(vec_low,  q_low);
        vec_high = _mm512_add_ps(vec_high, q_high);
    }
    float result = _mm512_reduce_add_ps(vec_low) + 2.0f * _mm512_reduce_add_ps(vec_high);
    return scale * result + offset;
}
\end{lstlisting}
\end{code*}

\begin{table}[t]
    \caption{The dot product $\left\langle \vect{x}, \vect{y} \right\rangle$ can be efficiently computed when one of the operands is binary, see \zcref{code:ash_dot_product}. A reference implementation of the dot product between unquantized vectors can be found in \zcref{sec:dot_product}. Latency and throughput are measured in CPU cycles and CPU cycles per instruction, respectively. One horizontal sum~\cite{intel_intrinsics_guide__mm512_reduce_add_ps_2025} across the 16 float32 elements completes the calculation after the FMA or the masked load.}
    \label{tab:binary_dot_product_efficient}

    \small
    \centering
    \begin{tblr}{
        colspec = {clccc},
        colsep = 3pt,
        rowspec = {|c|cc|ccc|},
    }
        Operands & Instructions & \# of calls & Latency & Throughput \\
        
        \SetCell[r=2]{c} $\vect{x},\vect{y} \in \Real^{16}$ &
        Load~\cite{intel_intrinsics_guide__mm512_loadu_ps_2025} & 2 & 8 & 0.5
        \\
        &
        FMA~\cite{intel_intrinsics_guide__mm512_fmadd_ps_2025} & 1 & 4 & 0.5 \\
        
        \SetCell[r=3]{c} \shortstack{$\vect{x} \in \{ 0, 1 \}^{16}$ \\ $\vect{y} \in \Real^{16}$}
        & Load mask~\cite{intel_intrinsics_guide__load_mask16_2025} & 2 & - & 1 \\
        & Masked load~\cite{intel_intrinsics_guide__mm512_maskz_loadu_ps_2025} & 2 & 8 & 0.5 \\
        & Add~\cite{intel_intrinsics_guide__mm512_add_ps_2025} & 2 & 4 & 0.5
    \end{tblr}
\end{table}

\section{Learning ASH representations}
\label{sec:optimization}

We now proceed to present our algorithm to learn the ASH parameters from data.
Following~\cite{gong_iterative_2011}, we parametrize $\mat{W}$ as $\mat{W} = \mat{R} \mat{P}$, where $\mat{P} \in \operatorname{St} (d, D)$ and $\mat{R} \in \operatorname{SO} (d)$ is a rotation matrix.
The matrix $\transpose{\mat{P}}$ is given by the top $d$ eigenvectors of $\sum_{i=1}^{n} \tilde{\vect{x}}_i \transpose{\tilde{\vect{x}}_i}$ (we assume that $n > d$ throughout this section so that the involved matrices are not rank deficient). Plugging this form for $\mat{W}$ into \zcref{prob:reconstruction_loss} and simplifying the constant terms, we obtain the new problem
\begin{align}
    \max_{\substack{
        \left\{ \vect{v}_i \in \set{V}_{b}^{d} \right\}_{i=1}^n
        \\
        \mat{R} \in \operatorname{SO} (d)
    }}
    \sum_{i=1}^{n} \norm{\vect{v}_i}{2}^{-1} \left\langle \mat{P} \tilde{\vect{x}}_i , \transpose{\mat{R}} \vect{v}_i \right\rangle
    \label{prob:learning_loss}
    .
\end{align}
We solve this problem by alternate minimization.
Let $\operatorname{cosSim}$ be the cosine similarity between two vectors.
The solution for an individual $\vect{v}_i$ with $\mat{R}$ fixed is
\begin{equation}
    \argmax_{\vect{v} \in \set{V}_{b}^{d}} \operatorname{cosSim} \left( \vect{v}, \mat{R} \mat{P} \tilde{\vect{x}}_i \right)
    .
\end{equation}
This matches exactly our definition of $\operatorname{quant}_b$, obtaining a formal derivation for the definition in \zcref{eq:quant}.
The problem for $\mat{R}$ with $\left\{ \vect{v}_i \right\}_{i=1}^n$ fixed is
\begin{equation}
    \max_{\mat{R} \in \operatorname{SO} (d)} \traceone{\mat{R} \mat{M}} .
    \label{eq:learning_W}
\end{equation}
where $\mat{M} = \mat{P} \left( \sum_{i=1}^{n} \norm{\vect{v}_i}{2}^{-1} \cdot \tilde{\vect{x}}_i \transpose{\vect{v}_i} \right)$. 
This corresponds to an orthogonal Procrustes problem, whose solution is $\mat{U} \transpose{\mat{V}}$ where $\mat{U} \mat{S} \transpose{\mat{V}}$ is the singular value decomposition of $\mat{M}$. Alternatively, the iterative Newton-Schulz method can be used to solve this problem and has gained popularity for deep learning using GPUs~\cite{jordan_muon_2024,pethick_training_2025}.
We proceed in an alternating fashion, applying the steps:
\begin{enumerate}
    \item $(\forall i)$ compute $\vect{v}_i^{(t)} \gets \operatorname{quant}_b \left( \mat{R}^{(t)} \mat{P} \tilde{\vect{x}}_i \right)$;
    \item $\mat{R}^{(t+1)} \defeq \mat{U} \transpose{\mat{V}}$ where $\mat{U} \mat{S} \transpose{\mat{V}}$ is the singular value decomposition of $\mat{P} \left( \sum_{i=1}^{n} \norm{\vect{v}_i}{2}^{-1} \cdot \tilde{\vect{x}}_i \transpose{\vect{v}_i} \right)$.
\end{enumerate}
The matrix $\mat{R}^{(0)}$ is initialized as $\mat{U}' \transpose{\mat{V}'}$ where $\mat{U}' \mat{S}' \transpose{\mat{V}'}$ is the singular value decomposition of a random $d \times d$ matrix with entries sampled from a standard normal distribution. The iterations converge to a local minimum since each minimization step decreases the objective function of the corresponding subproblem.
Finally, after $T$ iterations, we set $\mat{W} = \mat{R}^{(T)} \mat{P}$.

Note that for binary quantization ($b=1$), we recover the seminal ITQ learning algorithm~\cite{gong_iterative_2011,gong_iterative_2013} as
\begin{equation}
    \argmax_{\vect{v} \in \set{V}_{1}^{d}} \operatorname{cosSim} \left( \vect{v}, \mat{R} \mat{P} \tilde{\vect{x}}_i \right)
    =
    \operatorname{sign} \left( \mat{R} \mat{P} \tilde{\vect{x}}_i \right) .
\end{equation}

Although lightweight, the proposed learning algorithm involves multiple passes over the training data. As $\mat{M}$ in \zcref{eq:learning_W} is essentially a cross covariance matrix, it will quickly converge to its expected value as the number of samples grows. Thus, throughout our experiments, we only use $10D$ training vectors, ensuring an oversampling of $10\times$.

\section{Comparison with other quantizers}
\label{sec:comparisons}

In this section, we present detailed comparisons of the similarity computations in ASH and three popular alternatives: the standard PQ \cite{jegou_product_2011}, and the state-of-the-art EDEN \cite{vargaftik_eden_2022} and TurboQuant \cite{zandieh_turboquant_2025}.

\textbf{ASH versus PQ.}
PQ \cite{jegou_product_2011} divides the $D$ dimensions into $M < D$ segments (we assume $D/M \in \Nat$ for simplicity). A vector $\vect{x} \in \Real^D$ is divided into segments $\vect{x}^{(1)}_i \in \Real^{D/M}, \dots, \vect{x}^{(M)}_i  \in \Real^{D/M}$. In PQ, to create codes with $B = b \cdot M$ bits, the quantizer takes an additive form
\begin{equation}
    \vect{x}
    \approx
    \operatorname{quant}_{\textsc{pq}}(\vect{x})
    \defeq
    \mat{W}_{\textsc{pq}}
    \begin{bmatrix}
        \operatorname{assign} \left( \vect{x}^{(1)} \right) \\
        \vdots \\
        \operatorname{assign} \left( \vect{x}^{(M)} \right) \\
    \end{bmatrix} ,
    \label{eq:pq_representation}
\end{equation}
where $\mat{W}_{\textsc{pq}} \in \Real^{D \times (2^b \cdot M)}$ is a block-diagonal matrix with $M$ blocks $\mat{W}_{\textsc{pq}}^{(m)} \in \Real^{(D/M) \times 2^b}$ and $\operatorname{assign}(\vect{x}^{(m)}) \in 2^b$ is a one-hot binary vector (thus, $b$ bits are needed to represent it). Each block $\mat{W}_{\textsc{pq}}^{(m)}$ is learned from the set $\{ \vect{x}^{(m)}_i \}_{i=1}^{n}$ using a vector quantizer (i.e., k-means) with $2^b$ centroids and $\operatorname{bin}(\vect{x}_i^{(m)})$ is set using the index of the centroid closest to $\vect{x}^{(m)}_i$.

At search, the PQ similarity table $T$ is computed once for each query $\vect{q}$ and stores the non-zero blocks $\transpose{\mat{W}_{\textsc{pq}}^{(m)}} \vect{q}^{(m)}$ of $\transpose{\mat{W}_{\textsc{pq}}} \vect{q}$.
The similarity to the indexed vector $\vect{x}$ is approximated by PQ as
\begin{equation}
    \langle \vect{q}, \vect{x} \rangle \approx \langle \vect{q}, \operatorname{quant}_{\textsc{pq}}(\vect{x}) \rangle  = \sum_m \langle \transpose{\mat{W}_{\textsc{pq}}^{(m)}} \vect{q}^{(m)} , \operatorname{assign}(\vect{x}^{(m)}) \rangle .
    \label{eq:pq_similarity}
\end{equation}
Since $\operatorname{assign}(\vect{x}^{(m)})$ is a binary one-hot vector, each summand only involves a single gather operation \cite{andre_cache_2015}, similarly to the masked-add in \zcref{eq:masked-add}.

In general, $2^b \cdot M > D > d$, making the PQ gather more expensive than the $d$-dimensional masked-add in \zcref{eq:masked-add}.
Let us build an example with typical numbers. Let $D=1024$. We set $b=8$ to have byte-aligned PQ codes and $M=64$ or $M=128$ are reasonable practical values.
In these examples, the PQ similarity table occupies 64KB or 128KB at four bytes per entry (float32).
Even when compressing each entry to one byte \cite{andre_quicker_2021}, the PQ similarity table occupies 16KB or 32KB.
On the other side, the query vector used in \zcref{eq:masked-add} only occupies 4KB when $d=D$ and $C=1$. Considering that modern CPUs have 112KB of L1 cache per core, this is a significant reduction.
As detailed in \zcref{sec:experiments}, ASH already outperforms PQ in terms of search accuracy with $C=1$ and this lead is further widened when $C > 1$.

We also consider the computational aspect of the SIMD operations behind \zcref{eq:pq_similarity} in PQ against those behind \zcref{eq:masked-add} in ASH. Although mathematically they appear similar, their efficiency differs drastically when using AVX instructions, as observed for example in \zcref{tab:gather_comparison} when using the latest AVX-512.
For PQ, SIMD allows in-memory table lookups using gather instructions~\cite{intel_intrinsics_guide__mm512_i32gather_ps_2025}. While these instructions support 16-element gather operations, only a fraction of the underlying memory accesses can be performed concurrently, which limits the throughput and latency of the instruction. For ASH, the fixed-precision dot product is performed efficiently (as a masked load for $b=1$), involving a single memory access.

\begin{table}[t]
    \caption{Both PQ and ASH involve gather-like operations to compute similarities, as defined in \zcref{eq:pq_similarity,eq:masked-add}. However, these gathers are of a different computational nature: PQ uses an in-memory irregular gather while ASH uses a masked load with $b=1$. This can be observed in the efficiency of the underlying AVX-512 intrinsics. Latency and throughput are measured in CPU cycles and CPU cycles per instruction, respectively.}
    \label{tab:gather_comparison}

    \small
    \centering
    \begin{tblr}{
        colspec = {ccc},
        rowspec = {|c|cc|},
    }
        Operands & Instructions & Latency & Throughput \\
        
        PQ & Gather~\cite{intel_intrinsics_guide__mm512_i32gather_ps_2025} & 30 & 9.75 \\
        
        ASH ($b=1$) & Masked load~\cite{intel_intrinsics_guide__mm512_maskz_loadu_ps_2025} & 8 & 0.5 \\
    \end{tblr}
\end{table}

\textbf{ASH versus EDEN and TurboQuant.}
EDEN \cite{vargaftik_eden_2022} and TurboQuant \cite{zandieh_turboquant_2025} start by rotating each input vector $\vect{x}$ using a random rotation matrix $\mat{R} \in \operatorname{SO} (d)$ and encode each dimension of $\mat{R} \vect{x}$ using a $b$-bit Lloyd-Max quantizer \cite{lloyd_least_1982,max_quantizing_1960} to create codes with $B = Db$ bits (EDEN uses an additional scalar $s$ per vector that we omit from this footprint calculation for simplicity). The EDEN and TurboQuant quantizers take an additive form
\begin{equation}
    \vect{x}
    \approx
    \operatorname{quant}_{\textsc{ed/tq}}(\vect{x})
    \defeq
    s \cdot 
    \begin{bmatrix}
        \transpose{\vect{w}_{\textsc{LM}}} \operatorname{assign} \left( (\mat{R} \vect{x})_{1} \right) \\
        \vdots \\
        \transpose{\vect{w}_{\textsc{LM}}} \operatorname{assign} \left( (\mat{R} \vect{x})_{D} \right) \\
    \end{bmatrix}
    ,
\end{equation}
where $\vect{w}_{\textsc{LM}} \in \Real^{2^{b}}$ is a vector that contains each value of the one-dimensional Lloyd-Max grid with $b$-bits and $\operatorname{assign}(\vect{x}) \in 2^{b}$ is a one-hot binary vector (thus, $b$ bits are needed to represent it). The MSE version of TurboQuant, which yields its best performance, takes $s = 1$. For EDEN, a few different $s$ options were proposed in \cite{vargaftik_drive_2021,vargaftik_eden_2022}: following an empirical evaluation, we take $s = \norm{\vect{x}}{2} / \norm{\operatorname{quant}_{\textsc{ed/tq}}(\vect{x})}{2}$ as it yields the best results (EDEN has a small header but we omit it in this discussion).

For search, EDEN and TurboQuant approximate the similarity between a query $\vect{q}$ and the indexed vector $\vect{x}$ as
\begin{equation}
    \langle \vect{q}, \vect{x} \rangle
    \approx
    \langle \vect{q}, \operatorname{quant}_{\textsc{ed/tq}}(\vect{x}) \rangle
    =
    s \sum_{j=1}^D q_j \cdot \transpose{\vect{w}_{\textsc{LM}}} \operatorname{assign} \left( (\mat{R} \vect{x})_{j} \right) .
\end{equation}
Since $\operatorname{assign}(\vect{x}^{(m)})$ is a binary one-hot vector, each summand only involves a single gather operation \cite{andre_cache_2015}, similarly to \zcref{eq:masked-add}.

Unlike PQ, the search-time look-up table $\vect{w}_{\textsc{LM}}$ is very small, containing $2^b$ values. Because of this small size, the corresponding gather can be computed in-register using SIMD instructions. Once this gather is complete, the cost of the remaining SIMD operations is equivalent to that of ASH.

EDEN and TurboQuant assume that the data is isotropic. For a set of input vectors $\set{X} = \{ \vect{x}_i \}_{i=1}^{n}$, EDEN and TurboQuant assume that $\vect{x}_i / \norm{\vect{x}_i}{2}$ and $\vect{x}_i$, respectively, are sampled from a uniform distribution on the $D$-dimensional sphere $S^{D-1}$. However, these assumptions do not hold for modern vectors produced by embedding models, see \zcref{tab:spherical_assumption}. Hence, the optimality bounds provided by both methods, which are valid for data following their respective assumptions, are not directly applicable to real-world embedding vectors. In practice, this means that only a fraction of the $2^{Db}$ possible encodings is used, reducing the effectiveness of these methods. ASH, in turn, by using the landmarks to re-center and normalize the data, produces a more uniform distribution on the sphere, which helps drive up its search accuracy.

\begin{table}[t]
    \caption{Vectors produced by embedding models are not isotropically distributed. We take the target set $\set{X} = \{ \vect{x}_i \}_{i=1}^{n}$ and compute $\min_{ij} \operatorname{cosSim}( \vect{x}_i, \vect{x}_j )$ (the minimum cosine similarity) and $\norm{\bar{\vect{\mu}}}{\infty}$, where $\bar{\vect{\mu}} = n^{-1} \sum_i \vect{x}_i$ is the empirical mean. For isotropic data, these values should be close to -1 and 0, respectively, but they are not for standard embedding vectors.}
    \label{tab:spherical_assumption}

    \small
    \centering
    \begin{tblr}{
        colspec = {lSS},
        rowspec = {|Q|QQQQQ|},
    }
        Dataset & {$\displaystyle \min_{ij} \operatorname{cosSim}( \vect{x}_i, \vect{x}_j )$} & {$\norm{\bar{\vect{\mu}}}{\infty}$} \\
        ada002-100k & -0.104 & 0.659 \\
        openai-1536-100k & -0.200 & 0.064 \\
        openai-3072-100k & -0.175 & 0.053 \\
        gecko-100k & 0.221 & 0.123 \\
        nv-qa-v4-100k & -0.021 & 0.160 \\
    \end{tblr}
    
\end{table}

\textbf{ASH versus RaBitQ.}
RaBitQ \cite{gao_rabitq_2024,gao_practical_2025} is a particular instantiation of our model where $D = d$, $b=1$, $C=1$, and $\mat{W} = \mat{R}$ for a random orthogonal matrix $\mat{R} \in \Real^{D \times D}$. As such, both models have many similarities from a computational and efficiency perspective. However, ASH enables reducing the dimensionality, learning the projection from data. This has two immediate consequences. First, we can target higher compression ratios with ASH (i.e., lower values of $B$). Second, as pointed out in \zcref{sec:ash_error}, reducing the dimension while expanding the bitrate $b$ per dimension enables regimes with better tradeoffs of speed and accuracy.

\textbf{ASH versus LOPQ.}
Locally-optimized product quantization (LOPQ) \cite{kalantidis_locally_2014} is one of the most powerful additive quantizers and shares key features with ASH. LOPQ clusters the data with a vector quantizer (i.e., a coarse quantizer \cite{jegou_product_2011}). Then, for each cluster, the residuals are encoded using PQ augmented with a rotation matrix. The objective of the additional rotation is to split the space along the principal axes of the data within each cluster. For each cluster, the optimization becomes
\begin{equation}
    \min_{\mat{R}, \mat{W}_{\textsc{pq}}}
    \sum_{\vect{x}}
    \norm{\vect{x} - \mat{R} \operatorname{quant}_{\textsc{pq}}(\vect{x})}{2}^2
    \quad\text{s.t.}\quad
    \mat{R} \in \operatorname{St} (D, D) ,\,
    \mat{W}_{\textsc{pq}} \in \set{D} ,
    \label{eq:lopq}
\end{equation}
where $\operatorname{quant}_{\textsc{pq}}$ and $\mat{W}_{\textsc{pq}}  \in \Real^{D \times (2^b \cdot M)}$ are defined in \zcref{eq:pq_representation} and $\set{D}$ is the set of block-diagonal matrices with $M$ blocks in $\Real^{(D/M) \times 2^b}$. The problem can be optimized by alternating optimization, which involves alternating between $M$ k-means problems to find $\mat{W}_{\textsc{pq}}$ and computing the SVD of $\sum_{\vect{x}} \vect{x} \, \transpose{\operatorname{quant}_{\textsc{pq}}(\vect{x})}$ to find $\mat{R}$. In the LOPQ paper, the authors acknowledge the practicality limitations of this expensive optimization \cite{kalantidis_locally_2014}.

In ASH, we also use a coarse quantizer as the first step, learn a common rotation matrix $\mat{R}$ for all clusters (similar to OPQ \cite{ge_optimized_2014}), and swap the PQ representation with scalar quantization. We originally tried using multiple rotation matrices in ASH, one per cluster, but opted for using a single one for two main reasons. First, learning would become much slower with multiple rotation matrices, having to solve $C$ independent optimization problems. Second, and most importantly, similarity computations would gain a significant overhead as we would need to perform $C$ matrix multiplications with $C$ different matrices $\mat{W}$. In ASH, we perform a single matrix multiplication with the common matrix $\mat{W}$.

\textbf{ASH versus LeanVec.}
LeanVec \cite{tepper_leanvec_2024} uses dimensionality reduction with scalar quantization and can handle in-distribution and out-of-distribution queries. The in-distribution version is the most similar technique to ASH, and we will only refer to it henceforth. The dimensionality reduction in LeanVec is done with an SVD. Its scalar quantization uses LVQ \cite{aguerrebere_similarity_2023}, which quantizes each vector $\vect{x}$ individually by computing the quantization range $[\min(\vect{x}), \max(\vect{x})]$.

The main drawback in LeanVec is LVQ. The min-max scaling inflates norms in unpredictable ways, i.e., $\norm{\vect{x}}{2} \leq \norm{\operatorname{quant}_{\textsc{lvq}}(\vect{x})}{2}$, which causes distortions in dot product similarities. This effect is mild for $b=4$ and above, but more pronounced for lower bitrates. Additionally, LeanVec only incorporates the quantization as a post-processing step, which means that it cannot refine the initial PCA using the quantizer, as demonstrated in \zcref{sec:optimization}.

\section{Experimental Results}
\label{sec:experiments}

\begin{table}[t]
    \caption{Datasets used for our experimental results. We denote the dimensionality,
    processed database size, and query count by $D$, $n$, and $q$, respectively.}
    \label{tab:datasets}
    
    \centering
    \small
    \begin{tblr}{
        colspec = {lS[table-format=4]S[table-format=8]S[table-format=5]},
        rowspec = {|Q|QQQQQ|QQQQQQ|},
        row{1} = {c},
    }
        Dataset & {$D$} & {$n$} & {$q$} \\

        gecko-100k \cite{datastax_jvector_2026}              & 768  & 100000 & 10000 \\
        nv-qa-v4-100k \cite{datastax_jvector_2026}           & 1024 & 100000 & 10000 \\
        ada002-100k \cite{datastax_jvector_2026}             & 1536 & 100000 & 10000 \\
        openai-1536-100k \cite{qdrant_dbpedia-entities-openai3-text-embedding-3-large-1536-1m_2024}
        & 1536 & 100000 & 1000 \\
        openai-3072-100k \cite{qdrant_dbpedia-entities-openai3-text-embedding-3-large-3072-1m_2024}        & 3072 & 100000 & 1000 \\
        
        ada002-1m \cite{datastax_jvector_2026}               & 1536 & 982790 & 10000 \\
        cap-1m \cite{justicedao_caselaw_access_project_embeddings_nodate}                  & 1536 & 1000000 & 10000 \\
        cohere-1m \cite{coherelabs_msmarco-v2-embed-english-v3_nodate}               & 1024 & 1000000 & 10000 \\
        mpnet-1m \cite{olmer_wiki_mpnet_embeddings_nodate}                & 768  & 999812 & 10000 \\
        openai-1536-1m \cite{qdrant_dbpedia-entities-openai3-text-embedding-3-large-3072-1m_2024}          & 1536 & 999000 & 1000 \\
        openai-3072-1m \cite{qdrant_dbpedia-entities-openai3-text-embedding-3-large-3072-1m_2024}          & 3072 & 999000 & 1000 \\
    \end{tblr}
\end{table}

We use the datasets described in Table~\ref{tab:datasets}, which originate from different embedding models and types of data. Unless explicitly specified, we learn ASH from data using the techniques in \zcref{sec:optimization}.
Throughout this section, search accuracy is measured as 10-recall@R, where R is the number of retrieved elements.

\begin{figure*}[p]
    \centering

    \includegraphics[width=0.30\linewidth]{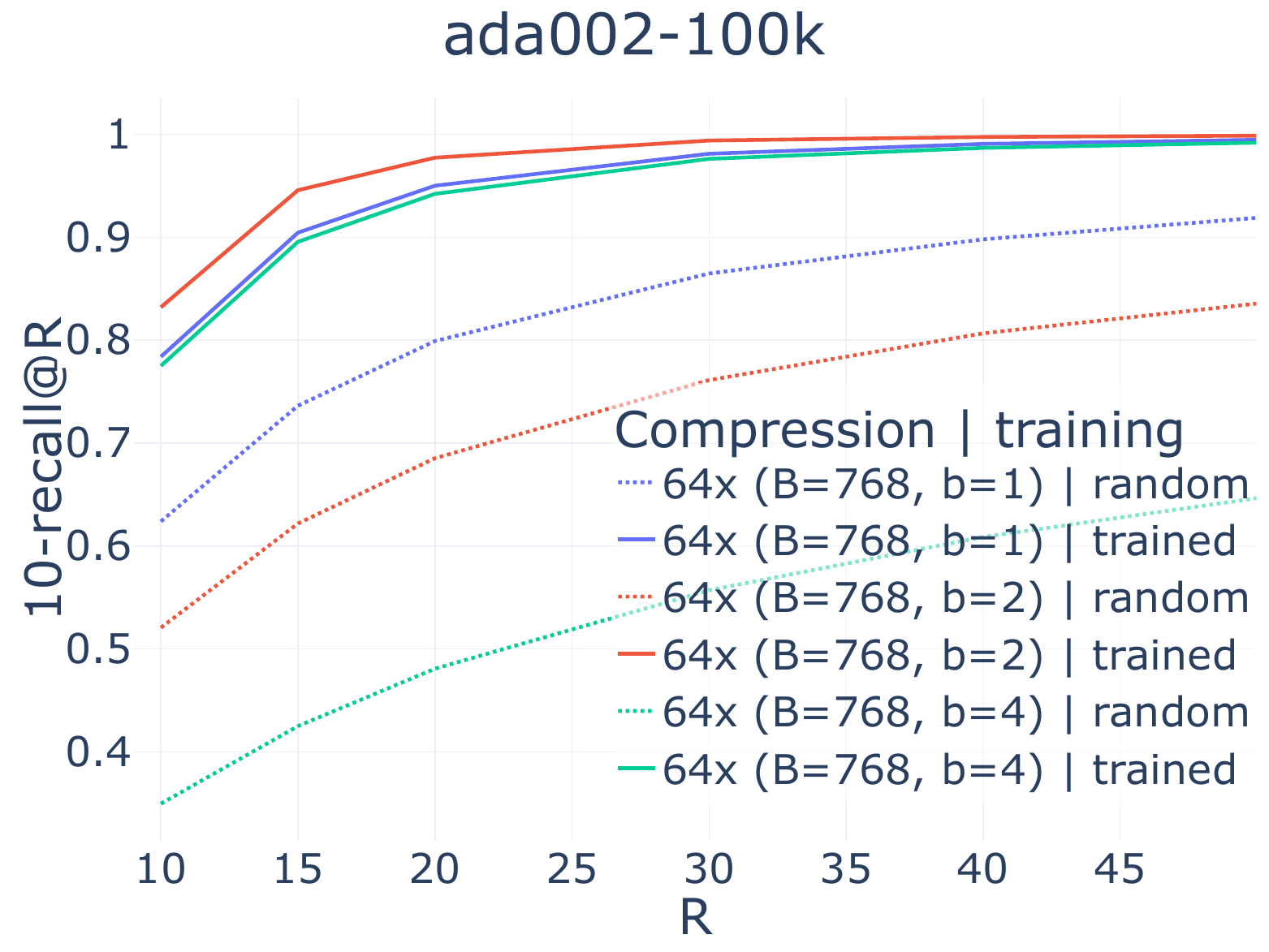}%
    \includegraphics[width=0.30\linewidth]{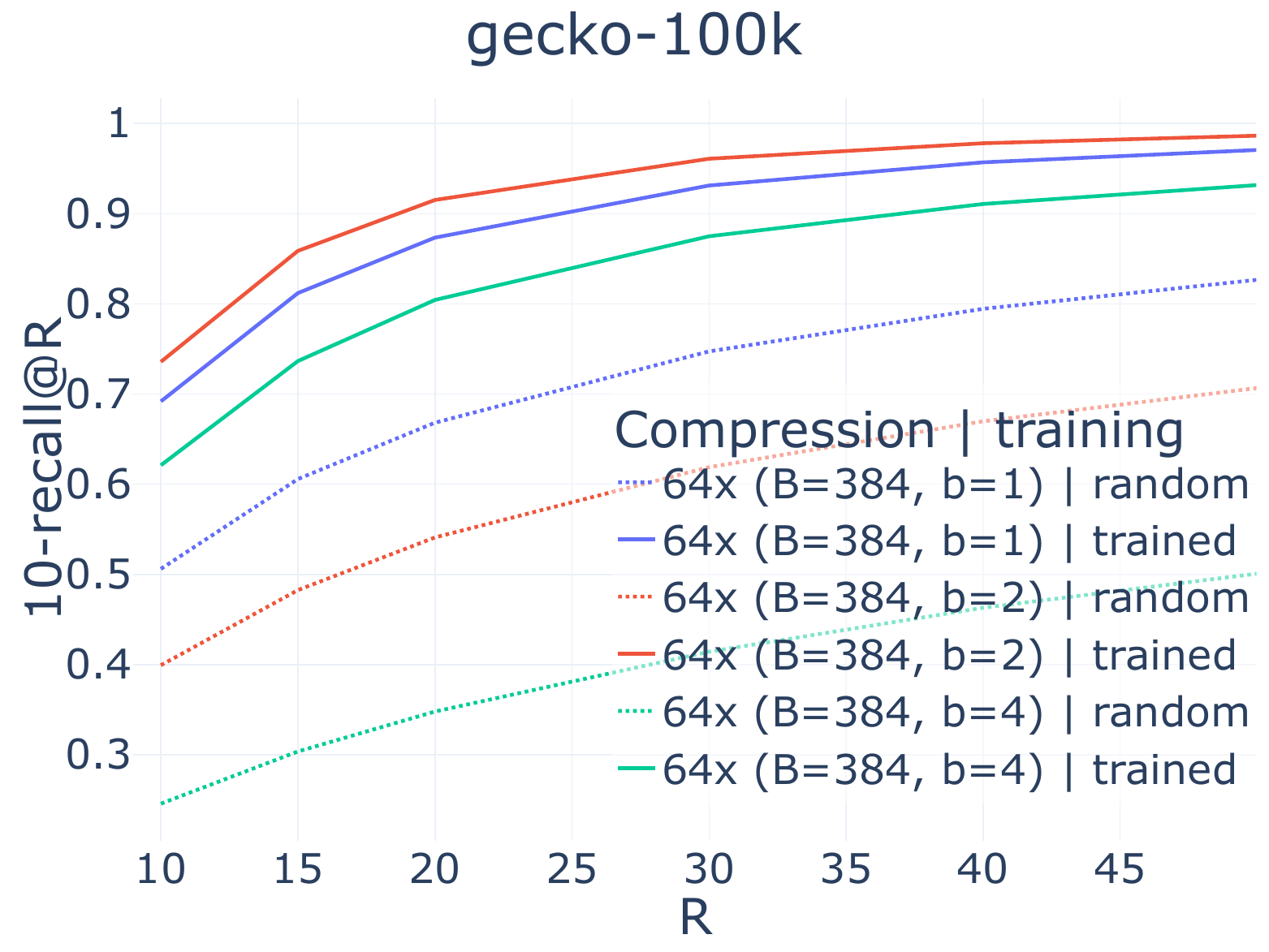}%
    \includegraphics[width=0.30\linewidth]{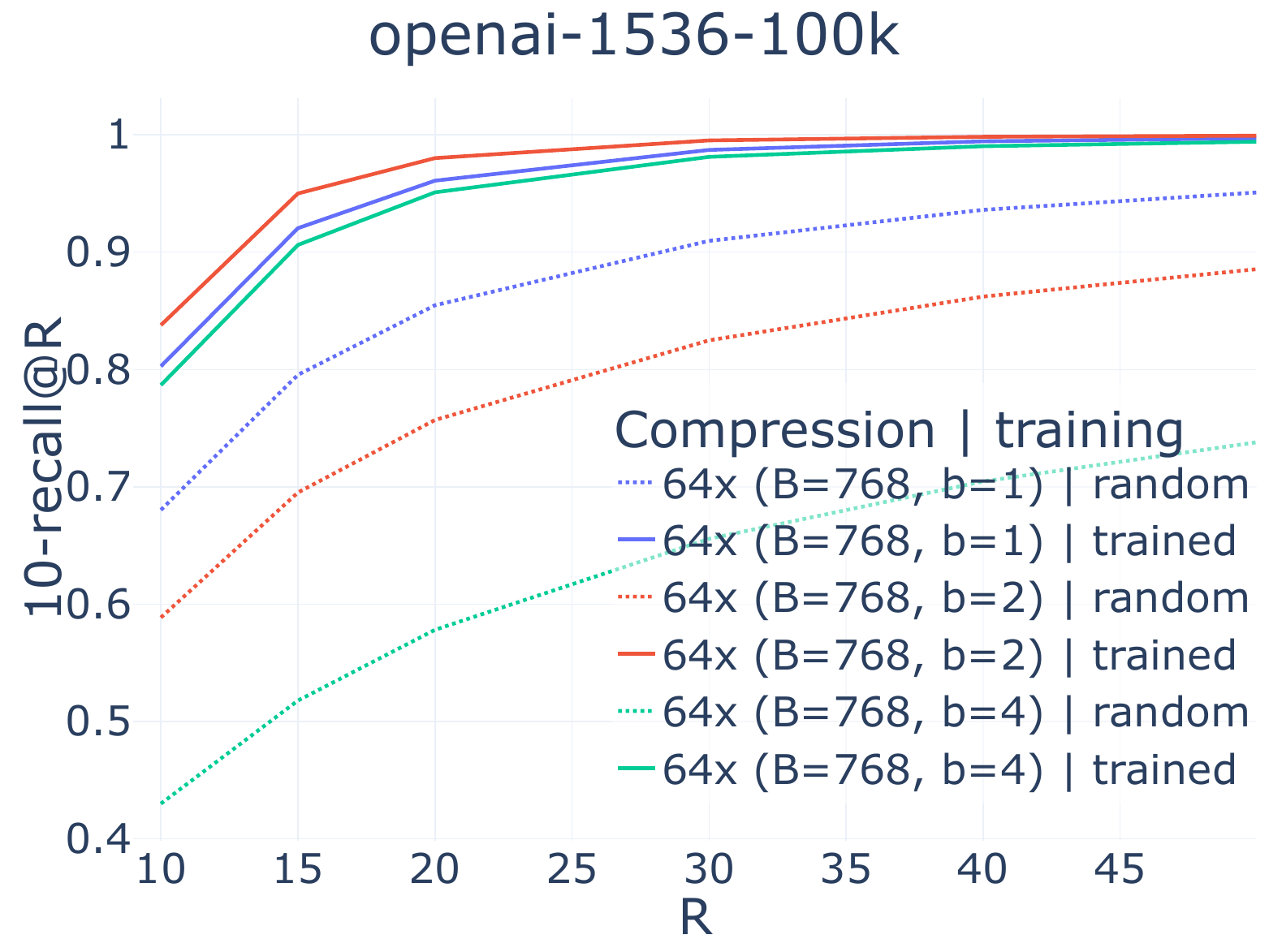}%
    
    \includegraphics[width=0.30\linewidth]{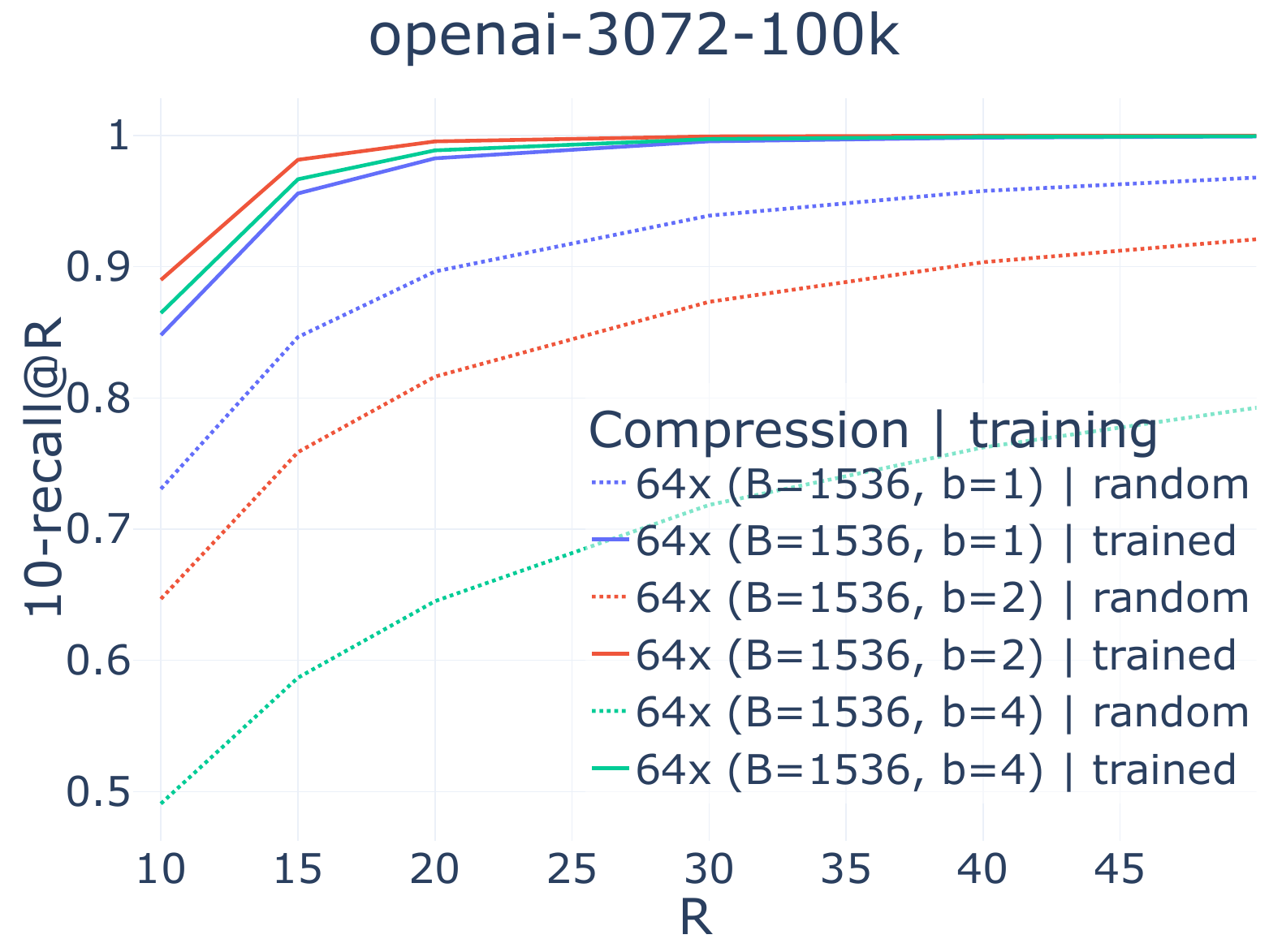}%
    \includegraphics[width=0.30\linewidth]{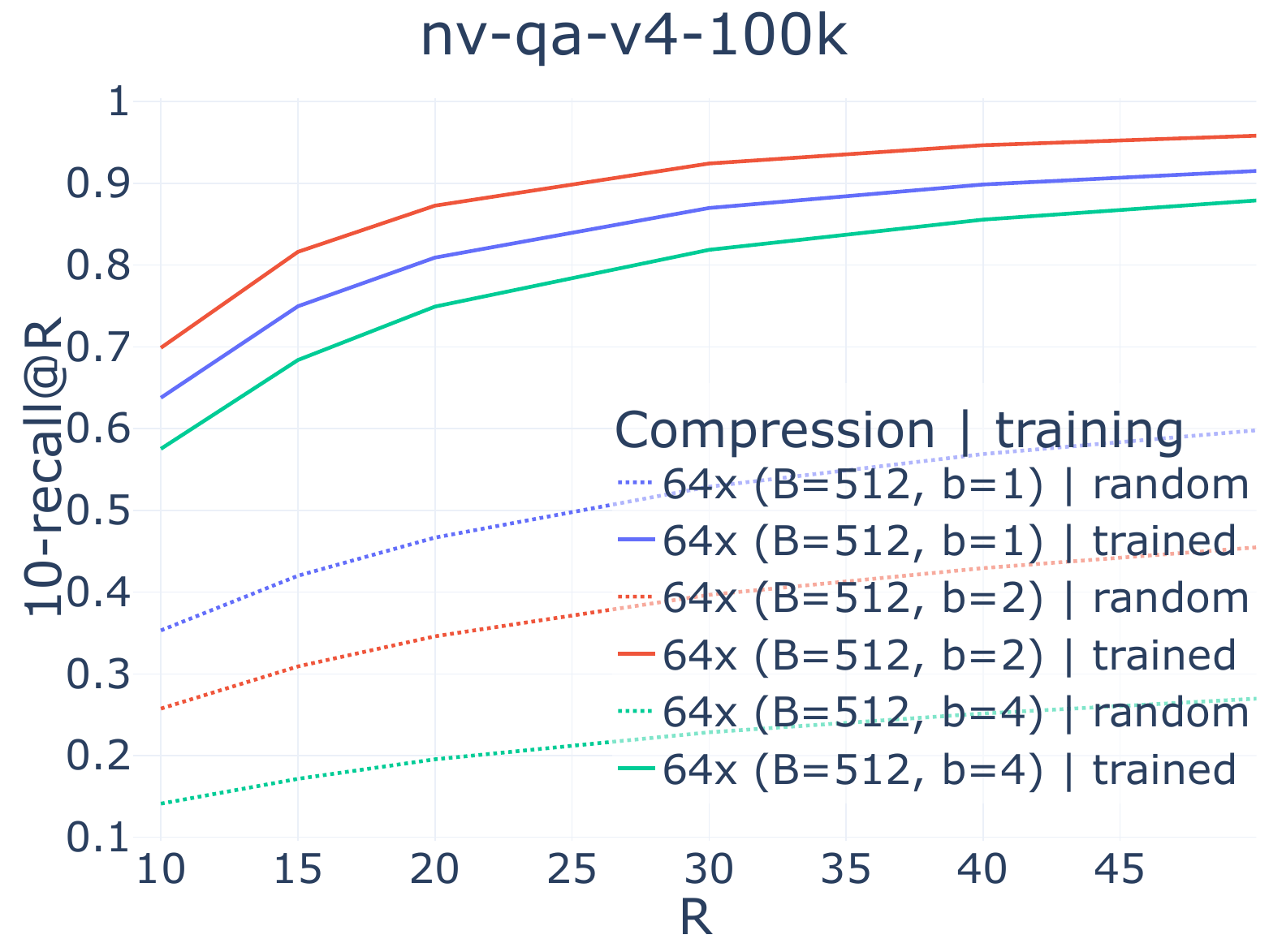}%

    \caption{Learning the projection matrix $\mat{W}$ leads to significant improvements in search accuracy (10-recall@R) for $B=D/2$. In ASH, increasing the bitrate $b$ with $B$ fixed means decreasing the target dimensionality $d$. When $D > d$, the advantage of the learned parameters becomes wider. 
    Notably, ASH with $b=2$ consistently beats $b=1$, meaning that reducing the dimensionality while increasing the bitrate pays off ($B=D$ and $B=D/4$ in \zcref{fig:ash_training_continued} of the appendix).}
    \label{fig:ash_training}
\end{figure*}

\begin{figure}
    \centering
    \includegraphics[width=0.3\linewidth]{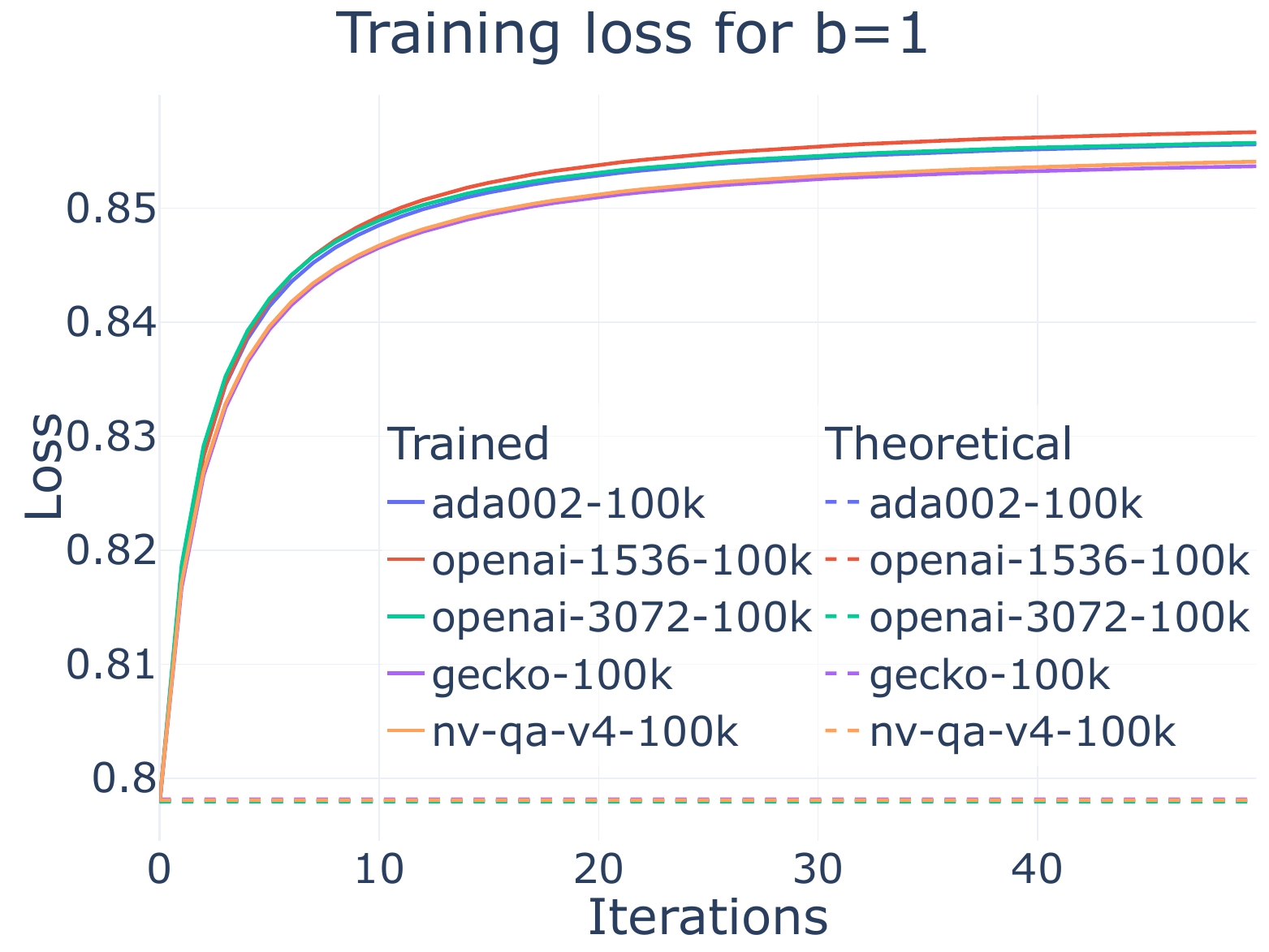}%
    \hspace{2em}
    \includegraphics[width=0.3\linewidth]{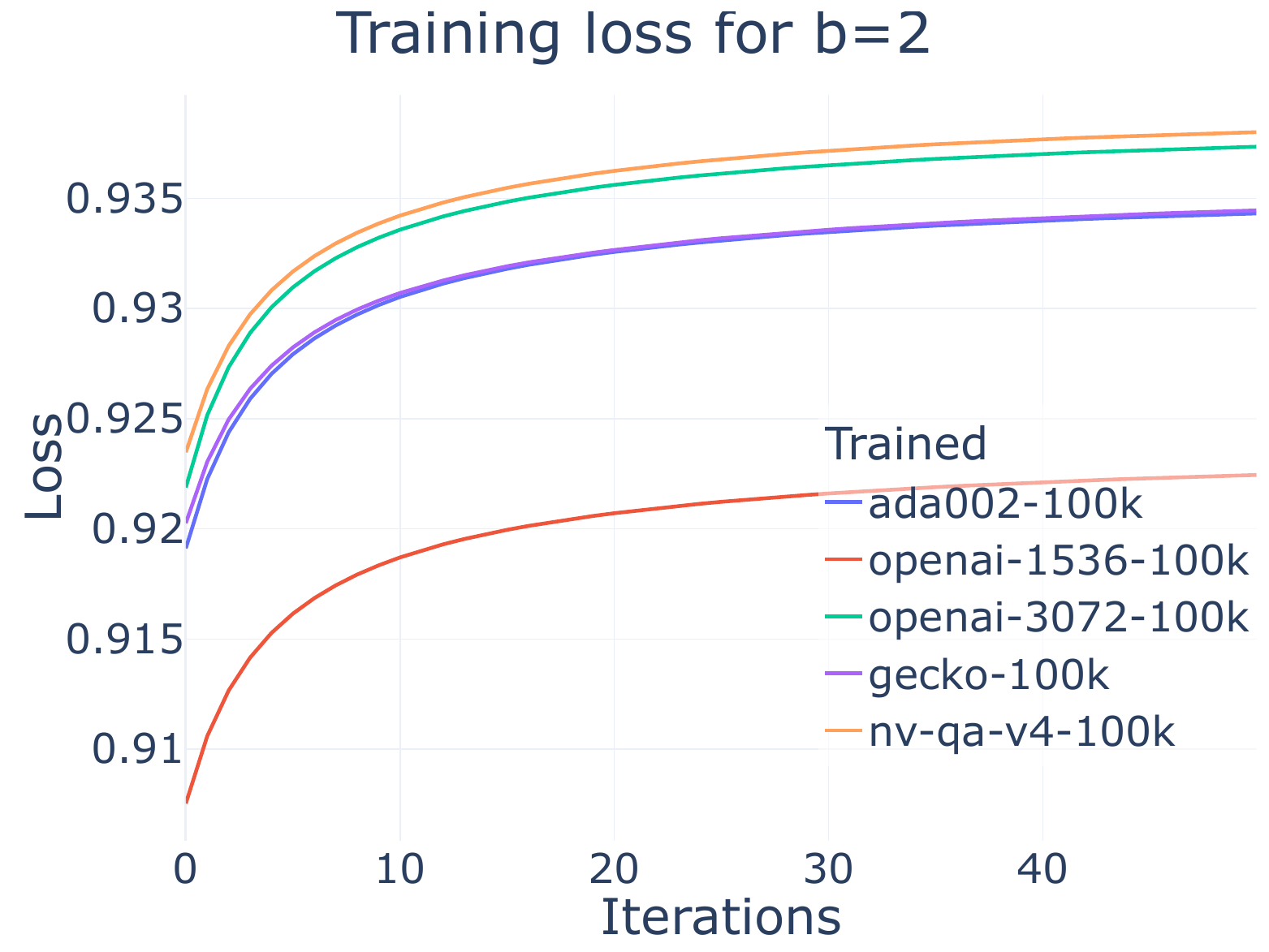}%
    
    \caption{The algorithm presented in \zcref{sec:optimization} converges as its iterations progress. Tracking the loss in \zcref{prob:learning_loss}, we observe that after 20-30 iterations, we start getting diminishing returns. For $B=D$ and $b=1$, we can compare the obtained results with the expected loss in \zcref{eq:expected_rabitq_loss} \cite{gao_rabitq_2024}, observing clear improvements.}
    \label{fig:trained_convergence}
\end{figure}

\textbf{The advantage of learning the quantizer.}
We start by showing that the parameterization and learning algorithm presented in \zcref{sec:optimization} produce better results than data-agnostic dimensionality reduction techniques. For this, we compare the results with those of a Johnson-Lindenstrauss projection, obtained by drawing a random matrix $\mat{W} \in \operatorname{St} (d, D)$. In \zcref{fig:ash_training}, we clearly observe that the learned regime outperforms the random one with gains that increase with the gap between $D$ and $d$.
Furthermore, we observe in \zcref{fig:trained_convergence} that the learning algorithm brings clear improvements over its SVD-plus-random initialization.

When $b=1$, the reconstruction error is $\| \vect{x} - \operatorname{quant}_1(\mat{W} \vect{x}) \|_2^2 = 2 - 2 \langle \vect{x} , \operatorname{quant}_1(\mat{W}\vect{x}) \rangle$ for $\vect{x} \in S^{D-1}$.
Additionally, when $D = d$, $C=1$, and $\mat{W} = \mat{R}$ for a random orthogonal matrix $\mat{R} \in \operatorname{SO} (D)$, the expectation of the dot product $\langle \vect{x} , \operatorname{quant}_1(\mat{W}\vect{x}) \rangle$ over $\mat{R}$ is \cite{gao_rabitq_2024}
\begin{equation}
    \expectationFromDist{\langle \vect{x} , \operatorname{quant}_1 (\mat{W}\vect{x}) \rangle}{\mat{R}}
    =
    2 \sqrt{D / \pi} \frac{\Gamma(D/2)}{(D-1) \Gamma((D-1)/2)} ,
    \label{eq:expected_rabitq_loss}
\end{equation}
where $\Gamma$ is the Gamma function. Its value is approximately 0.798 for $D \approx 1000$ (we include its plot in \zcref{fig:rabitq_expected_loss} of the appendix).
In the left plot of \zcref{fig:trained_convergence}, we show that in the RaBitQ \cite{gao_rabitq_2024} regime ($D = d$ and $C=1$), ASH provides clear improvements over RaBitQ in terms of reconstruction error and search accuracy.

\begin{figure}[t]
    \centering
    \includegraphics[width=0.3\linewidth]{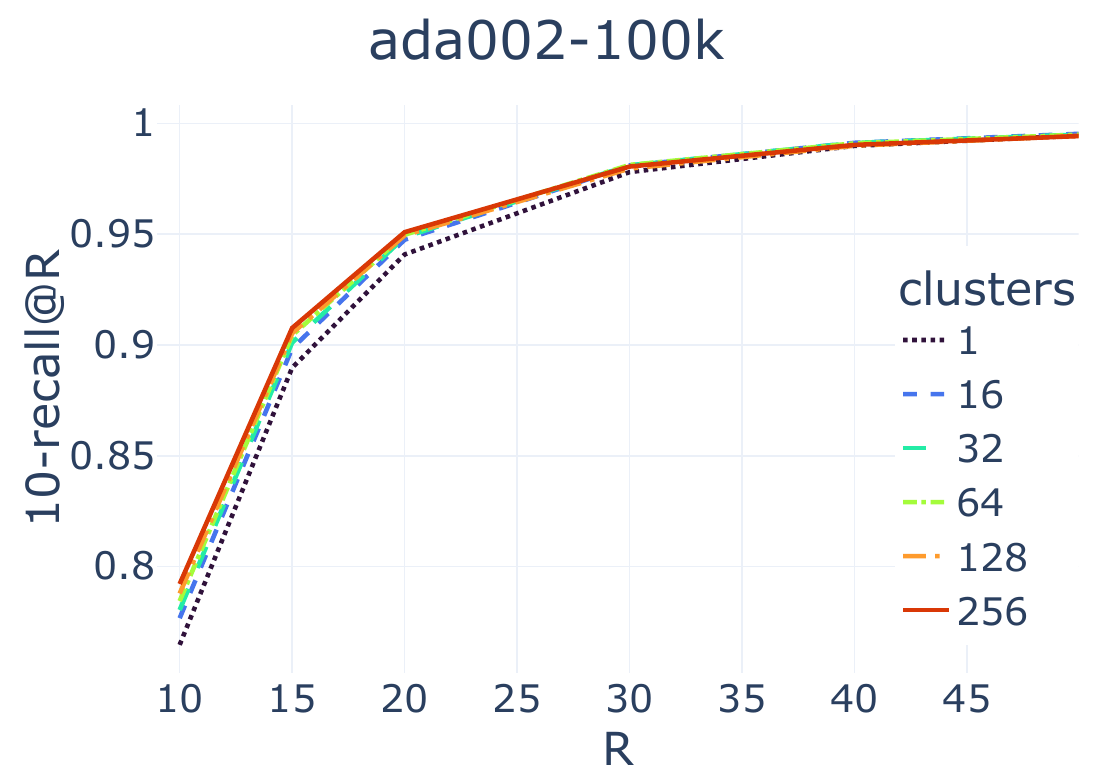}%
    \hspace{2em}
    \includegraphics[width=0.3\linewidth]{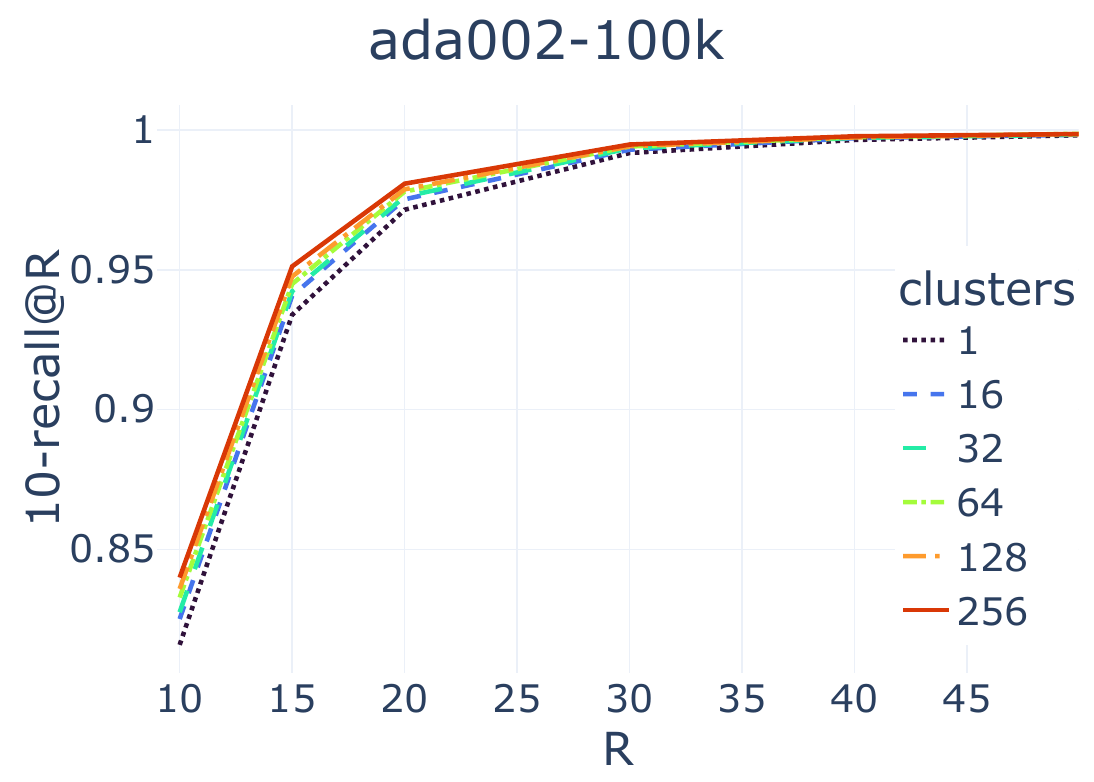}%

    \caption{The search accuracy (10-recall@R) increases with the number of ASH landmarks, defined in \zcref{eq:c_star} for $B = D / 2$, $b=1$ (left), and $b=2$ (right). Additional datasets are included in \zcref{fig:ash_groups_continued} of the appendix.}
    \label{fig:ash_groups}
\end{figure}

\textbf{The number of landmarks.}
The search accuracy of ASH improves with an increasing number of landmarks, as observed in \zcref{fig:ash_groups}. As mentioned earlier, this local data centering is a form of coarse quantization, which is known to whiten the error and improve the quantization \cite{jegou_product_2011,tepper_gleanvec_2024}. In \zcref{fig:ash_groups}, we observe that beyond $C=64$, we start hitting diminishing returns in terms of search accuracy.

\begin{figure}
    \centering
    \includegraphics[width=0.25\linewidth]{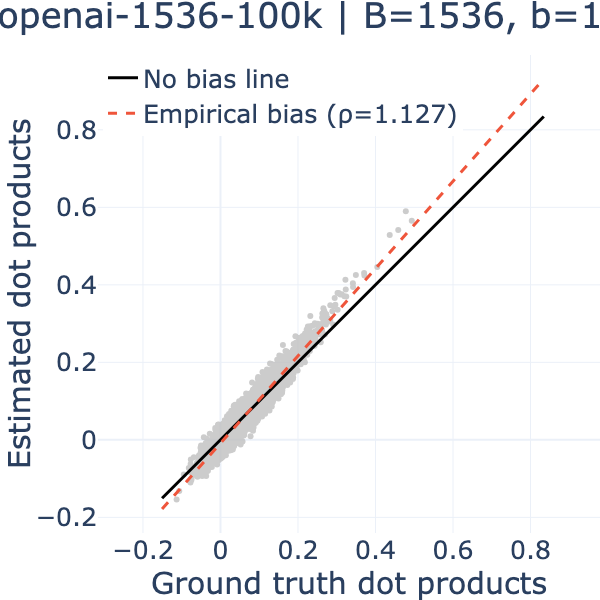}%
    \includegraphics[width=0.25\linewidth]{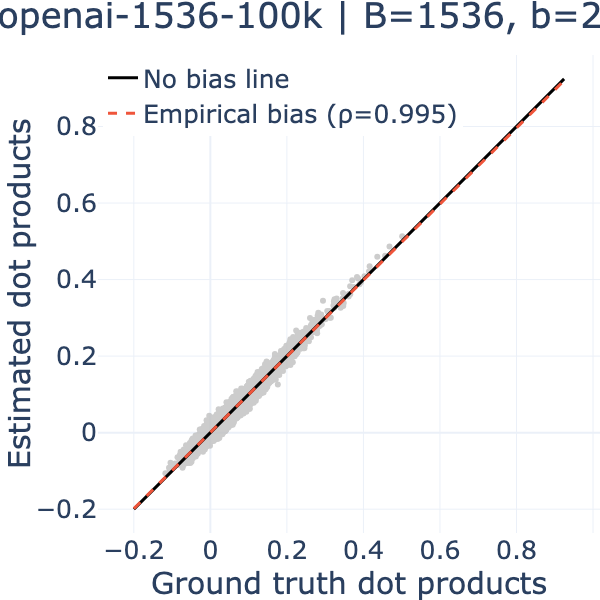}%
    \includegraphics[width=0.25\linewidth]{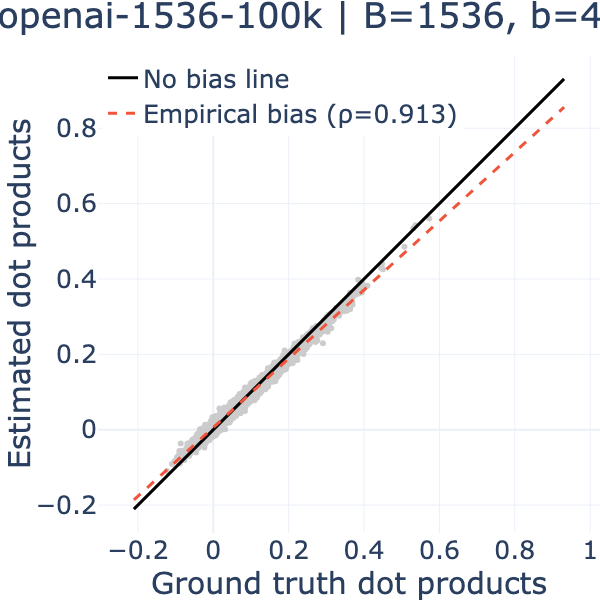}%
    
    \caption{The ASH estimator has a slight bias, see \zcref{eq:regression_coefficient_bias}, when comparing the ground truth $\langle \vect{q}, \vect{x} \rangle$ with the estimated $\langle \vect{q}, \operatorname{quant}(\vect{x}) \rangle$ (each pair $(\langle \vect{q}, \vect{x} \rangle, \langle \vect{q}, \operatorname{quant}(\vect{x}) \rangle)$ corresponds to a circle in the plot) using the dataset queries. The bias slope $\rho$ depends on the number of bits $B$ and the bitrate $b$ (in these examples $d = \lfloor (B - 32) / b \rfloor$); from left to right $b=1, 2, 4$. In this case, $B = D$. Additional datasets and configurations are included in \zcref{fig:ash_bias_continued,fig:ash_bias_continued2} of the appendix.}
    \label{fig:ash_bias}
\end{figure}

\textbf{The estimator bias.}
Interested in analyzing any systematic biases introduced by the ASH estimator, we plot the two-dimensional pairs $(x, y)$ formed by $x = \langle \vect{q}, \vect{x} \rangle$ and $y = \langle \vect{q}, \operatorname{quant}(\vect{x}) \rangle$ in \zcref{fig:ash_bias} for different values of $b$ and $d$. We can observe that the pairs follow a linear trend, which indicates that ASH did not introduce any aberrant distortions. 
We use the linear regression
\begin{equation}
    \min_{\rho, \beta}
    \sum_{\substack{
        \vect{q} \in \set{Q} \\
        \vect{x} \in \set{X}
    }}
    \norm{\rho \langle \vect{q}, \vect{x} \rangle + \beta - \langle \vect{q}, \hat{\vect{x}} \rangle}{2}^2
    \label{eq:regression_coefficient_bias}
\end{equation}
to unveil the linear trend in the distribution of the pairs.
In \zcref{fig:ash_bias}, the distribution exhibits a slight bias (its slope is different from 1). Interestingly, since ASH preserves the vector norms exactly, the bias comes entirely from the vector direction. The bias depends on $b$: for $b=1$ and $b=4$, we observe larger deviations than for $b=2$. For $b=1$, $\rho > 1$ while for $b=4$, $\rho < 1$. We leave the investigation of this interesting phenomenon for future work.

Finally, we point out that a linear bias does not affect MIP search as multiplying $\operatorname{quant}(\vect{x})$ by a global scalar does not change the order of the dot products. ANN powered by the Euclidean distance, for example, can be affected by a linear bias. In this case, the bias can be removed using the linear regression coefficients estimated from a small sample of query and index vectors (\textasciitilde 100 of each).

\textbf{Reducing the precision of the query.}
We study the necessity of keeping $\breve{\vect{q}}$, \zcref{eq:query_low_dim}, at full precision. For this, we downcast the values in $\breve{\vect{q}}$ to float16. In this lower-precision format, we can load 32 elements per iteration in \zcref{code:ash_dot_product} instead of just 16, increasing its efficiency. We evaluate the difference in search accuracy (10-recall@R) between both formats, as shown in \zcref{tab:ash_query_float16}. The differences are in the order of $10^{-5}$, which is negligible considering that recall is a quantity in $[0, 1]$. In the performance experiments, we use bfloat16 with similar effects.

\begin{table}[t]
    \caption{The maximum absolute 10-recall@R difference between queries in float32 or float16 format (in the latter case, the values are simply downcast) with ASH ($C=64$). The full 10-recall@R difference curves are available in \zcref{fig:ash_query_fp16} of the appendix. We can observe minute differences in a quantity that lies in $[0, 1]$.}
    \label{tab:ash_query_float16}

    \small
    \centering
    \begin{tblr}{
        colspec = {l *{2}{Q[si={table-format=1.1e1,round-mode=places},c]}},
        rowspec = {|Q|QQQQQ|},
    }
        Dataset & {$b=1$} & {$b=2$} \\
        ada002-1536-100k & 4e-5 & 4e-5 \\
        openai-v3-large-1536-100k & 4e-5 & 6e-5 \\
        openai-v3-large-3072-100k & 4e-5 & 3e-5 \\
        gecko-768-100k & 4e-5 & 6e-5 \\
        nv-qa-v4-100k & 1.79e-4 & 2.3e-4 \\
    \end{tblr}
\end{table}

\begin{figure*}[p]
    \centering
    
    \includegraphics[width=0.3\linewidth]{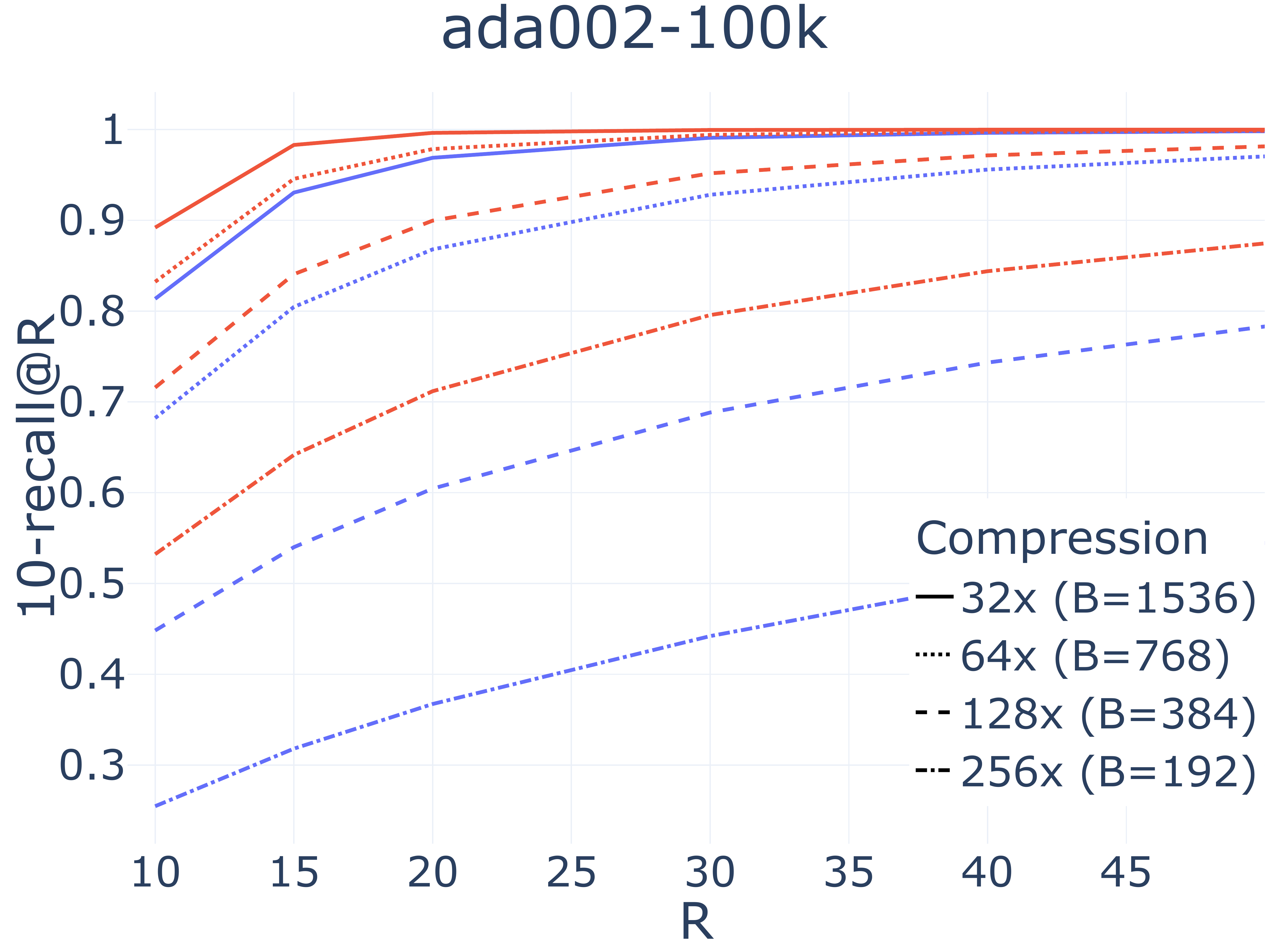}%
    \includegraphics[width=0.3\linewidth]{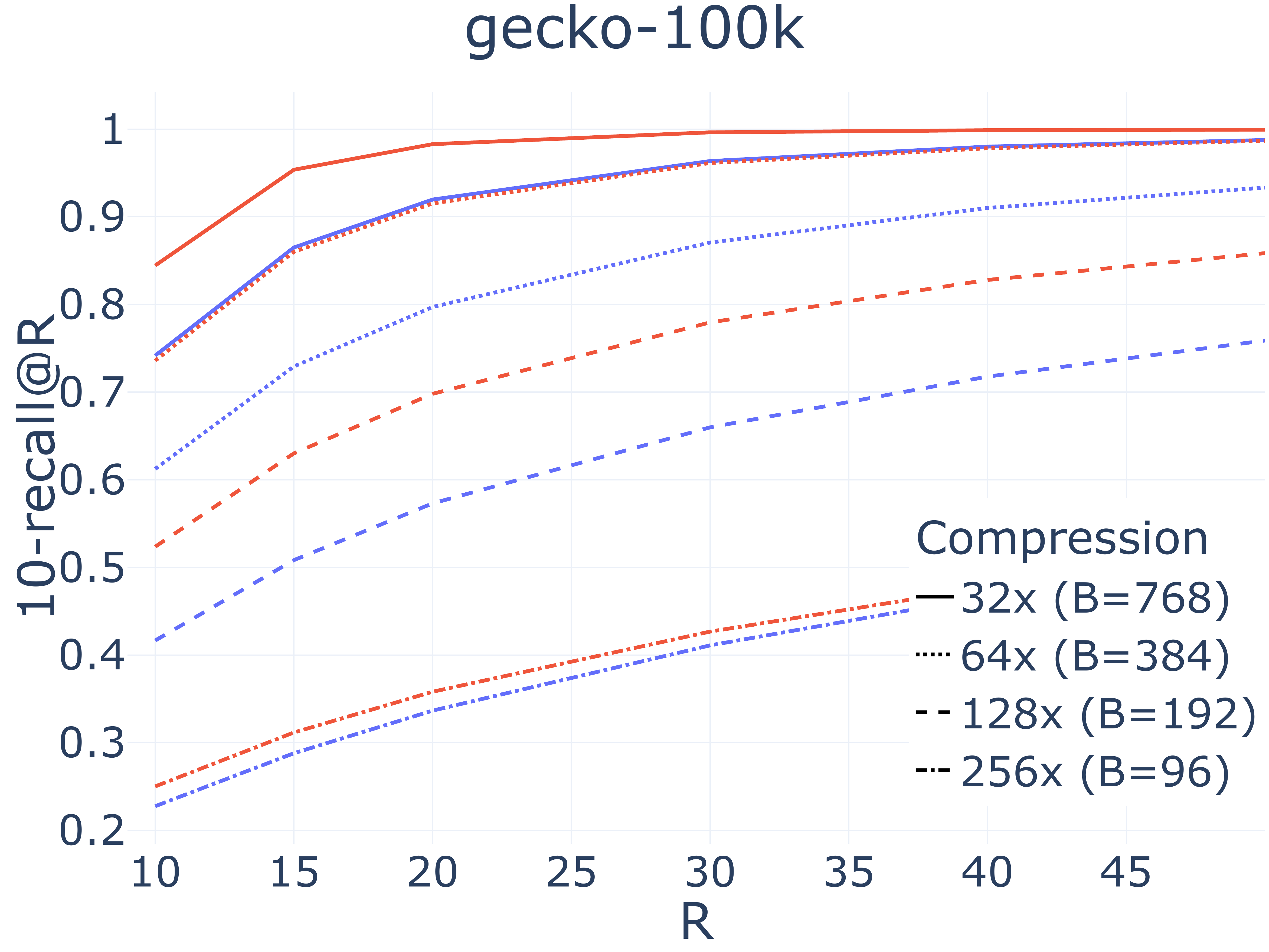}%
    \includegraphics[width=0.3\linewidth]{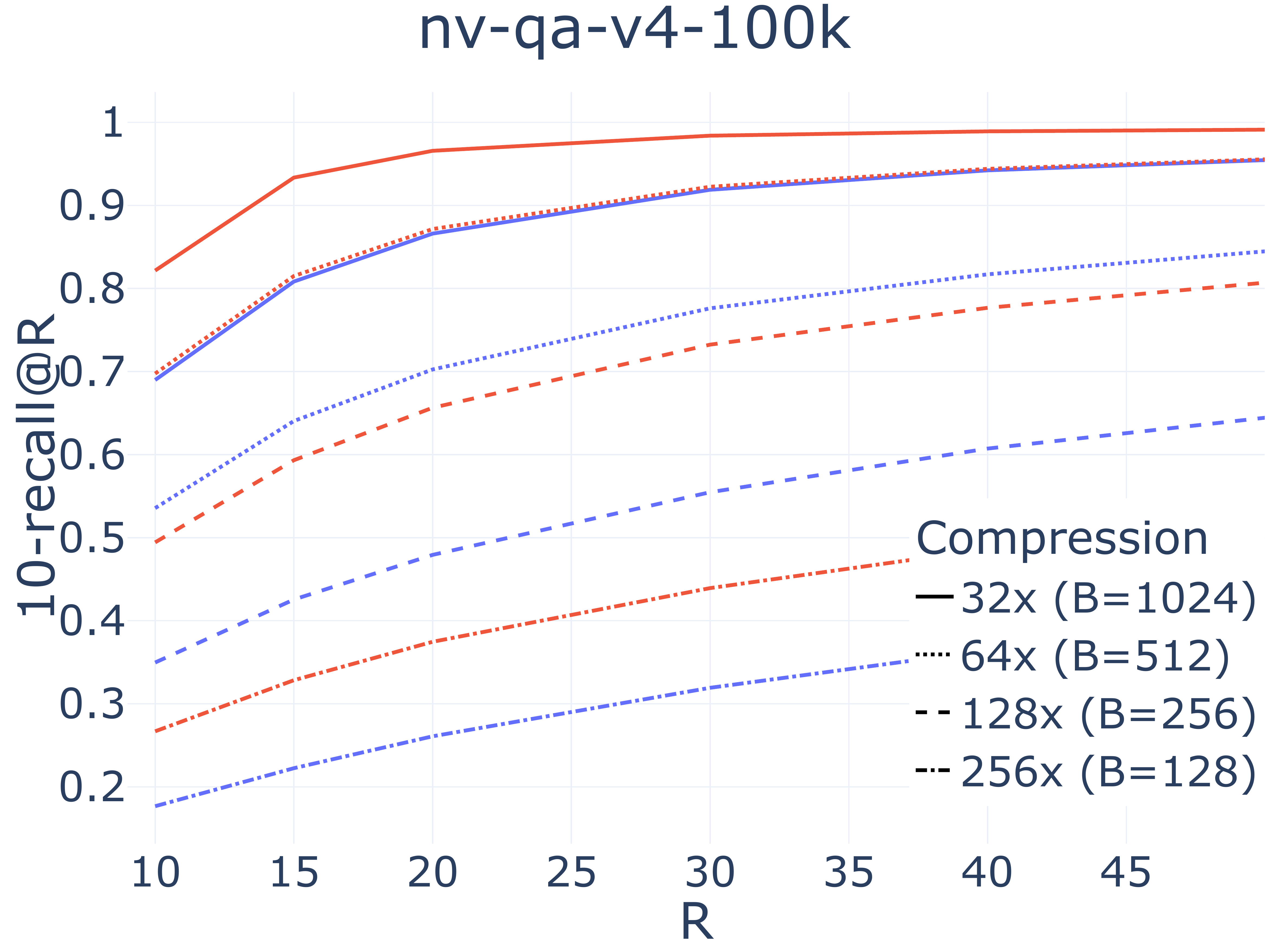}%
    
    \includegraphics[width=0.3\linewidth]{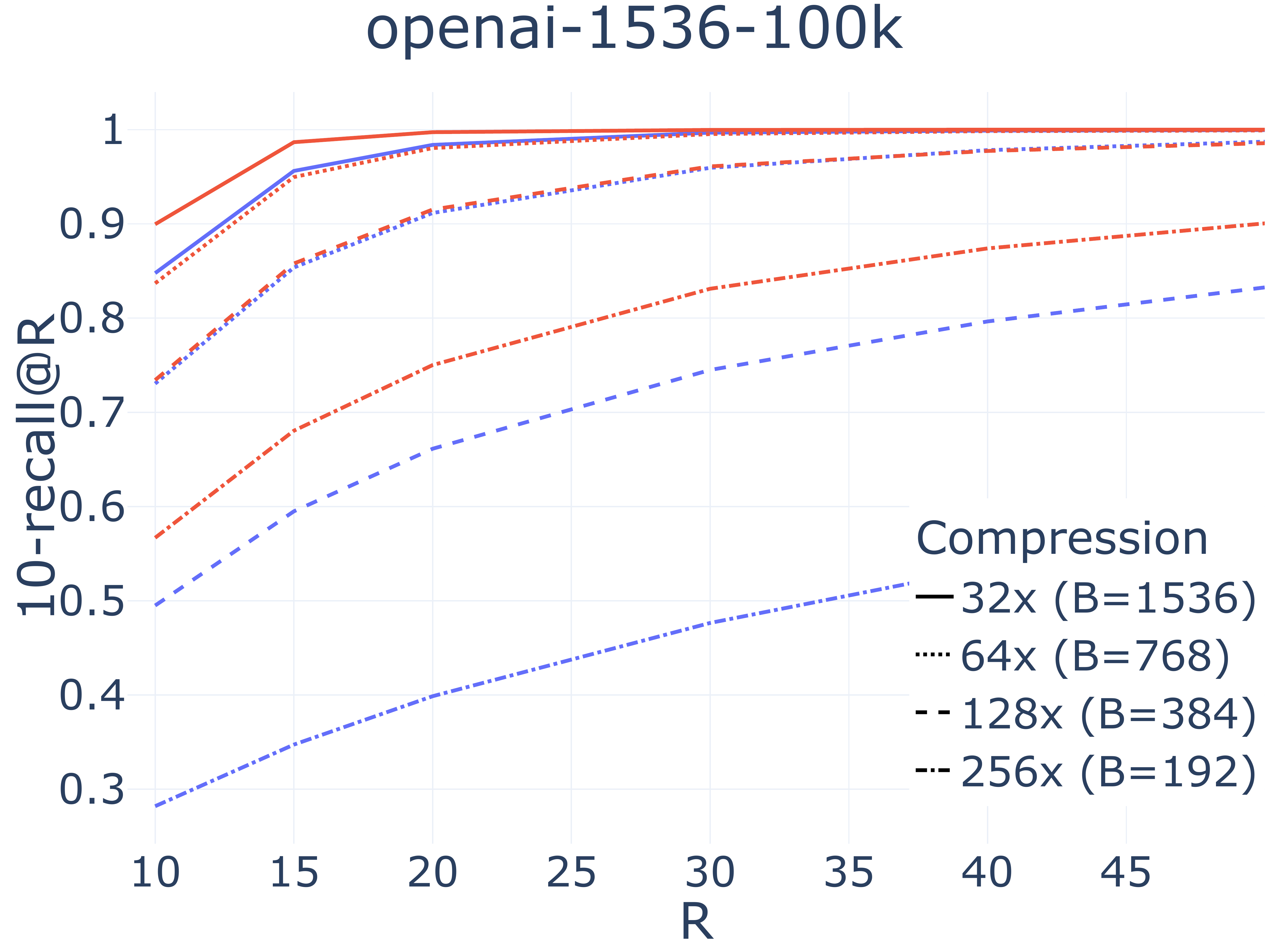}%
    \includegraphics[width=0.3\linewidth]{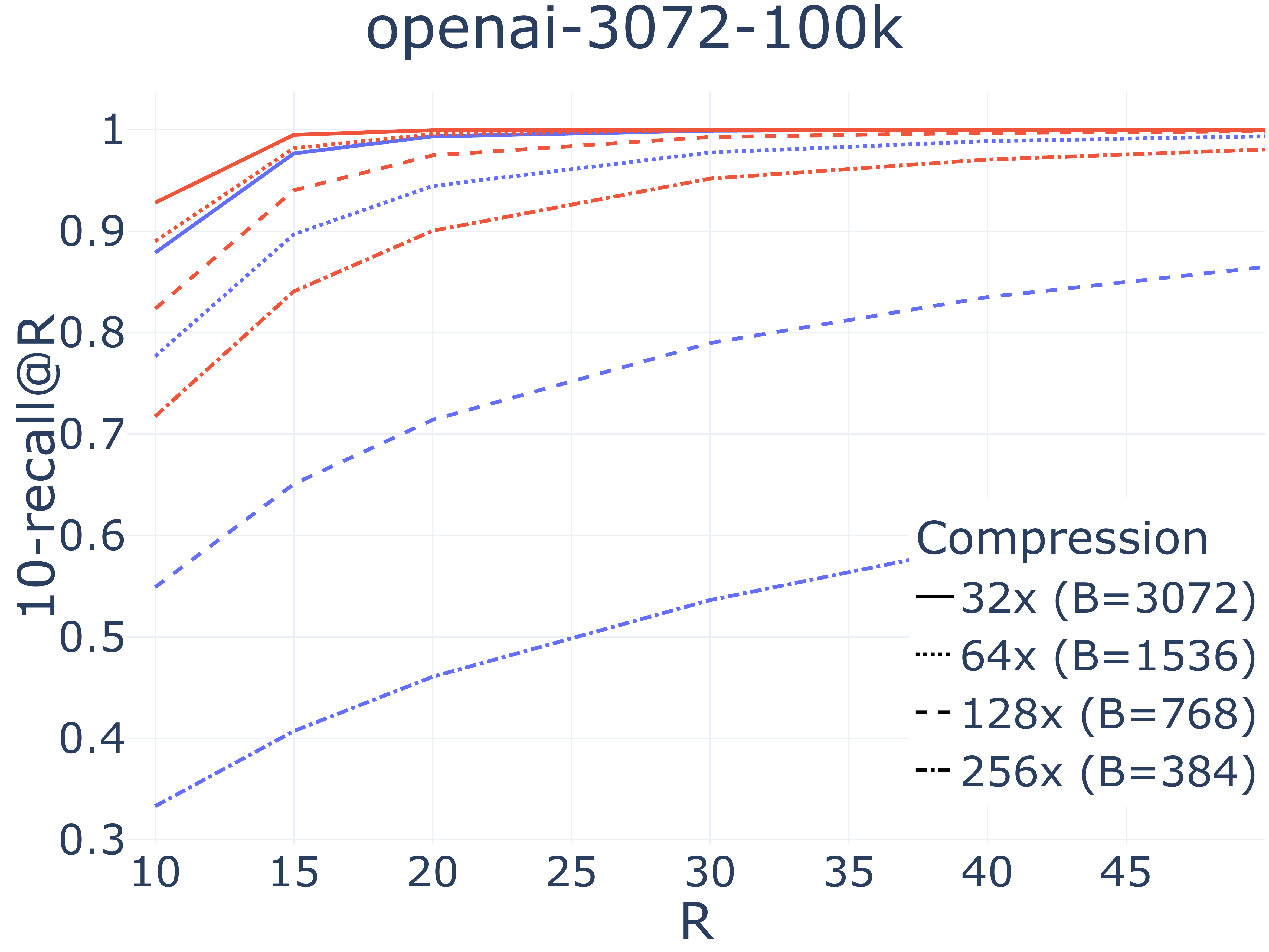}%

    \includegraphics[width=0.15\linewidth]{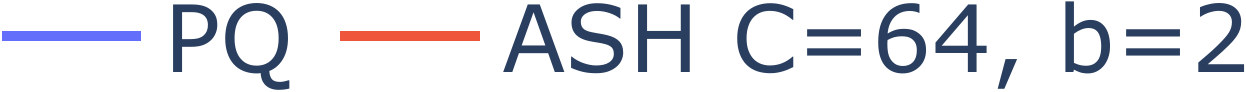}
    
    \caption{ASH outperforms PQ in search accuracy (10-recall@R). ASH is even competitive at several compression levels with PQ configurations that use twice the space (additional datasets in \zcref{fig:ash_vs_pq_continued} of the appendix).}
    \label{fig:ash_vs_pq}
\end{figure*}

\textbf{ASH versus PQ.} We show in \zcref{fig:ash_vs_pq} that ASH outperforms PQ \cite{jegou_product_2011} in search accuracy. In this comparison, ASH uses $C=64$ landmarks, whereas PQ does not employ a coarse quantizer, i.e., $C=1$. The effect of $C$ on ASH's accuracy is modest, as observed in \zcref{fig:ash_groups}, so our results remain valid even at matched $C=1$. ASH exhibits search accuracy higher than that of PQ at iso-compression, and in several cases, ASH is even competitive with PQ while using half the number of bits.

\begin{figure*}[p]
    \centering
    \includegraphics[width=0.3\linewidth]{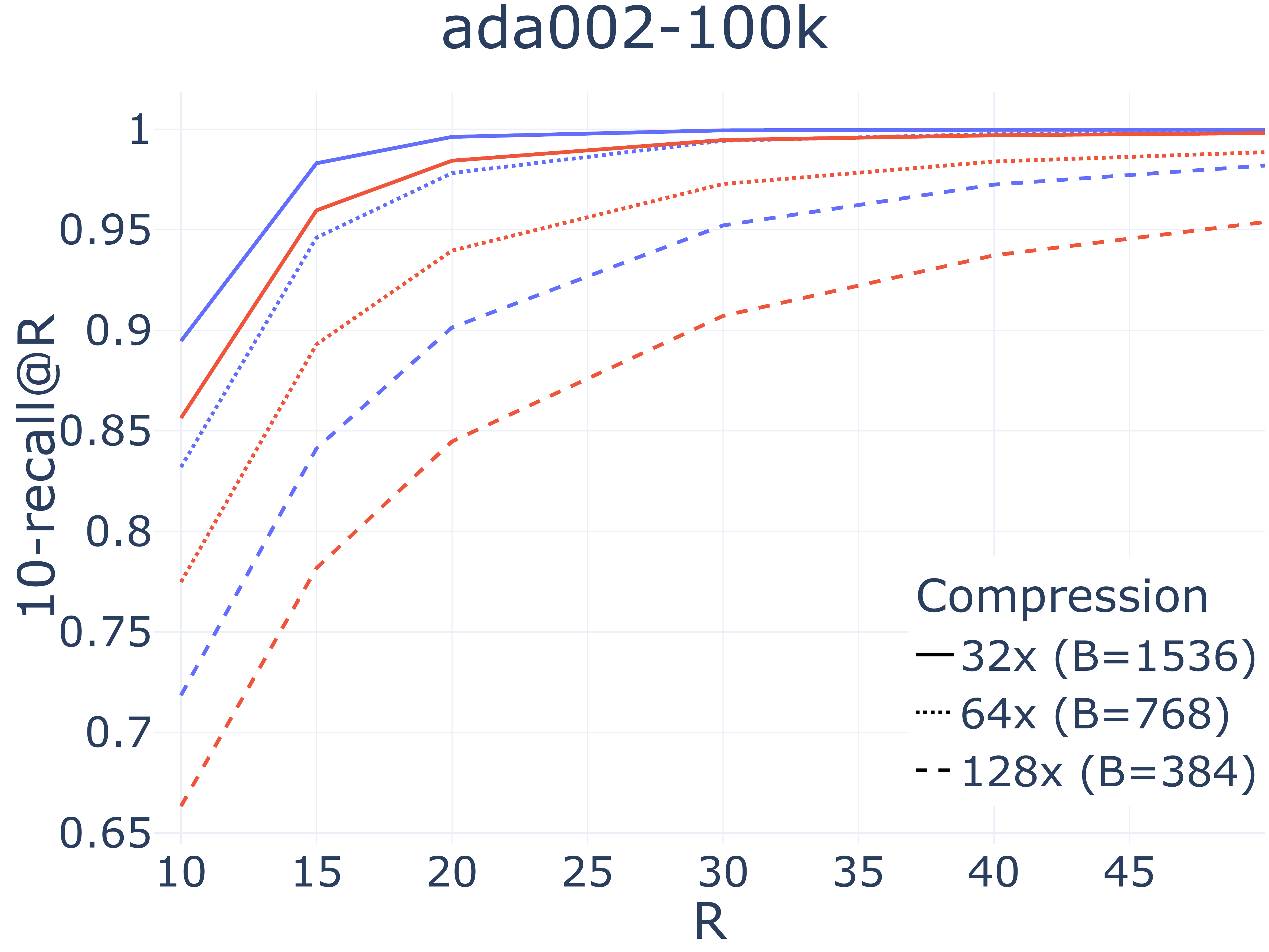}%
    \includegraphics[width=0.3\linewidth]{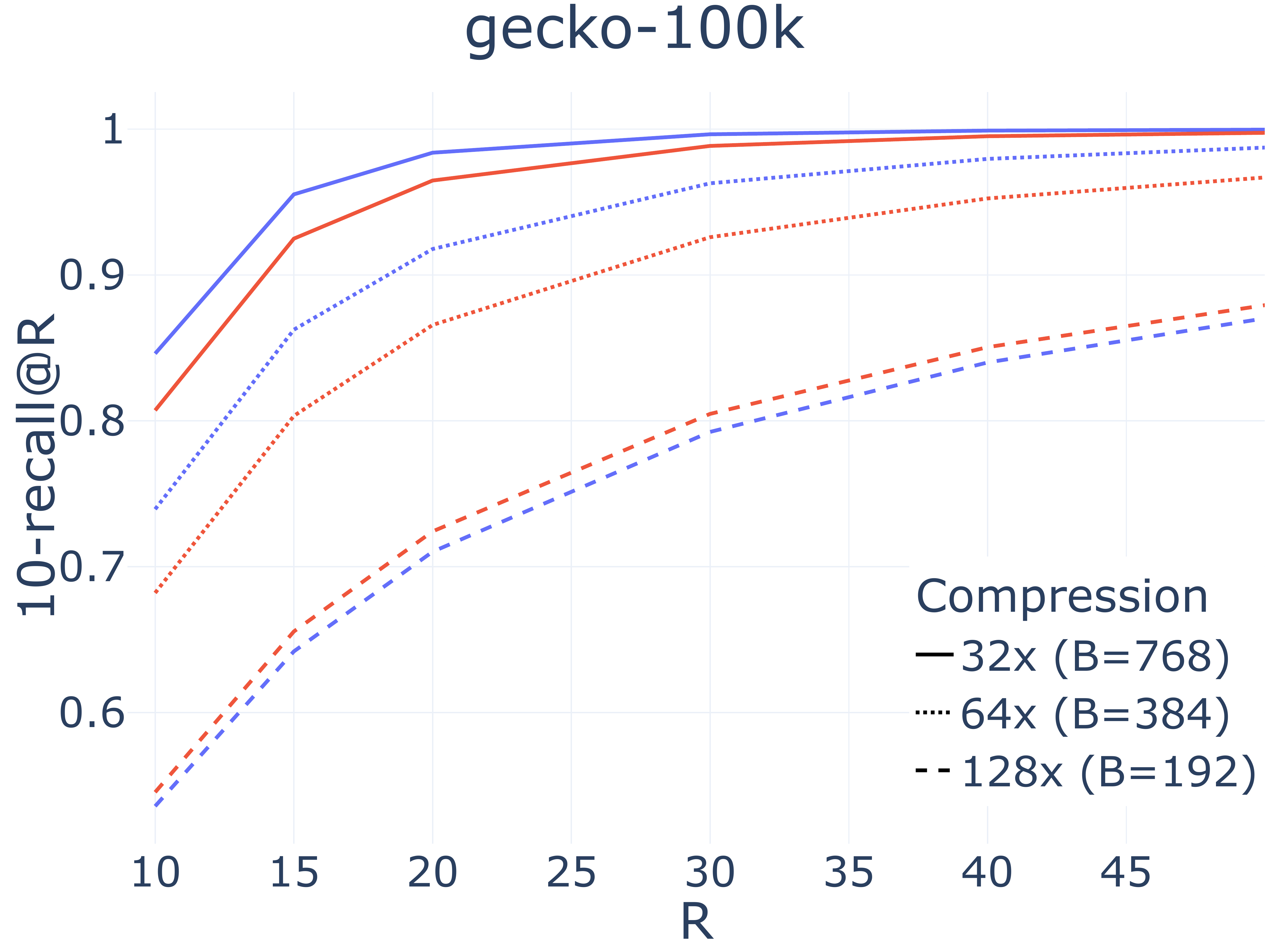}%
    \includegraphics[width=0.3\linewidth]{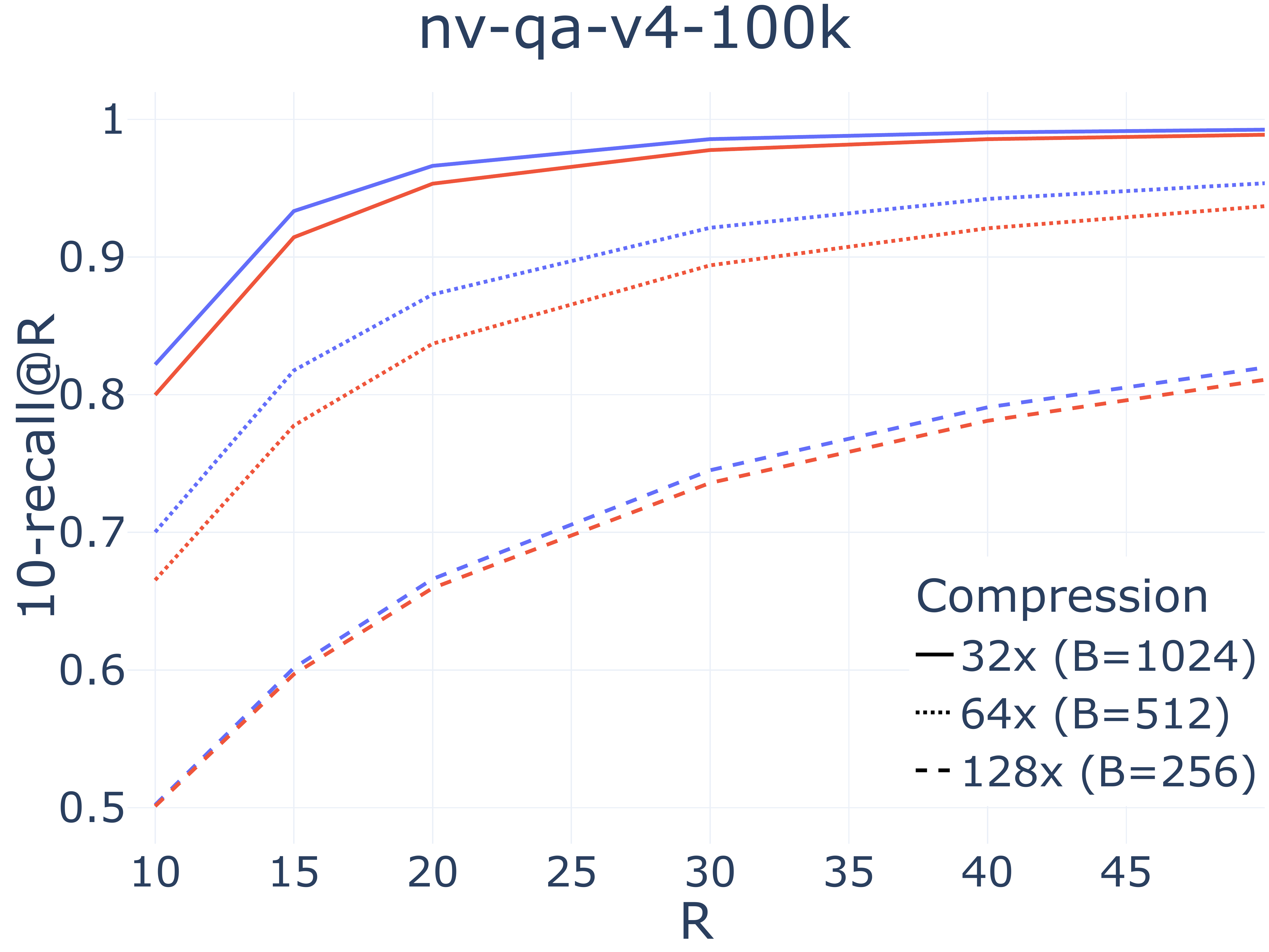}%
    
    \includegraphics[width=0.3\linewidth]{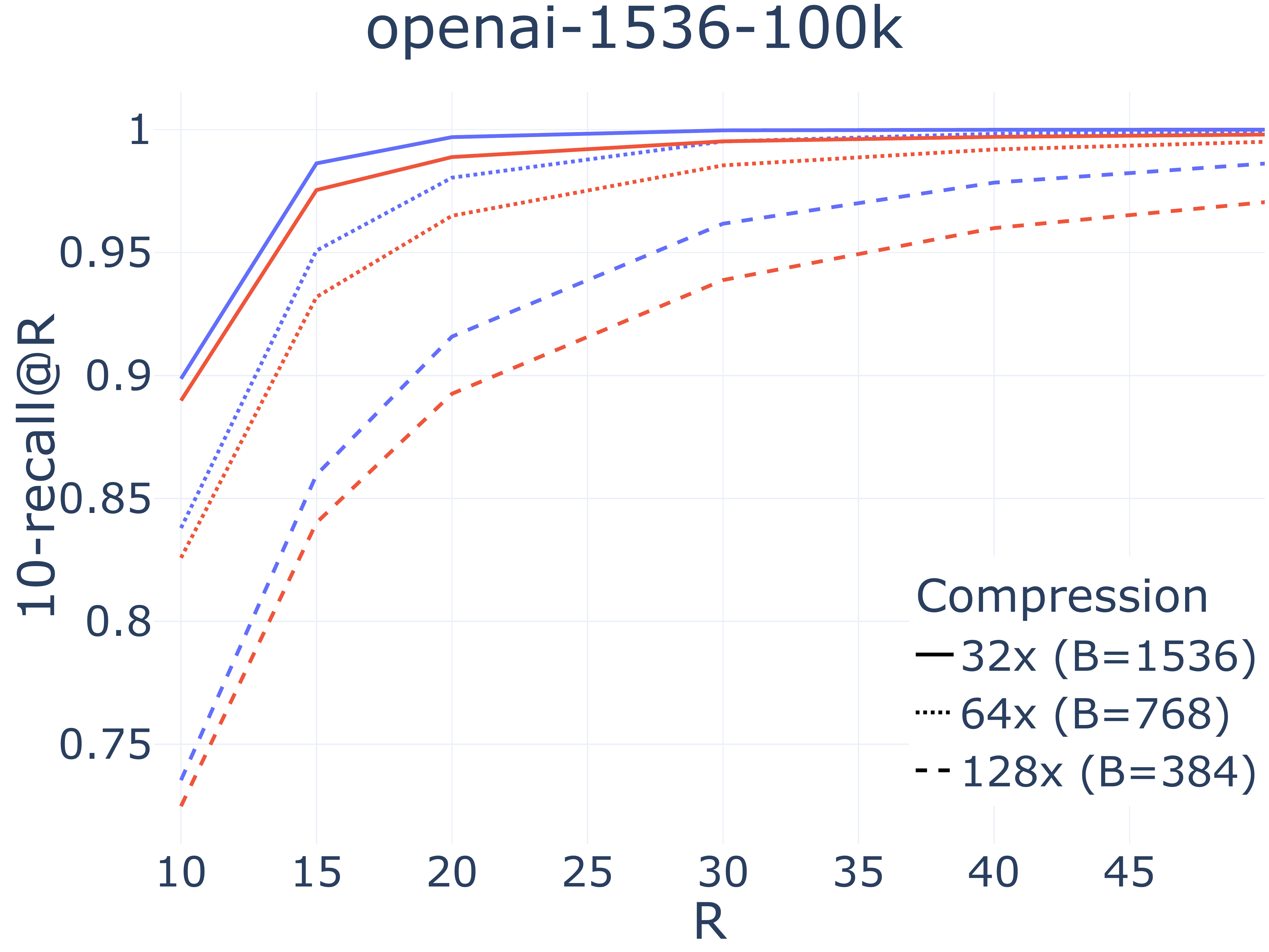}%
    \includegraphics[width=0.3\linewidth]{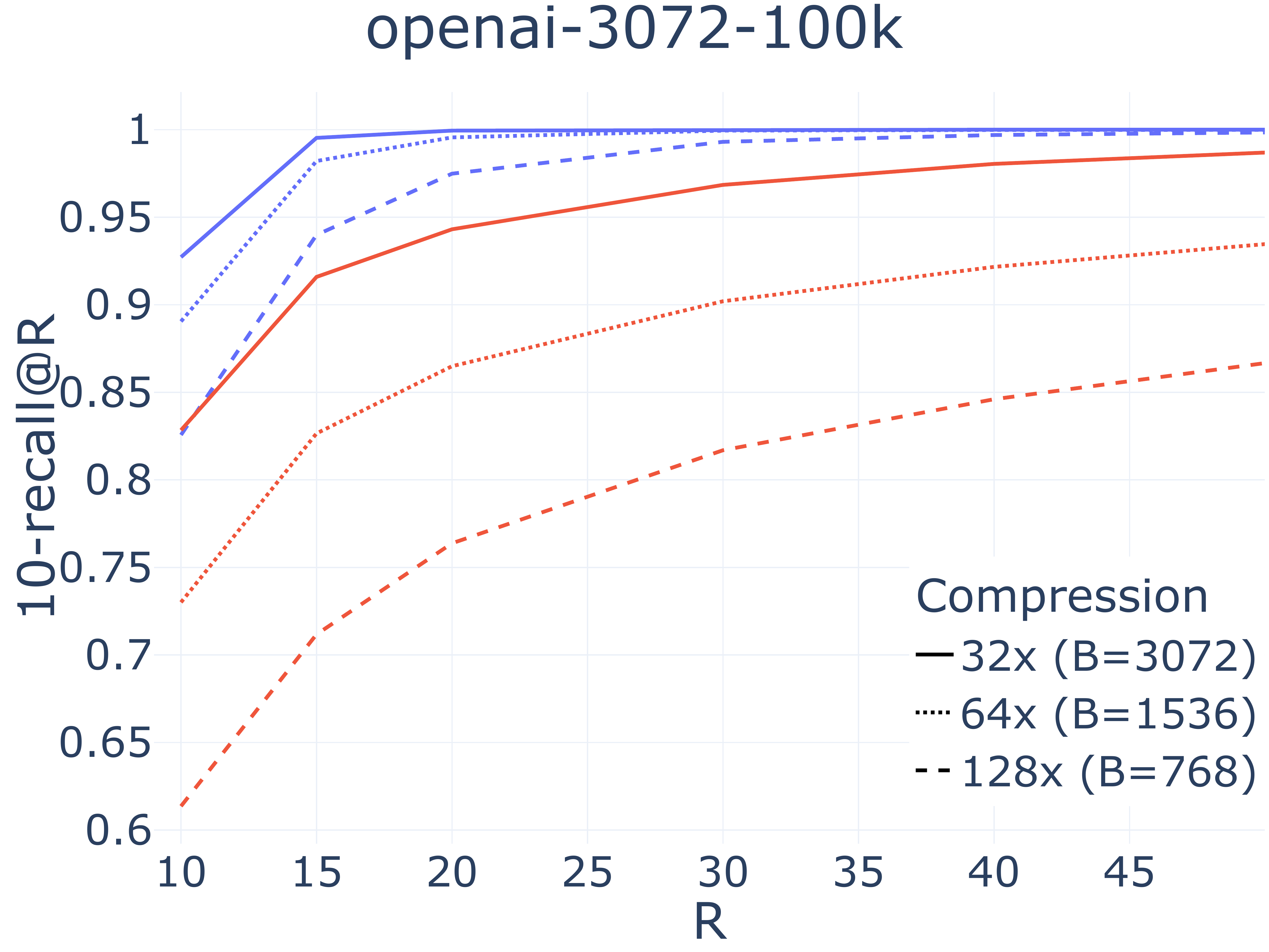}%

    \includegraphics[width=0.2\linewidth]{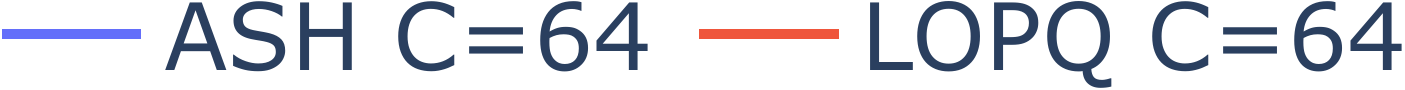}
    
    \caption{ASH outperforms LOPQ \cite{kalantidis_locally_2014} in search accuracy (10-recall@R). Additionally, ASH is computationally more efficient at learning the quantizer and computing similarities.}
    \label{fig:ash_vs_lopq}
\end{figure*}

\textbf{ASH versus LOPQ.}
We can observe in \zcref{fig:ash_vs_lopq} that ASH outperforms LOPQ in terms of search accuracy while being much faster to train, as discussed in \zcref{sec:optimization}. Finally, similarity computations with ASH are faster than with LOPQ (ASH is more efficient than PQ, which is in turn more efficient than LOPQ).

\begin{figure*}[p]
    \centering
    \includegraphics[width=0.3\linewidth]{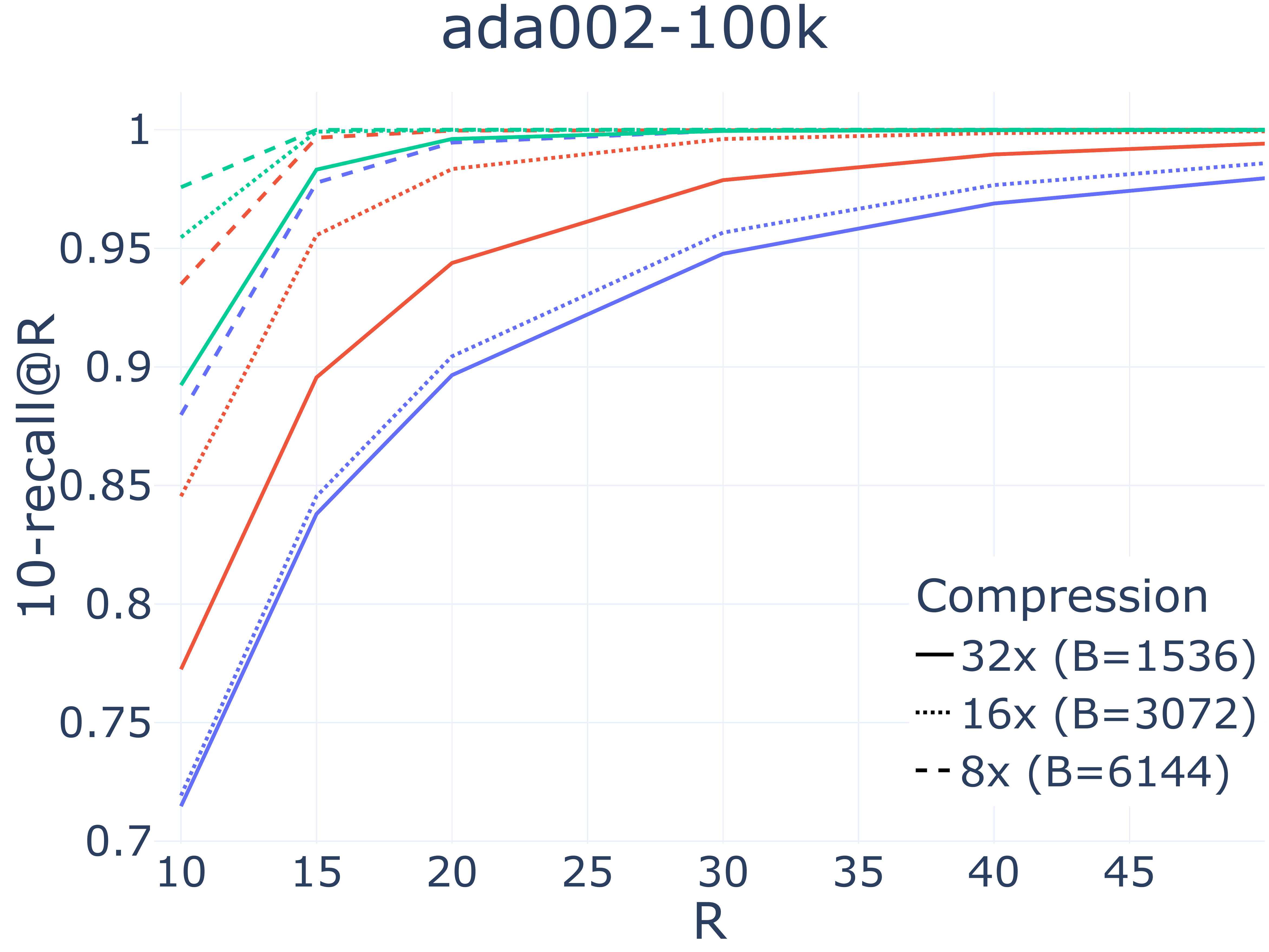}%
    \includegraphics[width=0.3\linewidth]{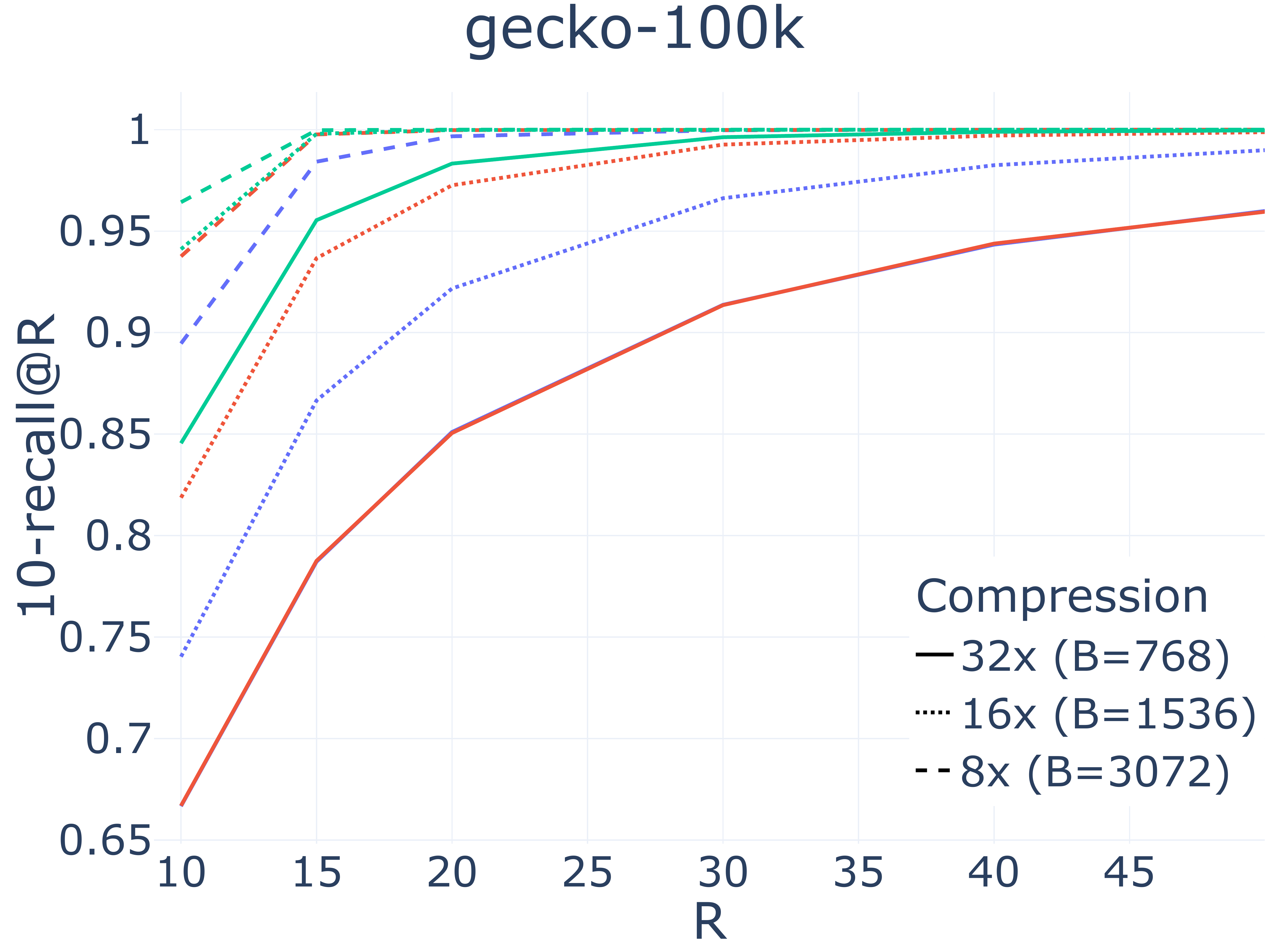}%
    \includegraphics[width=0.3\linewidth]{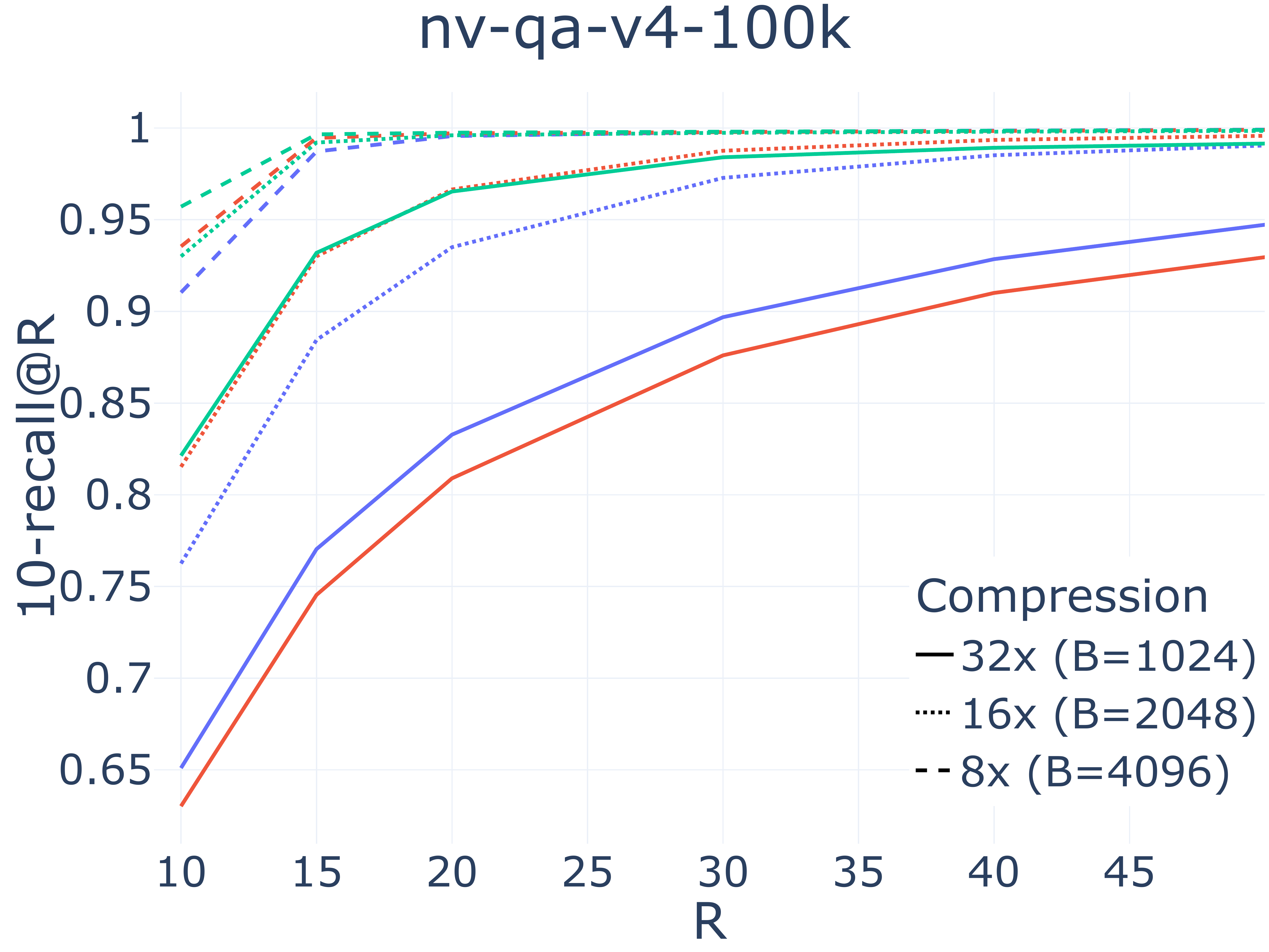}%
    
    \includegraphics[width=0.3\linewidth]{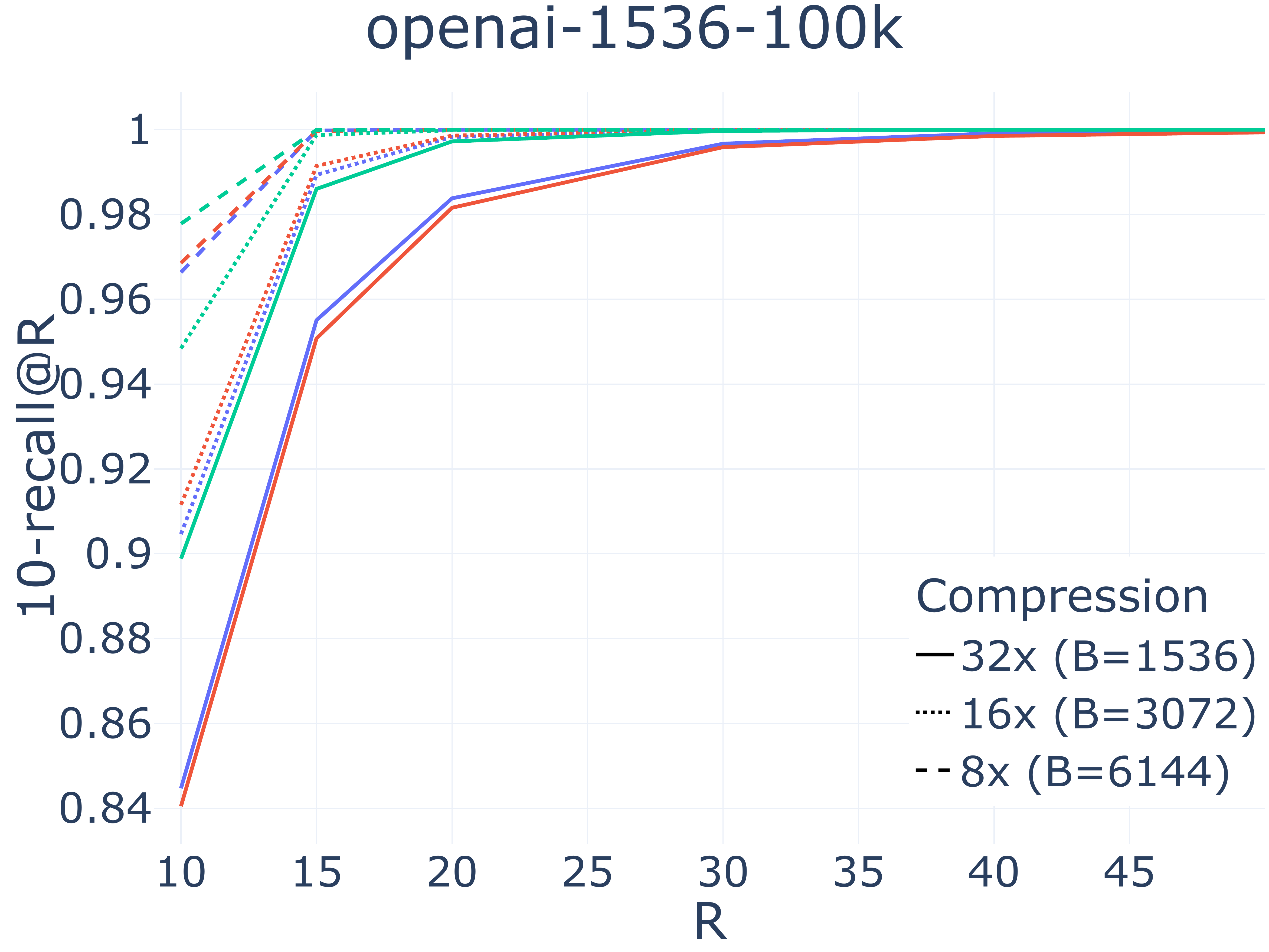}%
    \includegraphics[width=0.3\linewidth]{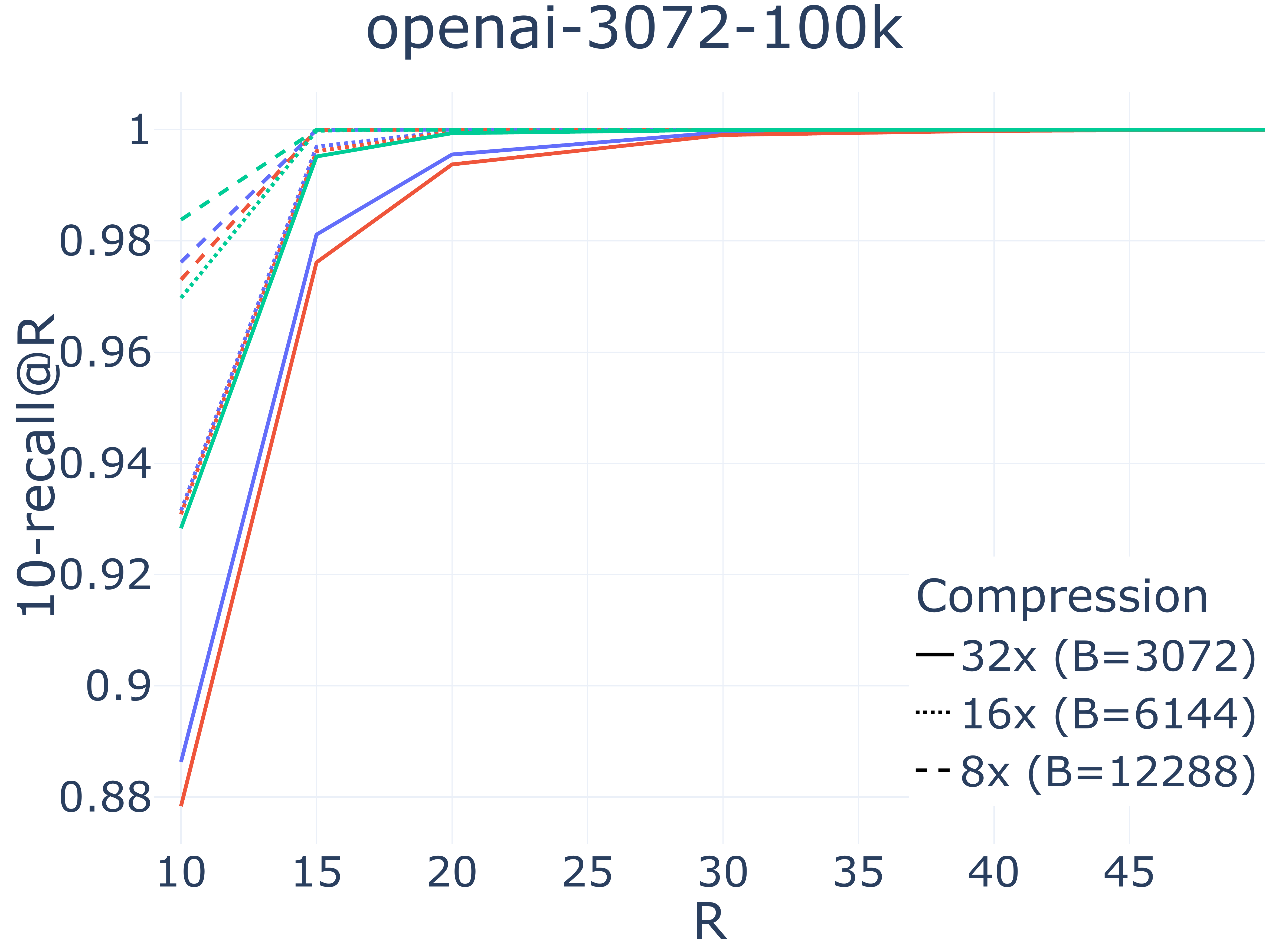}%

    \includegraphics[width=0.22\linewidth]{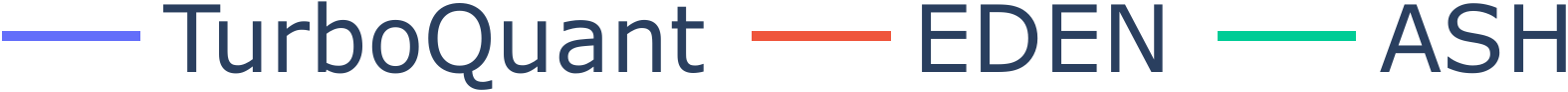}
    
    \caption{ASH outperforms EDEN \cite{vargaftik_eden_2022} and TurboQuant \cite{zandieh_turboquant_2025} in search accuracy (10-recall@R). Often, ASH is competitive at several compression levels with EDEN and TurboQuant configurations that use twice the space.}
    \label{fig:ash_vs_tq}
\end{figure*}

\textbf{ASH versus EDEN and TurboQuant.} We evaluate the performance of ASH with respect to these techniques, which are described in \zcref{sec:comparisons}. ASH shows a clear advantage over both methods in terms of search accuracy, as presented in \zcref{fig:ash_vs_tq}. In many cases, ASH performs competitively with EDEN and TurboQuant configurations that use twice the total number of bits.

\begin{figure*}[p]
    \centering
    \includegraphics[width=0.3\linewidth]{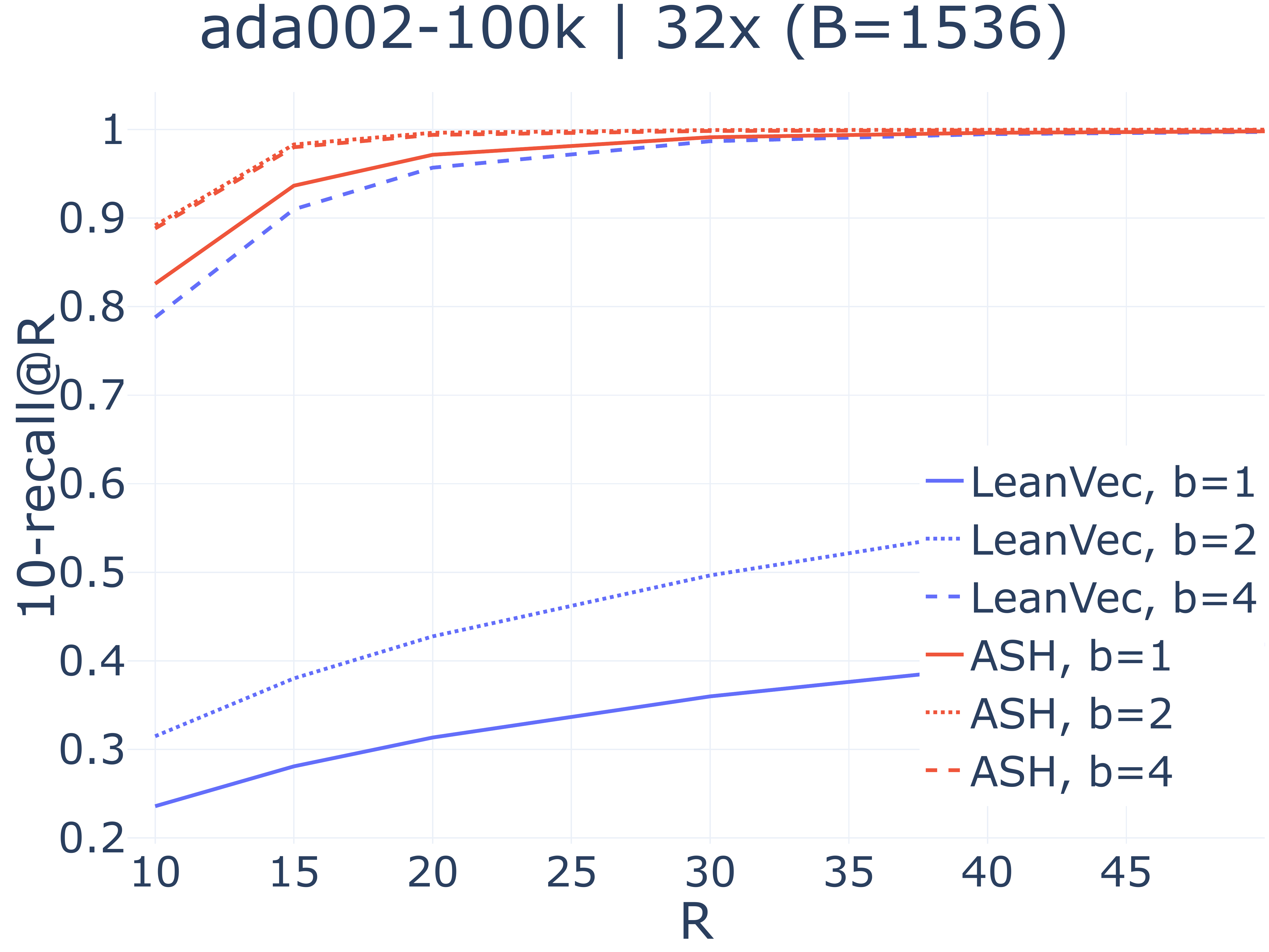}%
    \includegraphics[width=0.3\linewidth]{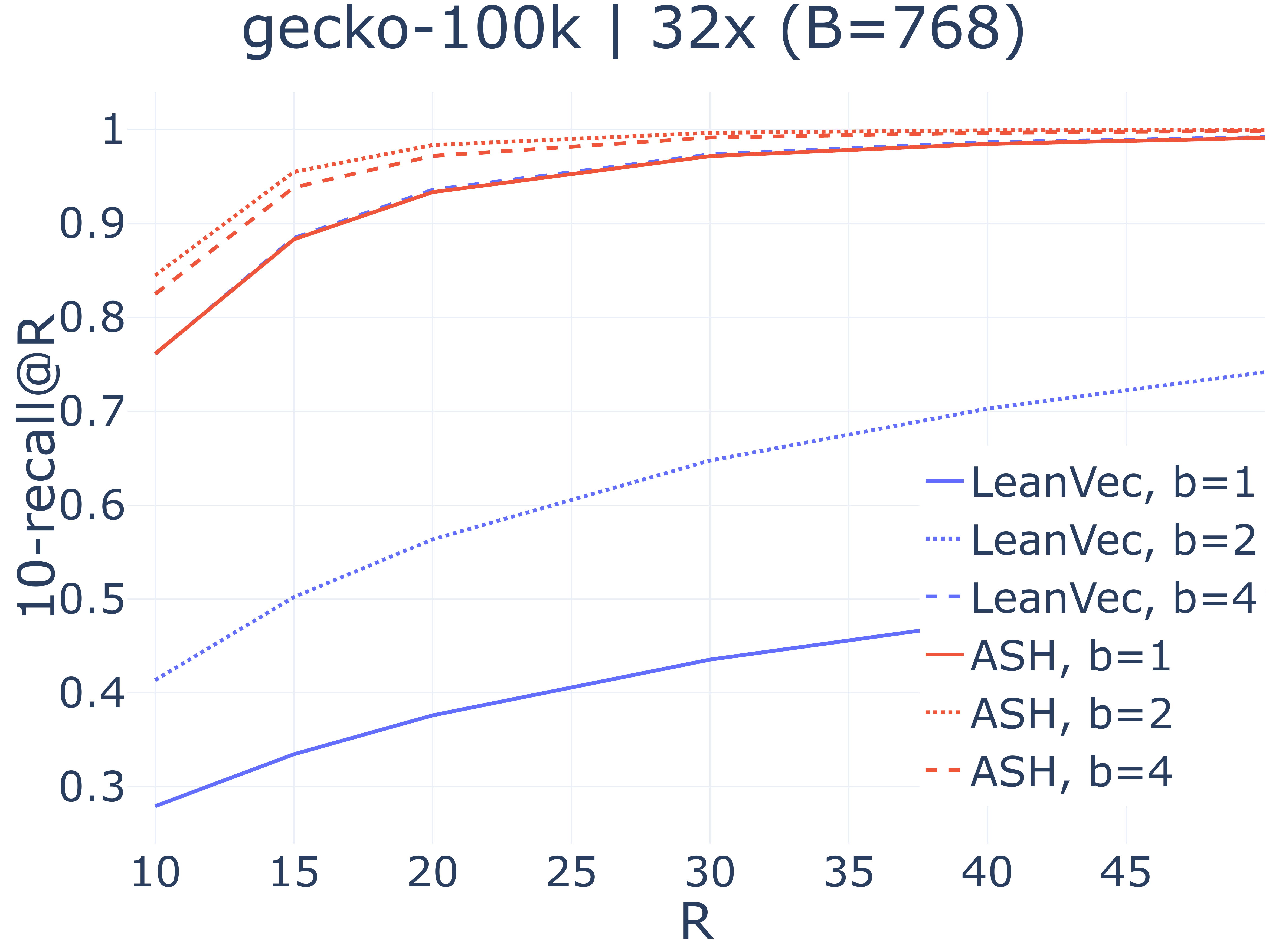}%
    \includegraphics[width=0.3\linewidth]{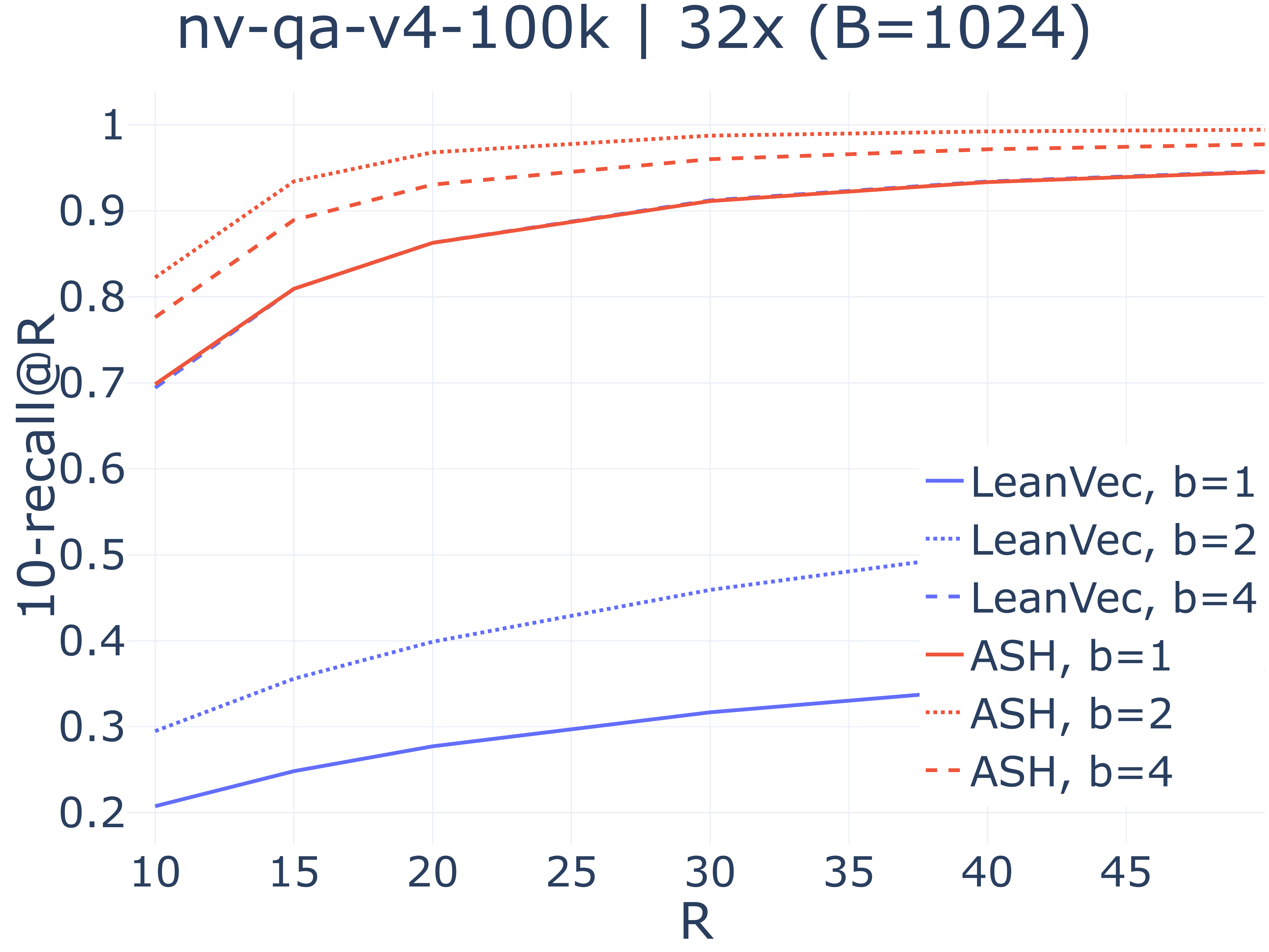}%
    
    \includegraphics[width=0.3\linewidth]{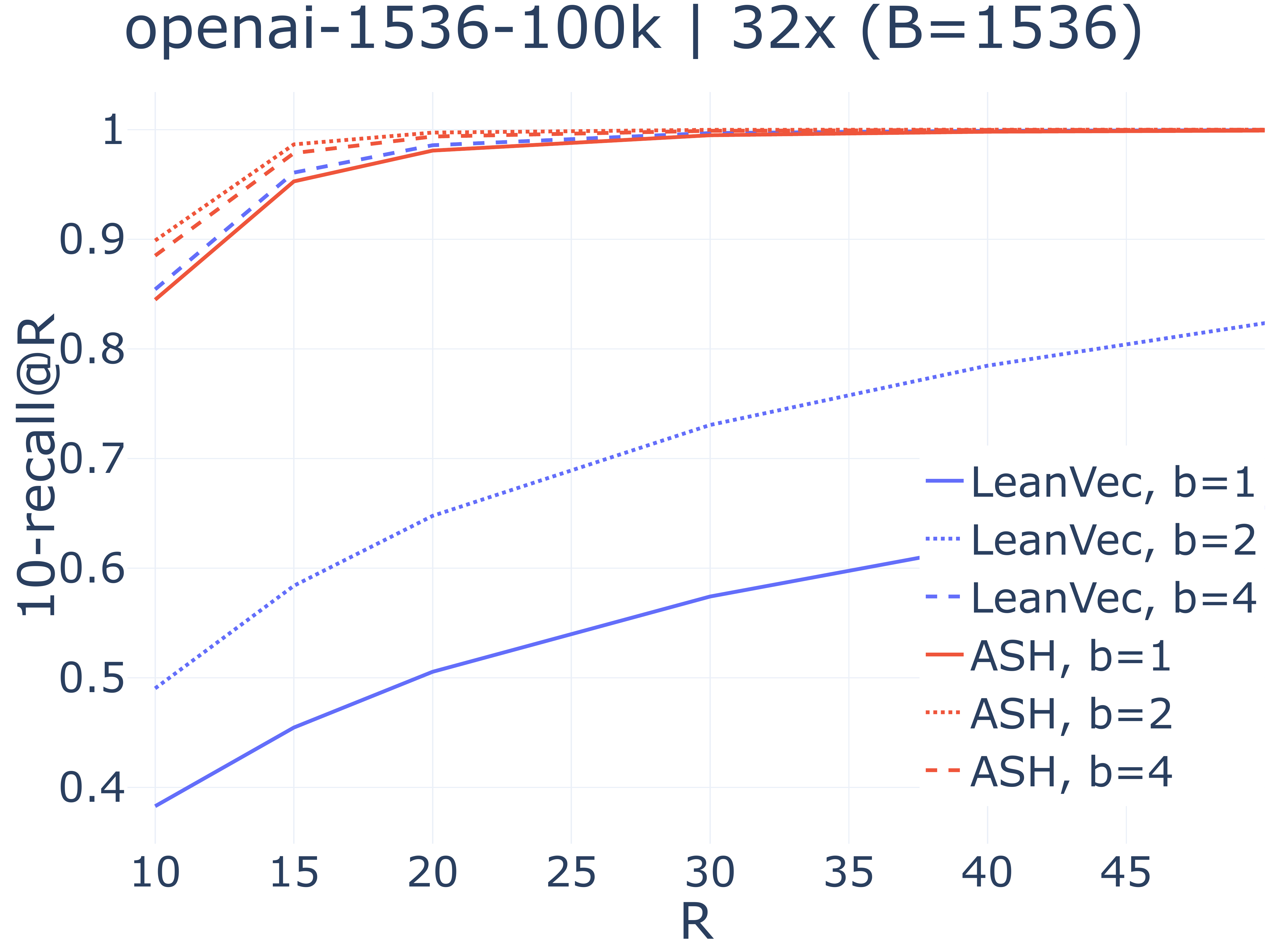}%
    \includegraphics[width=0.3\linewidth]{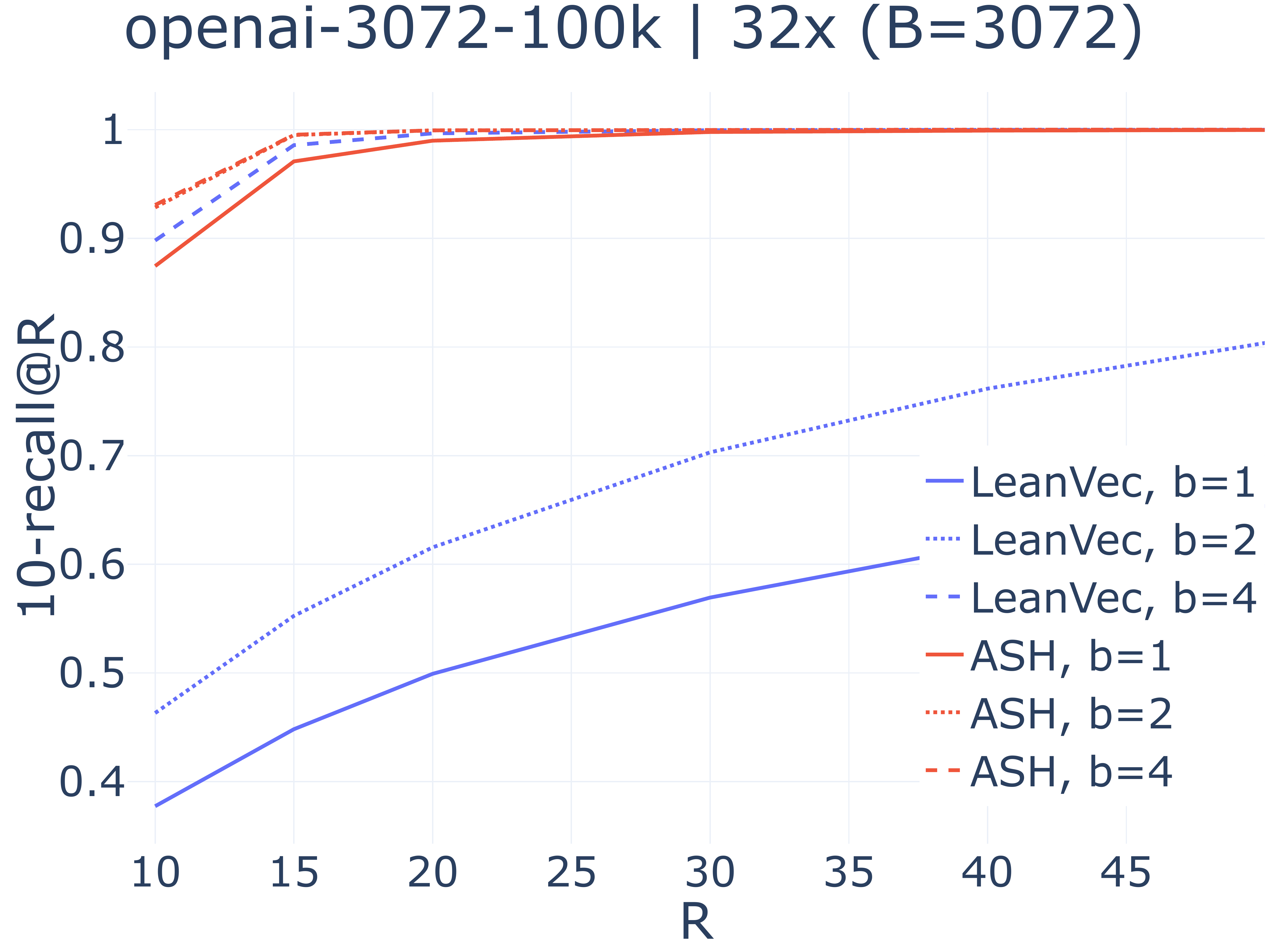}%
    
    \caption{ASH outperforms LeanVec \cite{tepper_leanvec_2024} in search accuracy (10-recall@R). ASH with $b=1$ is competitive with LeanVec with $b=4$, which uses four times more space (additional configurations in \zcref{fig:ash_vs_leanvec_continued} of the appendix).}
    \label{fig:ash_vs_leanvec}
\end{figure*}

\textbf{ASH versus LeanVec.} We evaluate the performance of ASH compared to its conceptually most similar technique, LeanVec \cite{tepper_leanvec_2024} (described in \zcref{sec:comparisons}). In \zcref{fig:ash_vs_leanvec}, we observe that ASH clearly outperforms LeanVec in search accuracy.

\textbf{Performance evaluations.} We implement an inverted file index (IVF) with ASH in C++ to evaluate indexing cost, single-thread throughput (QPS), and 10-recall@10 on modern embedding datasets, using IVF \texttt{nlist}=4096 for 1M-scale sets and $\approx4\sqrt{n}$ for larger runs shown in \zcref{fig:qps_recall_10M} of the appendix. IVF search sweeps \texttt{nprobe} to trace Pareto frontiers; IVF training uses at most $300k$ vectors.

\textbf{Baselines and configurations.} ASH uses $C=32$ landmarks trained for 5 iterations and $b\in\{2,4\}$ with $d=D/2$ for the main comparison; $b=1$ and $d\in\{D/2,D\}$ are included for indexing analysis. Projection training uses samples $10D$ and runs for at most 25 iterations with early stopping after 10 epochs (patience 3 epochs, loss improvement thresholds: $10^{-4}$ absolute, $2.5 \cdot 10^{-3}$ relative). PQ uses Faiss's \texttt{IndexIVFPQFastScan} (factory string \texttt{IVF\{nlist\},PQ\{M\}x4fs}) performing 4-bit PQ in packed batches with $4M=bd$ bits/vector \cite{douze_faiss_2026,meta_research_faiss_2026}. For RaBitQ, we use its official library \cite{vectordb-ntu_rabitq-library_2025} with $b\in\{1,2\}$, fast quantization enabled, and high-accuracy FastScan disabled (negligible difference when enabled).

The experiments run on a GCP \texttt{c4-standard-96-lssd} instance (Intel Xeon 6985P-C @ 2.30GHz, 48 physical cores, Debian Linux). ASH is compiled with GCC/C++20 using \texttt{-O3}, \texttt{-march=native}, OpenMP, AVX-512 kernels, and MKL/LAPACKE for projection training. Faiss-PQ uses Faiss~1.14.1 (cpu) \cite{meta_research_faiss_2026} through the Python API, loading the AVX512-SPR wrapper with OpenMP and MKL enabled; the RaBitQ library \cite{vectordb-ntu_rabitq-library_2025} is built with C++17, OpenMP, \texttt{-Ofast}, and \texttt{-march=native}. Index builds request 48 OpenMP threads pinned to 48 physical cores; all query timings are single-threaded.

\textbf{Indexing cost.} \zcref{tab:ash-indexing-times} reports ASH projection-training and base-encoding time. Projection training follows the $10D$ sample size and projection dimension. At equal code size, halving $d$ while doubling $b$ substantially reduces projection training: $b=2,d=D/2$ is $2.3$--$3.3\times$ faster than $b=1,d=D$, and $b=4,d=D/2$ is $4.0$--$5.3\times$ faster than $b=2,d=D$. Early stopping benefits $b=4$ with reduced $d$ the most, cutting roughly $30$--$50\%$ of iterations with negligible impact on recall. Encoding takes only a few seconds for 1M vectors and grows mainly with quantization bitwidth.

\begin{figure*}[p]
    \centering
    \includegraphics[width=0.33\linewidth]{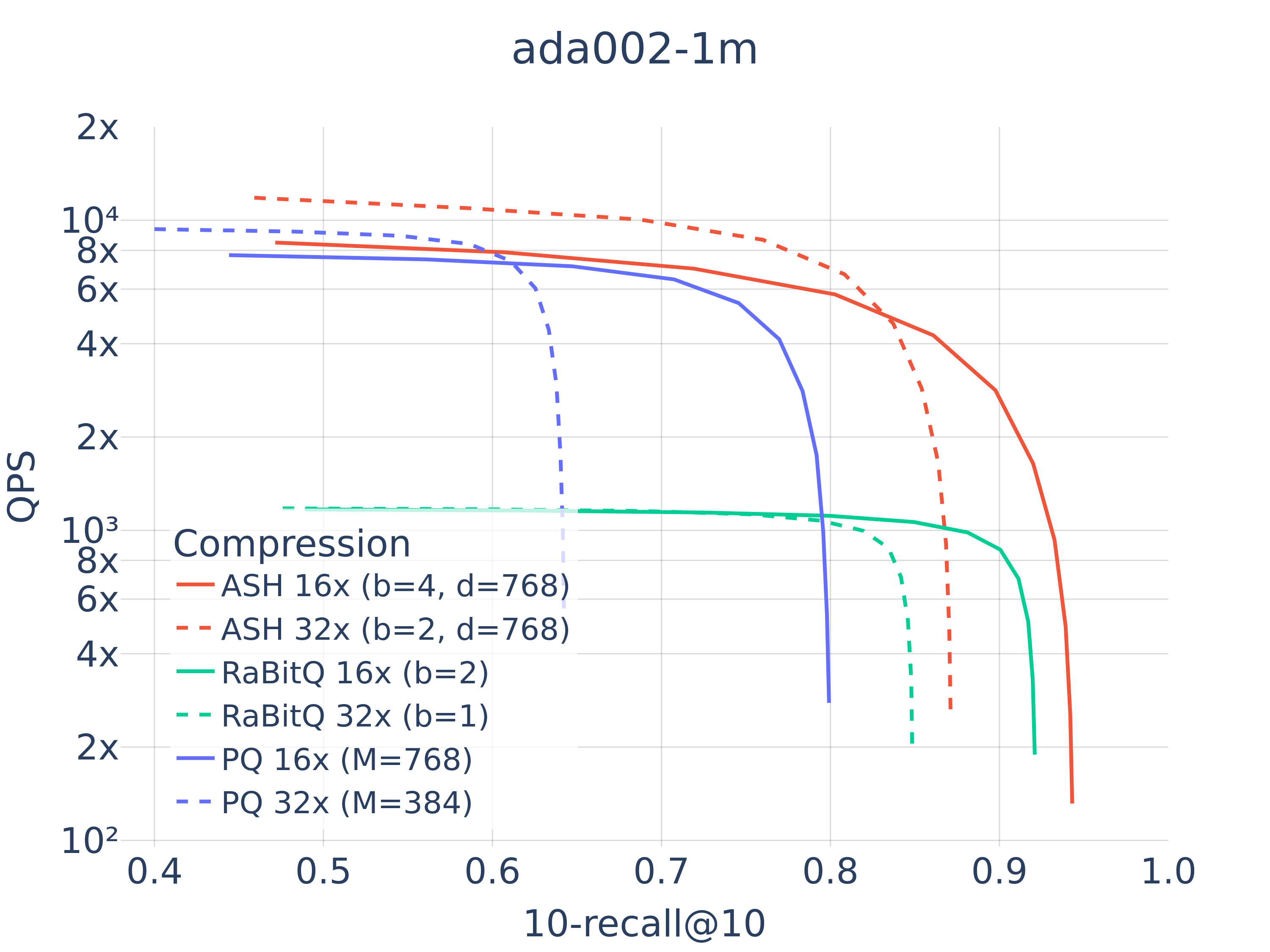}%
    \includegraphics[width=0.33\linewidth]{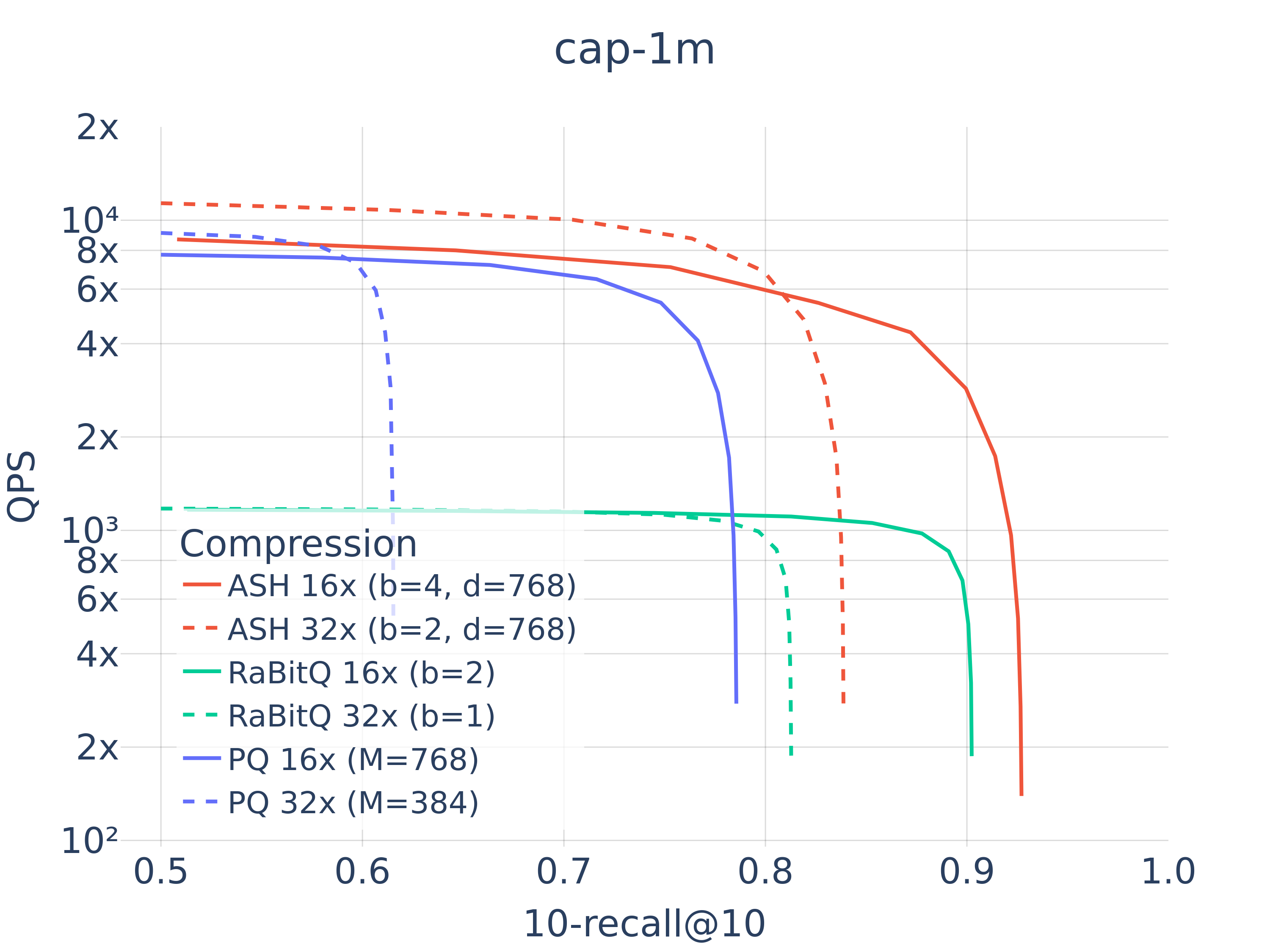}%
    \includegraphics[width=0.33\linewidth]{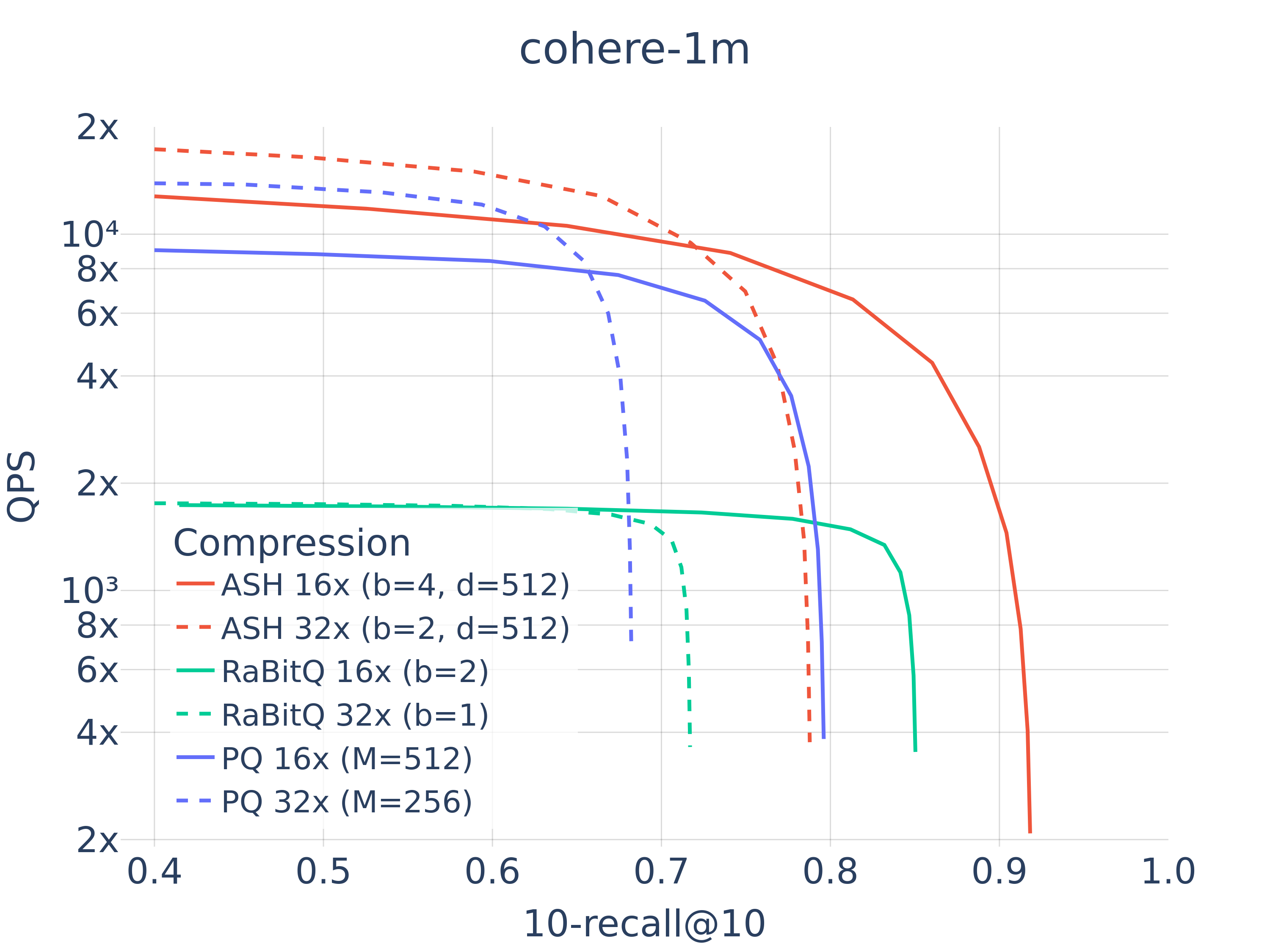}%
    
    \includegraphics[width=0.33\linewidth]{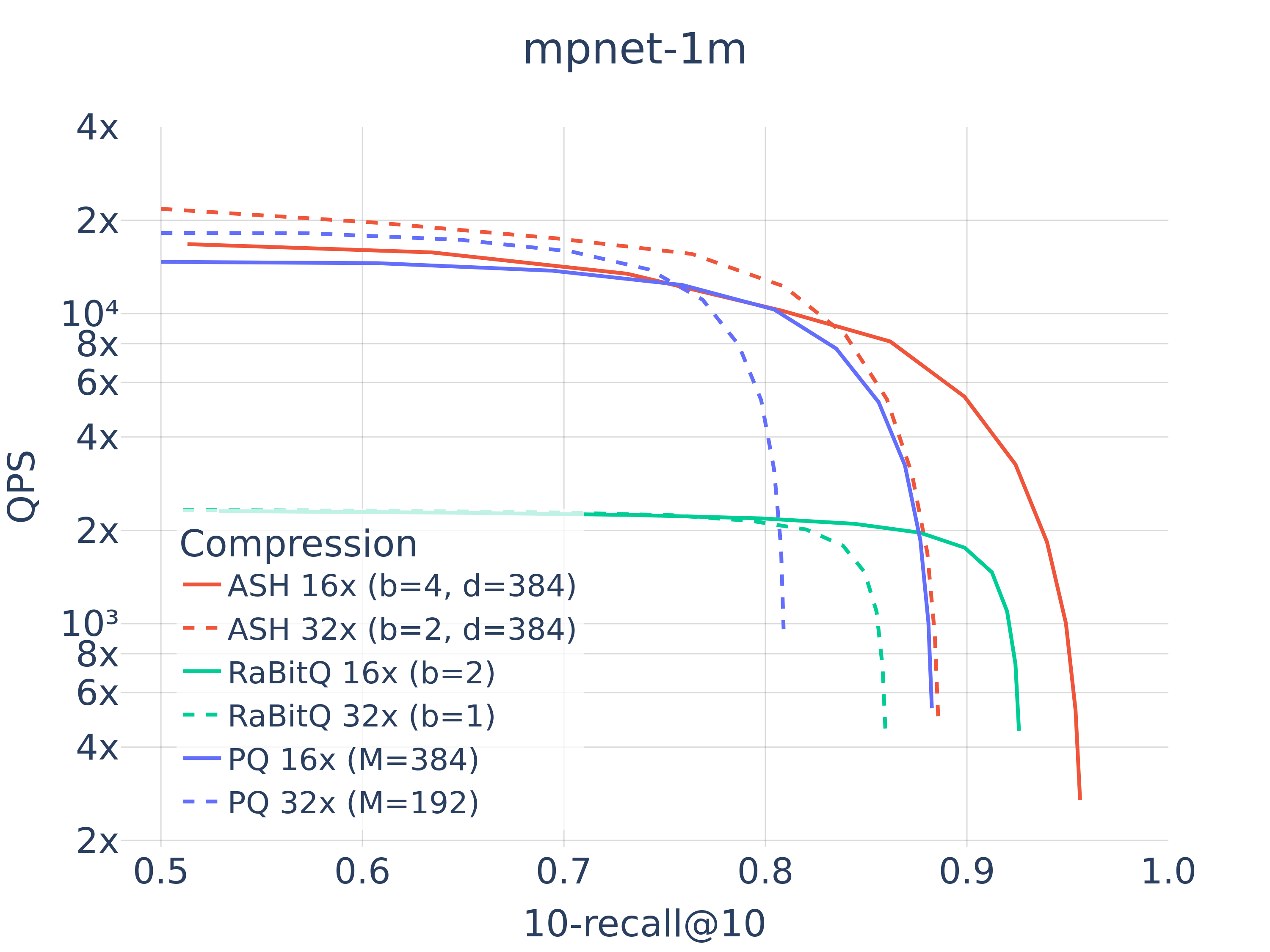}%
    \includegraphics[width=0.33\linewidth]{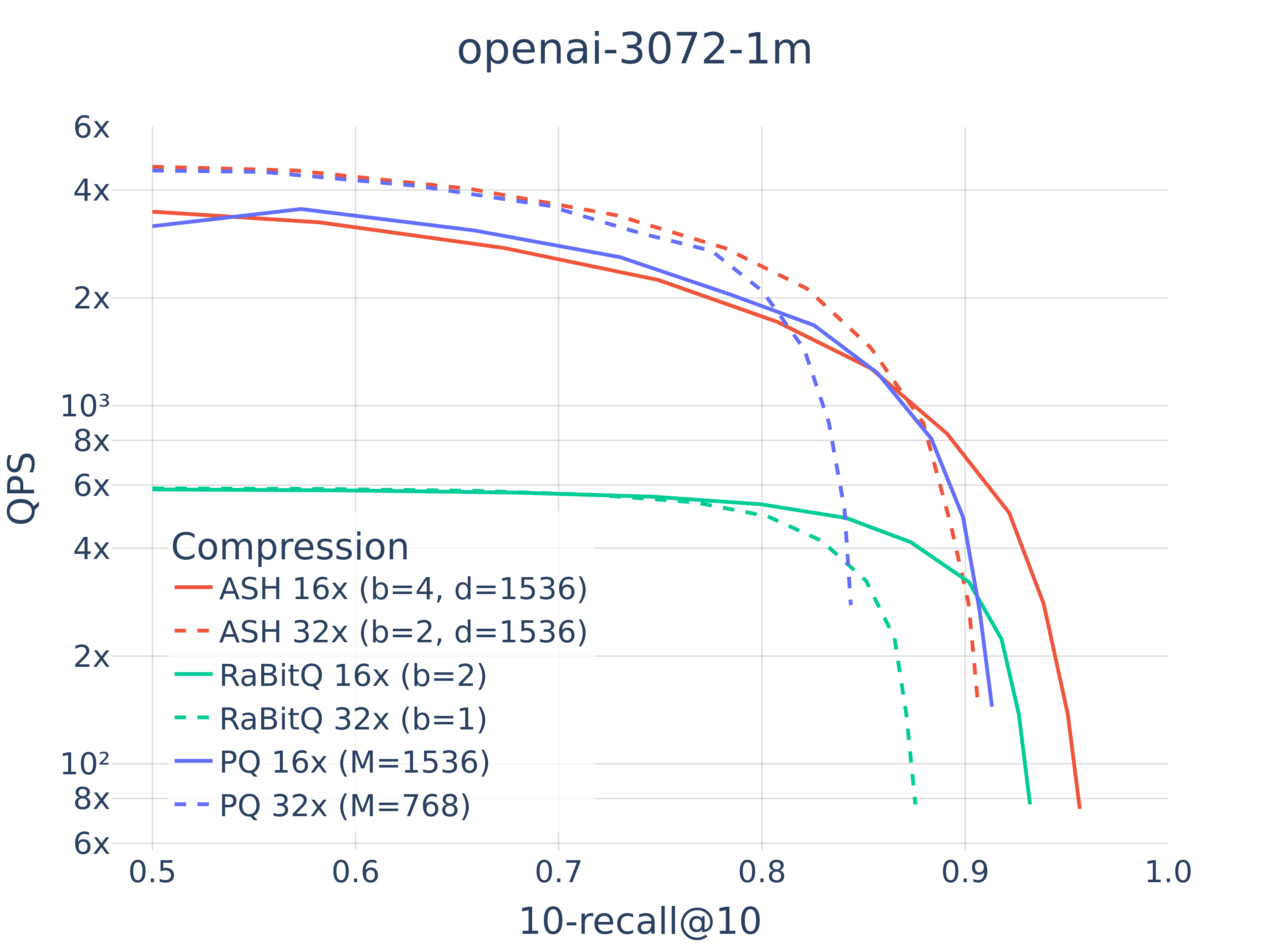}%
    \includegraphics[width=0.33\linewidth]{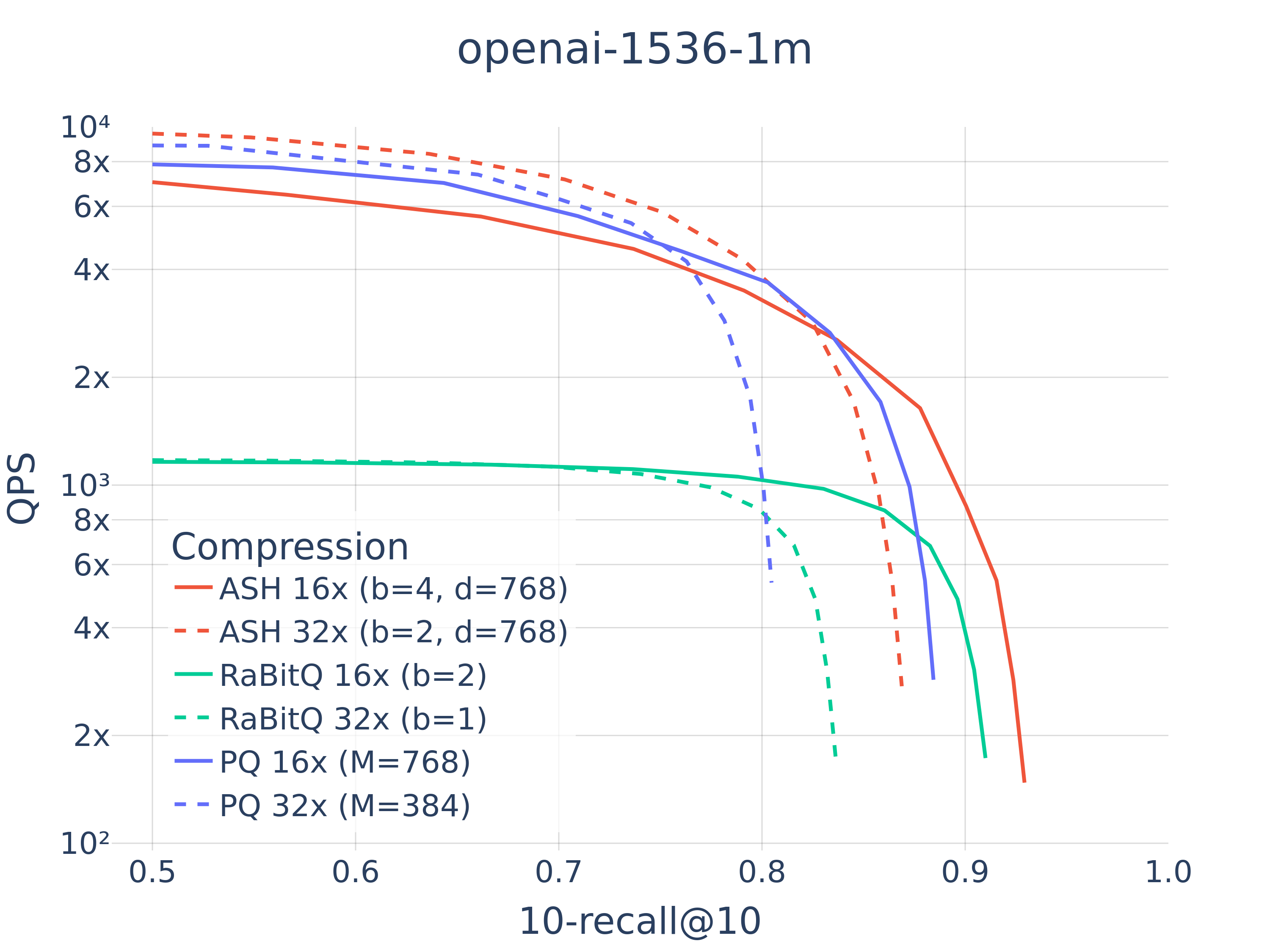}%
    
    \caption{ASH outperforms RaBitQ \cite{gao_rabitq_2024,gao_practical_2025} and PQ \cite{jegou_product_2011} in search accuracy (10-recall@R) and throughput (queries per second, QPS), clearly improving the Pareto frontier (larger datasets in \zcref{fig:qps_recall_continued} of the appendix).}
    \label{fig:qps_recall}
\end{figure*}

\textbf{Throughput and recall.}
\zcref{fig:qps_recall} shows that ASH improves the high-recall end of the Pareto frontier. At $32\times$ compression ($b=2,d=D/2$), ASH reaches 2.3--7.1 points higher terminal recall than RaBitQ and is over 12$\times$ faster at RaBitQ's highest-recall operating point. At $16\times$ compression ($b=4,d=D/2$), ASH improves terminal recall by 2.2--6.8 points and is over 4.6$\times$ faster at the corresponding RaBitQ endpoint. Compared with PQ FastScan, ASH improves recall across the frontier, with terminal gains of 6.3--22.9 points at $32\times$ and 4.3--14.4 points at $16\times$, with the same trend on additional datasets in the appendix. Additional experiments on Faiss's RaBitQ-FastScan (factory string: \texttt{RR\{D\},IVF\{nlists\},RaBitQfs\{b\}}) revealed higher max QPS (albeit still lower than both ASH and PQ) but a less favorable QPS vs recall trade-off in the high-recall regime.

\begin{table}[t]
    \caption{ASH training and encoding time (s). As detailed in \zcref{sec:optimization}, we use a training sample size of $10D$, which explains the training time trend. As in search, increasing $b$ while decreasing $d$ improves the efficiency of training (compare the iso-footprint cases $b=2, d=D/2$ and $b=1, d=D$).}
    \label{tab:ash-indexing-times}

    \small
    \centering
    \begin{tblr}{
        colspec = {l S[table-format=4] S[table-format=4] cccccc},
        colsep = 3pt,
        rowspec = {|QQ|QQ|QQ|QQ|QQ|},
        row{1} = {c},
    }
        \SetCell[r=2]{c} Dataset & \SetCell[r=2]{c} {$D$} & \SetCell[r=2]{c} {$d$} & \SetCell[c=2]{c} {$b=1$} && \SetCell[c=2]{c} {$b=2$} && \SetCell[c=2]{c} {$b=4$} \\
        &&& {Tr.} & {Enc.} & {Tr.} & {Enc.} & {Tr.} & {Enc.} \\
        \SetCell[r=2]{l} mpnet-1m & \SetCell[r=2]{c}  768 & 384 & 0.52 & 0.17 & 0.57 & 0.48 & 0.38 & 0.84 \\
        &  & 768 & 1.33 & 0.29 & 1.51 & 0.90 & 0.99 & 1.70 \\
        \SetCell[r=2]{l} cohere-1m & \SetCell[r=2]{c}  1024 & 512 & 0.89 & 0.24 & 1.03 & 0.66 & 0.70 & 1.19 \\
        &  & 1024 & 3.05 & 0.47 & 3.11 & 1.30 & 2.15 & 2.41 \\
        \SetCell[r=2]{l} ada002-1m & \SetCell[r=2]{c}  1536 & 768 & 1.79 & 0.47 & 2.05 & 1.08 & 1.41 & 1.84 \\
        &  & 1536 & 6.27 & 0.83 & 6.84 & 2.22 & 4.31 & 3.71 \\
        \SetCell[r=2]{l} openai-3072-1m & \SetCell[r=2]{c}  3072 & 1536 & 8.77 & 1.50 & 9.71 & 2.80 & 6.51 & 4.62 \\
        &  & 3072 & 32.41 & 2.86 & 34.25 & 5.71 & 20.96 & 8.99 \\
    \end{tblr}
\end{table}

\section{Conclusions}

The dominance of additive quantization methods has long rested on two pillars: an asymmetric similarity computation that keeps queries at full precision while quantizing only indexed vectors, and a fast, data-driven learning algorithm. Traditionally, learning-to-hash methods have remained binary and symmetric. In recent years, a number of asymmetric and data-agnostic scalar hashing techniques have emerged, offering great efficiency and accuracy.

In this work, we blend both worlds.
We presented ASH (Asymmetric Scalar Hashing), a data-driven encoder-decoder framework that brings the asymmetric learning paradigm to scalar quantization. Learning an orthonormal projection that reduces dimensionality before quantization allows ASH to increase the bitrate per dimension while reducing the reconstruction error, yielding higher accuracy than the best additive and scalar quantizers at iso-compression. The training algorithm converges rapidly, requiring only 20–30 iterations and a small training sample. ASH generalizes prior methods such as ITQ and RaBitQ as special cases.
Since ASH is a scalar quantizer, similarity computations can be implemented with great efficiency using SIMD instructions in both sequential and random access patterns. This makes ASH naturally suitable for both inverted index and graph-based vector search structures.

Extensive experiments confirm that ASH achieves state-of-the-art ANN recall across all compression regimes, outperforming alternatives by a wide margin. In several cases, ASH with half the total bits matches or exceeds the alternatives' accuracy, offering significant memory savings with no loss in recall.

In future work, we are interested in deeper architectures for asymmetric hashing, going beyond the linear projection functions used in this work. We also plan to study end-to-end learning schemes to jointly learn the ASH landmarks and the projection functions.

\bibliographystyle{plainnat}
\bibliography{AsymBQ_clean.bib}

\appendix

\counterwithin{figure}{section}
\counterwithin{table}{section}
\counterwithin{equation}{section}

\section{Additional similarity functions}
\label{sec:other_metrics}

The Euclidean distance is straightforward to incorporate in the current framework. We start by decomposing the distance $\norm{\vect{q} - \vect{x}_i}{2}^2$ as follows
\begin{align}
    \norm{\vect{q} - \vect{x}_i}{2}^2
    &=
    \norm{\vect{q} - \vect{\mu}_i^* + \vect{\mu}_i^* - \vect{x}_i}{2}^2
    \\
    &=
    \norm{\vect{q} - \vect{\mu}_i^*}{2}^2
    +
    \norm{\vect{x} - \vect{\mu}_i^*}{2}^2
    - 2
    \Big(
        \left\langle \vect{q} , \vect{x}_i \right\rangle
        - \left\langle \vect{\mu}_i^* , \vect{x}_i \right\rangle
        - \left\langle \vect{q} , \vect{\mu}_i^* \right\rangle
        + \norm{\vect{\mu}_i^*}{2}^2
    \Big) .
\end{align}
The terms $\norm{\vect{x} - \vect{\mu}_i^*}{2}^2$ and $\left\langle \vect{\mu}_i^* , \vect{x}_i \right\rangle$ are stored in the ASH payload (\zcref{tab:payload}). The terms $\norm{\vect{q} - \vect{\mu}_i^*}{2}^2$ and $\left\langle \vect{q} , \vect{\mu}_i^* \right\rangle$ are computed once for each query. We can pre-compute $\norm{\vect{\mu}_c}{2}^2$ once all $c$. Finally, to approximate $\left\langle \vect{q} , \vect{x}_i \right\rangle$, we use \zcref{eq:spherical_approach}.

The cosine similarity
\begin{equation}
    \operatorname{cosSim} (\vect{q} , \vect{x})
    \defeq
    \left\langle \vect{q} , \vect{x} \right\rangle
    / \left(
        \norm{\vect{q}}{2}
        \cdot
        \norm{\vect{x}_i}{2}
    \right)
    \label{eq:cosSim}
\end{equation}
requires a little more work.
The first and simpler approach is to add the value of $\norm{\vect{x}_i}{2}$ to the ASH header (\zcref{tab:payload}). This would increase our payload by 16 bits. Alternatively, we can use \zcref{eq:x_normalized} to write
\begin{align}
    \vect{x}_i
    &= \norm{\vect{x}_i - \vect{\mu}_i^*}{2} \hat{\vect{x}}_i + \vect{\mu}_i^* \nonumber \\
    &\approx d^{-1/2} \norm{\vect{x}_i - \vect{\mu}_i^*}{2} \transpose{\mat{W}} \left( \operatorname{quant}_b(\hat{\vect{x}}_i) \right) + \vect{\mu}_i^* .
\end{align}
From this approximation of $\vect{x}_i$ and using simple calculations, we can estimate its norm as
\begin{equation}
    \norm{\vect{x}_i}{2}^2
    \approx
        \norm{\vect{x}_i - \vect{\mu}_i^*}{2}^2
        + 2 d^{-1/2} \norm{\vect{x}_i - \vect{\mu}_i^*}{2}
        \left\langle \operatorname{quant}_b(\hat{\vect{x}}_i), \mat{W} \vect{\mu}_i^* \right\rangle
        + \norm{\vect{\mu}_i^*}{2}^2
        .
\end{equation}
Only the term $\left\langle \operatorname{bin}(\hat{\vect{x}}_i), \mat{W} \vect{\mu}_i^* \right\rangle$ requires being computed for each $i$. The remaining terms are global, can be computed once per query, or are already stored in the ASH payload.

\section{The symmetric case}
\label{sec:symmetric_dot_product}

Multiple similarity computations between vectors in $\set{X}$ are performed when building a graph index for $\set{X}$. To save memory, it is of critical importance to compute these similarities from quantized vectors efficiently and \emph{symmetrically}.
In this case, we set $C = 1$, further simplifying the computation (we drop the sub-indices to simplify the notation).
For $\vect{x}, \vect{y} \in \set{X}$, we write their dot product as
\begin{align}
    \langle \vect{x}, \vect{y} \rangle
    &=
    \langle \vect{x} - \vect{\mu} , \vect{y} - \vect{\mu} \rangle
    + \langle \vect{x} , \vect{\mu} \rangle
    + \langle \vect{y} , \vect{\mu} \rangle
    - \norm{\vect{\mu}}{2}^2 \\
    &=
    \norm{\vect{x} - \vect{\mu}}{2}
    \norm{\vect{y} - \vect{\mu}}{2}
    \left\langle \tilde{\vect{x}} , \tilde{\vect{y}} \right\rangle
    + \langle \vect{x} , \vect{\mu} \rangle
    + \langle \vect{y} , \vect{\mu} \rangle
    - \norm{\vect{\mu}}{2}^2 ,
    \label{eq:spherical_approach_symmetric}
\end{align}
where $\tilde{\vect{x}}, \tilde{\vect{y}}$ are defined as in \zcref{eq:x_normalized}.
Then, all quantities except $\left\langle \tilde{\vect{x}} , \tilde{\vect{y}} \right\rangle$ are stored as part of the encodings of $\vect{x}$ and $\vect{y}$, see \zcref{tab:payload}. We use the representation proposed in \zcref{eq:encoder_decoder} to approximate the dot product $\left\langle \tilde{\vect{x}} , \tilde{\vect{y}} \right\rangle$ as
\begin{flalign}
    \left\langle \tilde{\vect{x}} , \tilde{\vect{y}} \right\rangle
    &\approx
    \norm{g(\tilde{\vect{x}})}{2}^{-1}
    \norm{g(\tilde{\vect{y}})}{2}^{-1}
    \left\langle \transpose{\mat{W}} g(\tilde{\vect{x}}) , \transpose{\mat{W}} g(\tilde{\vect{y}}) \right\rangle
    \\
    &=
    \operatorname{cosSim} \left( \operatorname{quant}_b \left( \mat{W} \tilde{\vect{x}} \right) , \operatorname{quant}_b \left( \mat{W} \tilde{\vect{y}} \right) \right)
    .
    \label{eq:symmetric_approximation}
\end{flalign}
When plugging \zcref{eq:symmetric_approximation} in \zcref{eq:spherical_approach_symmetric}, all operations can be performed with integer FMAs and scalar computations using global constants or information available in the ASH header (see \zcref{tab:payload}).

Similarly as in the asymmetric case, we can further develop \zcref{eq:symmetric_approximation} for binary quantization ($b=1$) as
\begin{flalign}
    \left\langle \tilde{\vect{x}} , \tilde{\vect{y}} \right\rangle
    &\approx
    d^{-1}
    \left\langle \transpose{\mat{W}} g(\tilde{\vect{x}}) , \transpose{\mat{W}} g(\tilde{\vect{y}}) \right\rangle
    \\
    &=
    d^{-1}
    \left\langle g(\tilde{\vect{x}}) , g(\tilde{\vect{y}}) \right\rangle
    \\
    &=
    \begin{multlined}[t]
    4 d^{-1}
    \left\langle \operatorname{bin} \left( \mat{W} \tilde{\vect{x}} \right) , \operatorname{bin} \left( \mat{W} \tilde{\vect{y}} \right) \right\rangle
    - \\
    - 2 d^{-1} \left\langle \operatorname{bin} \left( \mat{W} \tilde{\vect{x}} \right) , \vect{1} \right\rangle
    - \\
    - 2 d^{-1} \left\langle \operatorname{bin} \left( \mat{W} \tilde{\vect{y}} \right) , \vect{1} \right\rangle
    + d^{-1}
    .
    \end{multlined}
    \label{eq:symmetric_approximation_binary}
\end{flalign}
The dot product $\left\langle \operatorname{bin} \left( \mat{W} \tilde{\vect{x}} \right) , \operatorname{bin} \left( \mat{W} \tilde{\vect{y}} \right) \right\rangle$ between binary vectors can be performed using a negated XOR. The dot product $\left\langle \operatorname{bin} \left( \cdot \right) , \vect{1} \right\rangle$ can be computed using a popcount on-the-fly or added to the payload. Note that these computations are similar to the classical use of hashing, where $\langle \vect{x}, \vect{y} \rangle \approx \left\langle \operatorname{bin} \left( \mat{W} \vect{x} \right) , \operatorname{bin} \left( \mat{W} \vect{y} \right) \right\rangle$ for a linear or nonlinear function $h$ \cite{indyk_approximate_1998}.

\section{Reference dot product code}
\label{sec:dot_product}

\begin{lstlisting}{c}
float dot_product(const float* query, const float* vector,
                  uint32_t d) {
  __m512 vec_result = _mm512_setzero_ps();
  for (uint32_t j = 0; j < d; j += 16) {
    __m512 v1 = _mm512_load_ps((void const*) (query + j));
    __m512 v2 = _mm512_load_ps((void const*) (vector + j));
    vec_result = _mm512_fmadd_ps(v1, v2, vec_result);
  }
  float result = _mm512_reduce_add_ps(vec_result);
  return result;
}
\end{lstlisting}

\section{Additional experimental results}

\begin{figure*}
    \centering
    \includegraphics[width=0.4\linewidth]{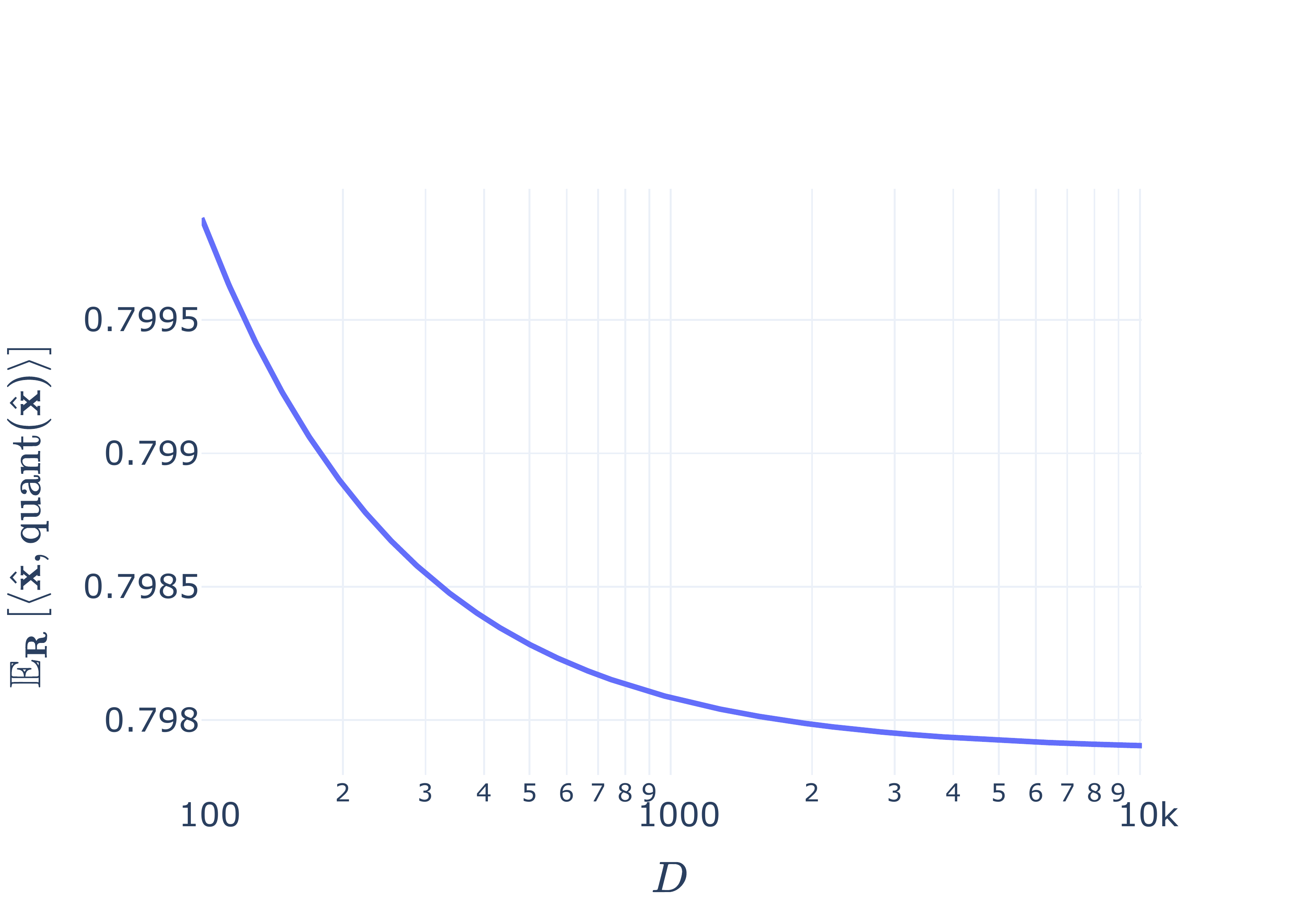}
    
    \caption{The expected dot product in \zcref{eq:expected_rabitq_loss} between a normalized vector $\vect{x}$ and its quantized and normalized version $\operatorname{quant}(\hat{\vect{x}})$ as a function of $D$ (in logarithmic scale).}
    \label{fig:rabitq_expected_loss}
\end{figure*}

\begin{figure*}
    \centering

    \includegraphics[width=0.30\linewidth]{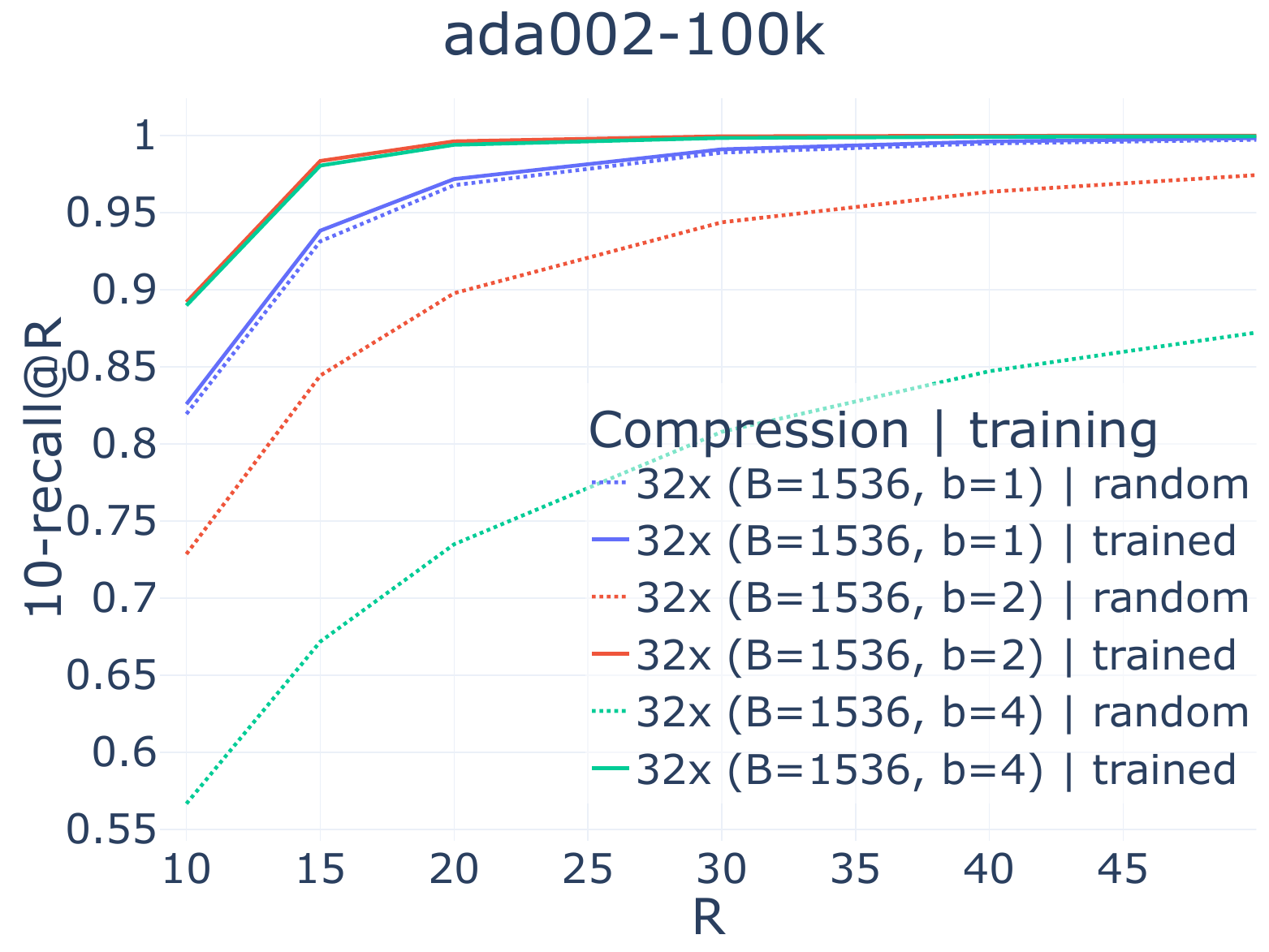}%
    \includegraphics[width=0.30\linewidth]{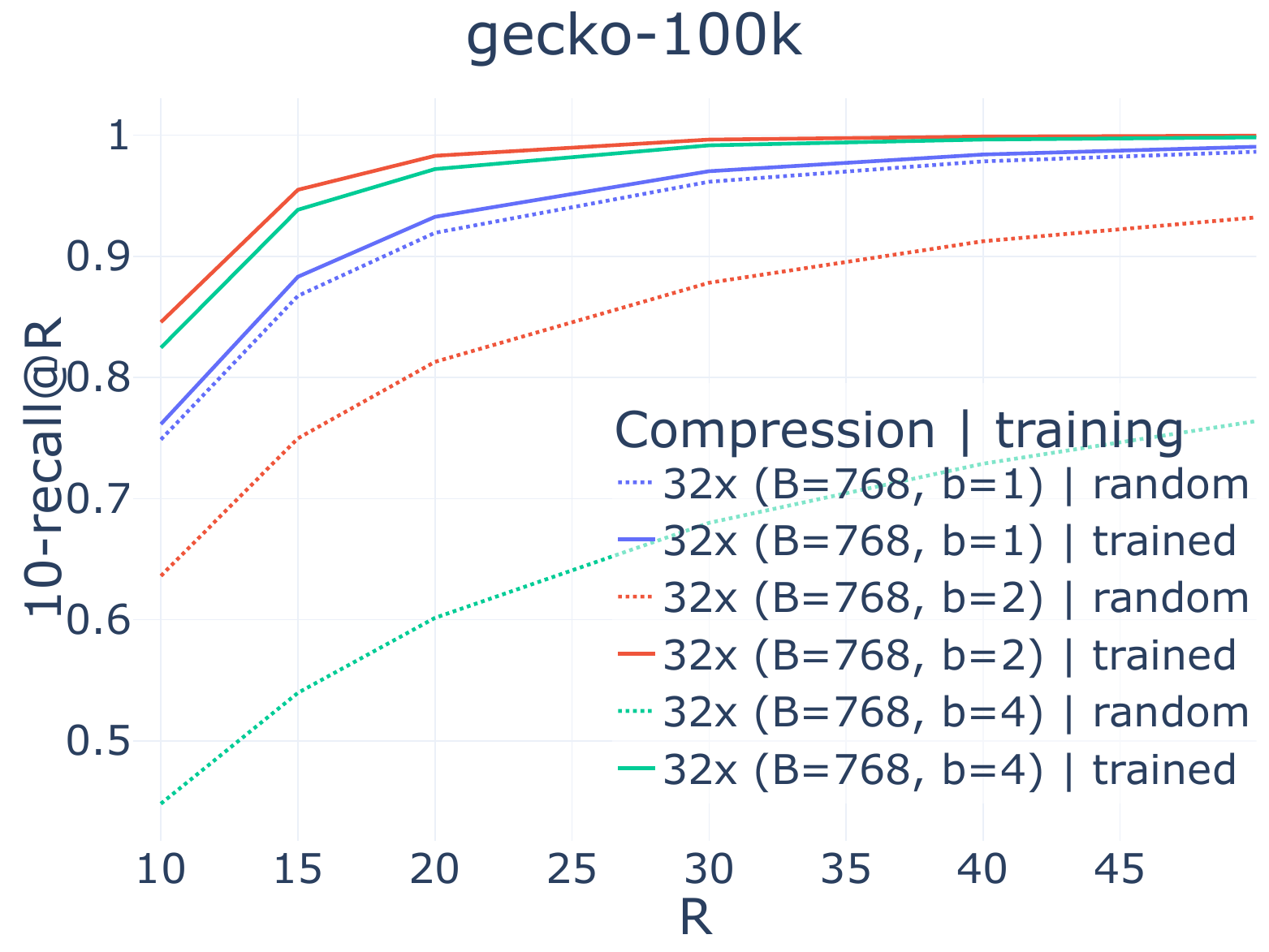}%
    \includegraphics[width=0.30\linewidth]{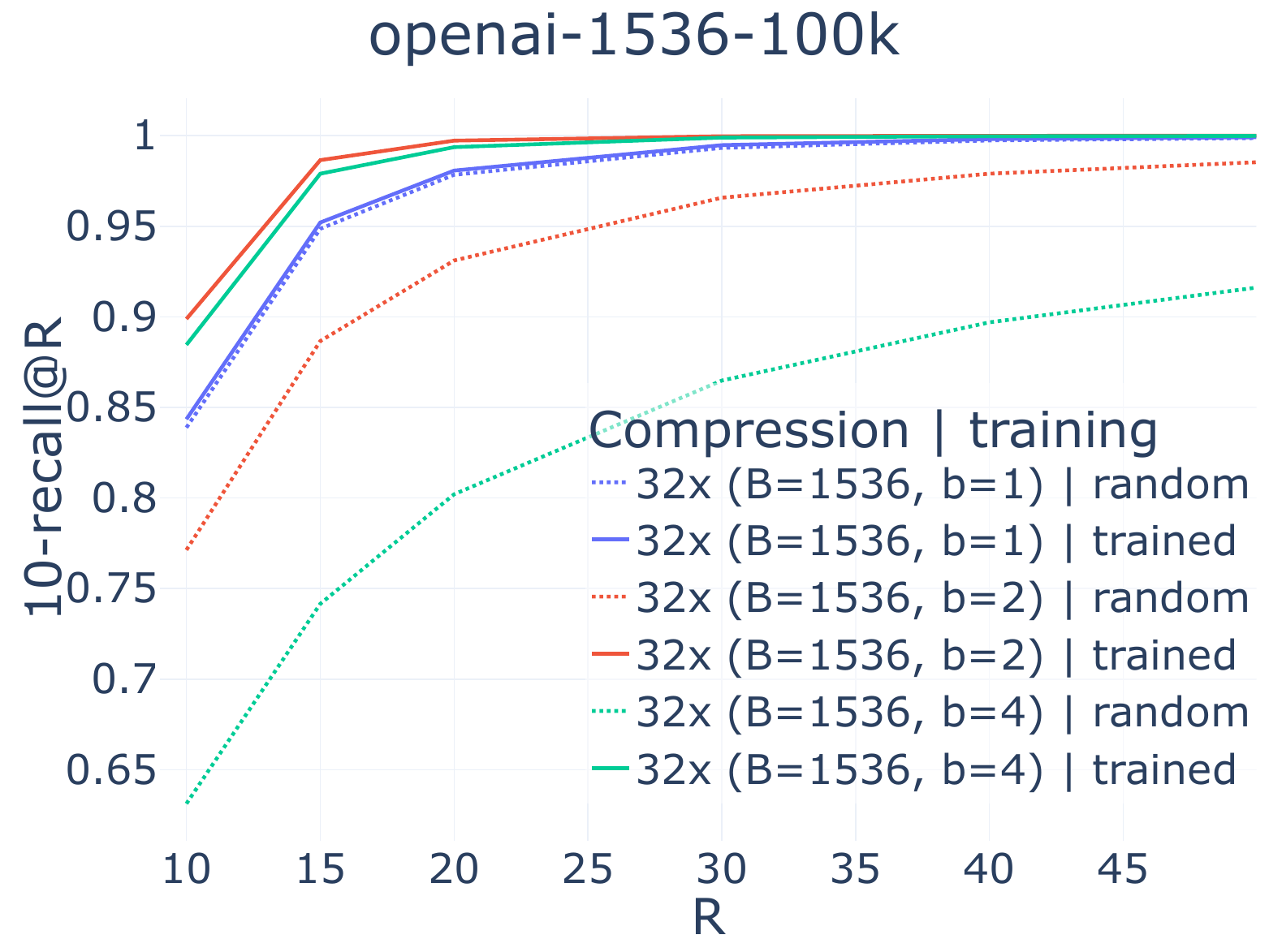}%
    
    \includegraphics[width=0.30\linewidth]{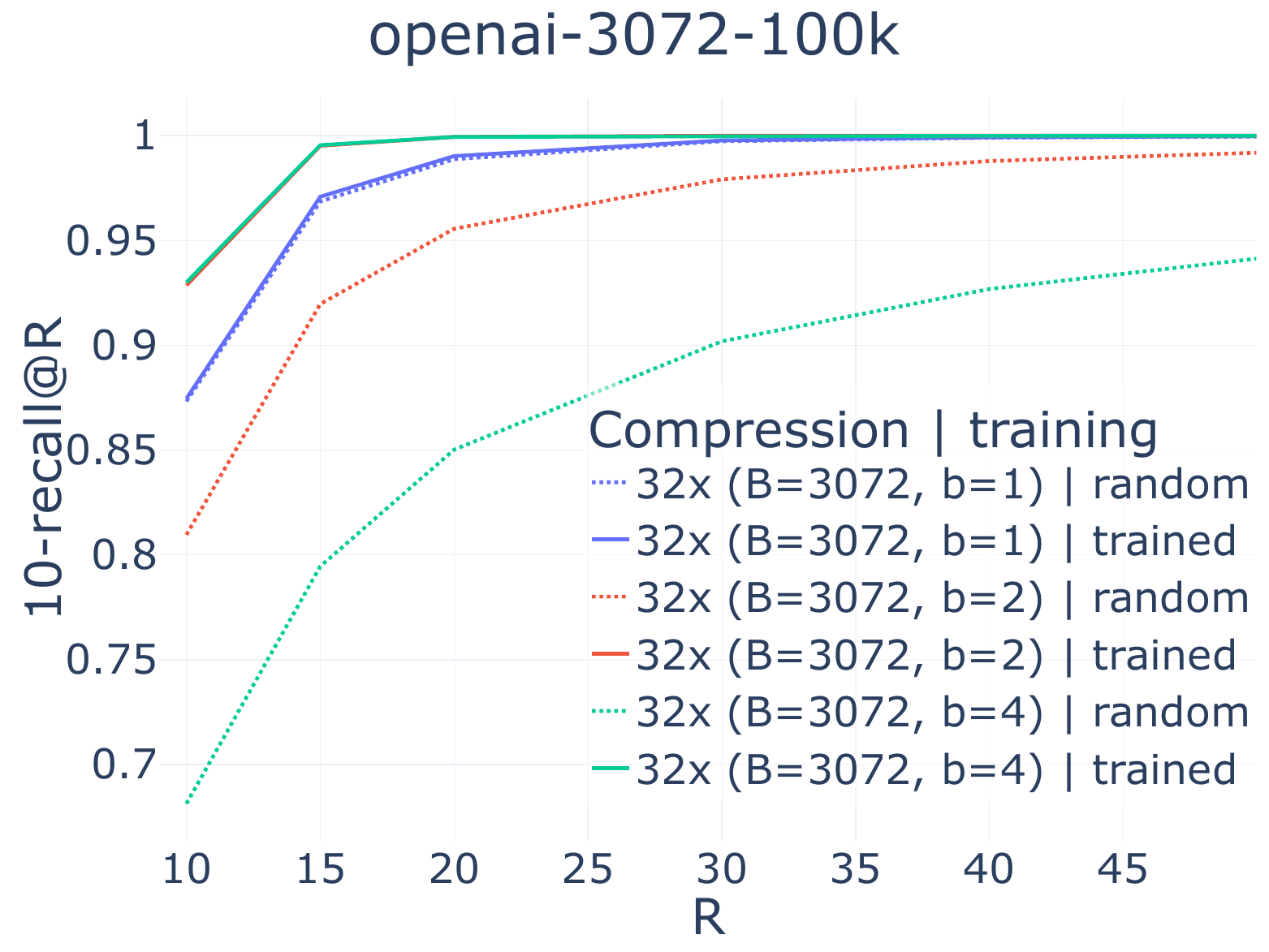}%
    \includegraphics[width=0.30\linewidth]{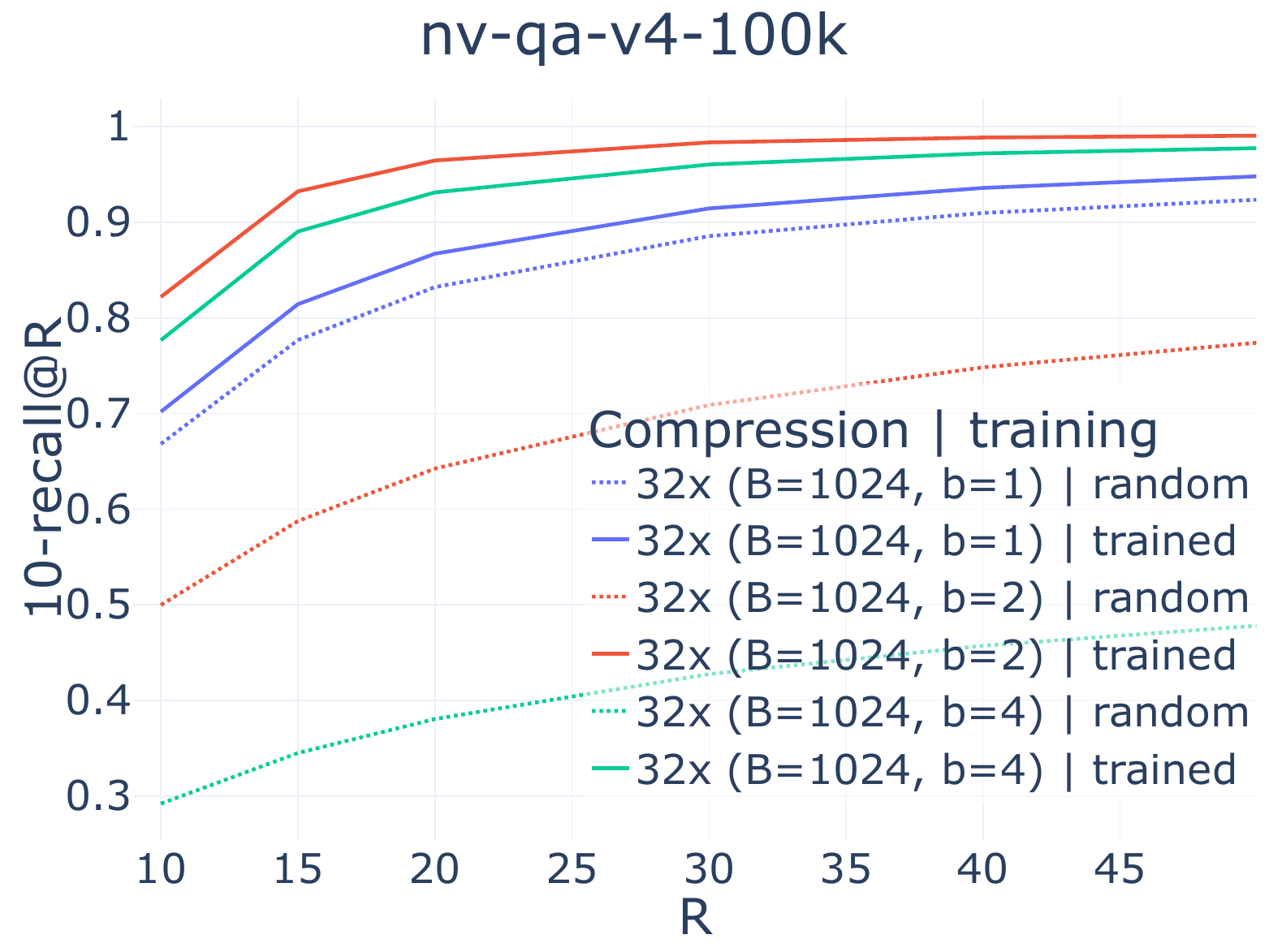}%

    \vspace{1em}
    
    \includegraphics[width=0.20\linewidth]{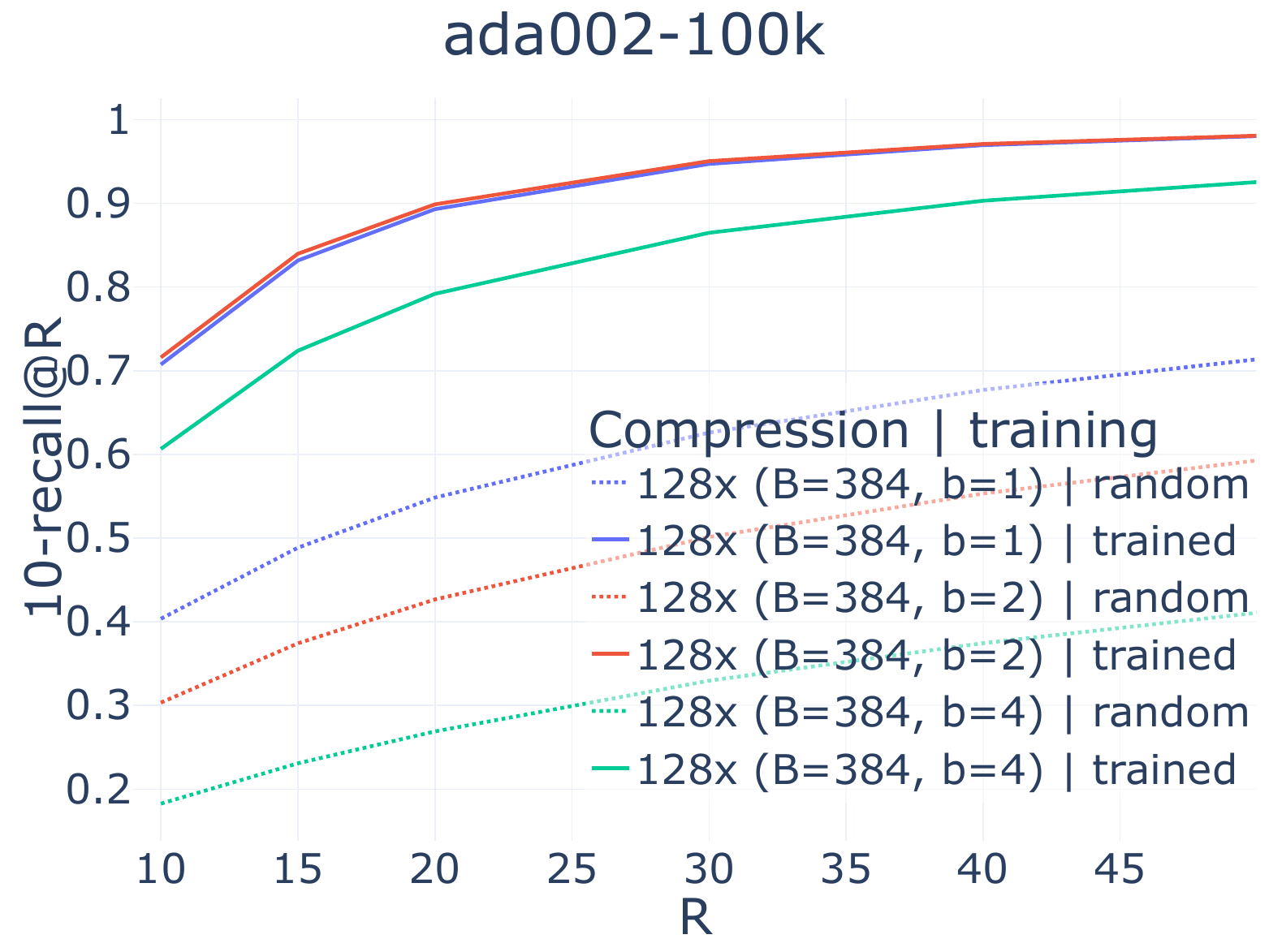}%
    \includegraphics[width=0.20\linewidth]{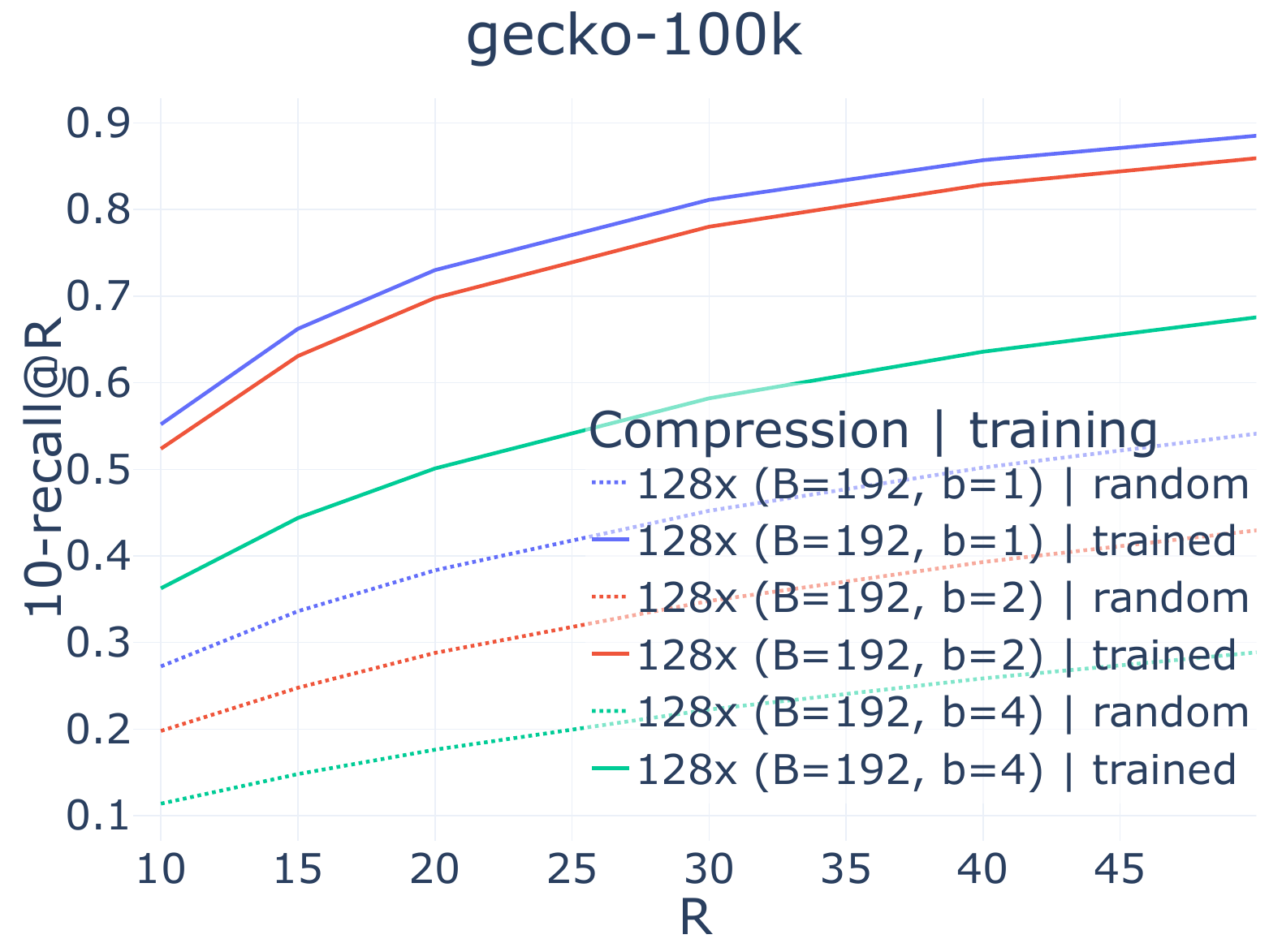}%
    \includegraphics[width=0.20\linewidth]{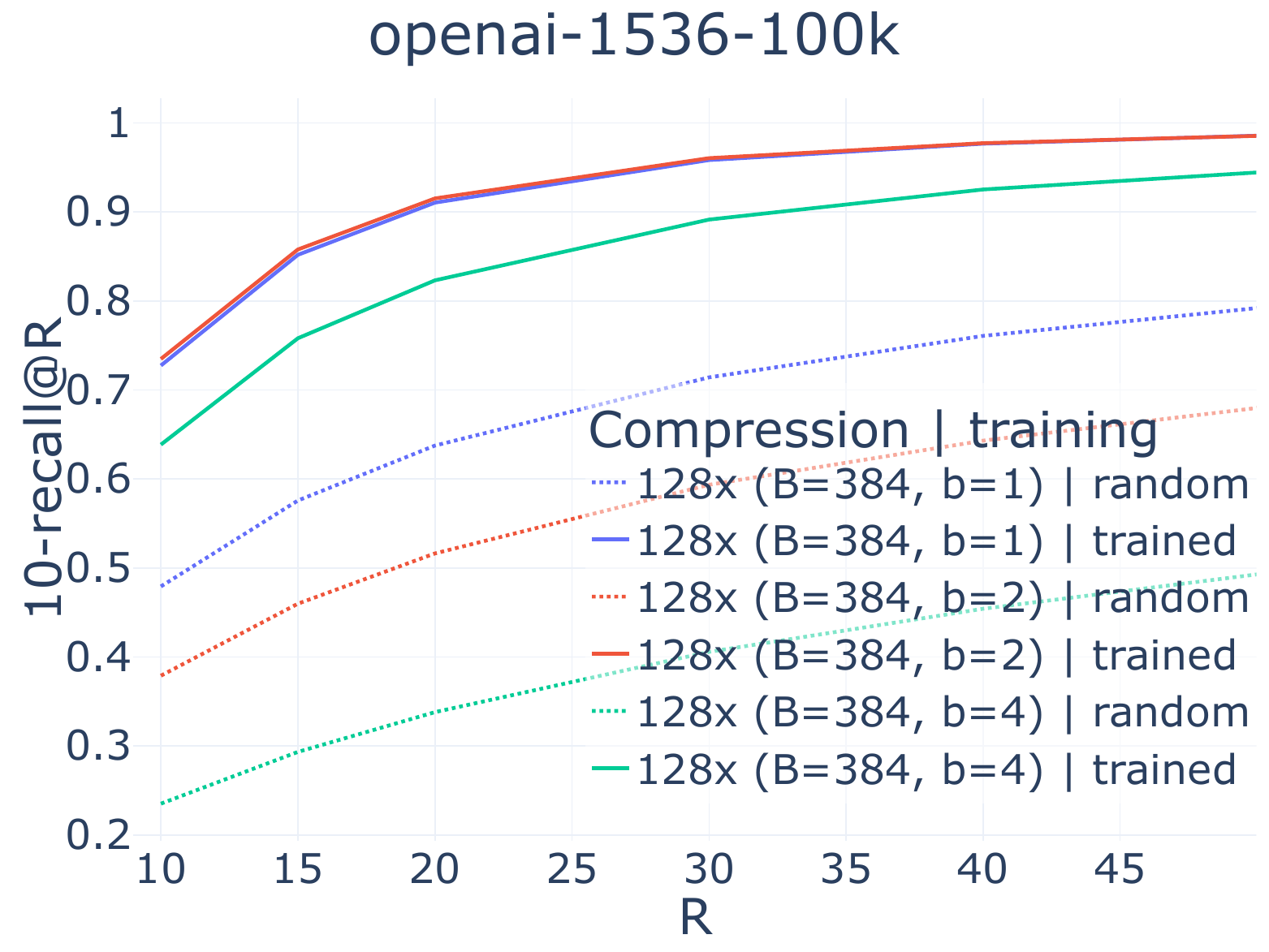}%
    \includegraphics[width=0.20\linewidth]{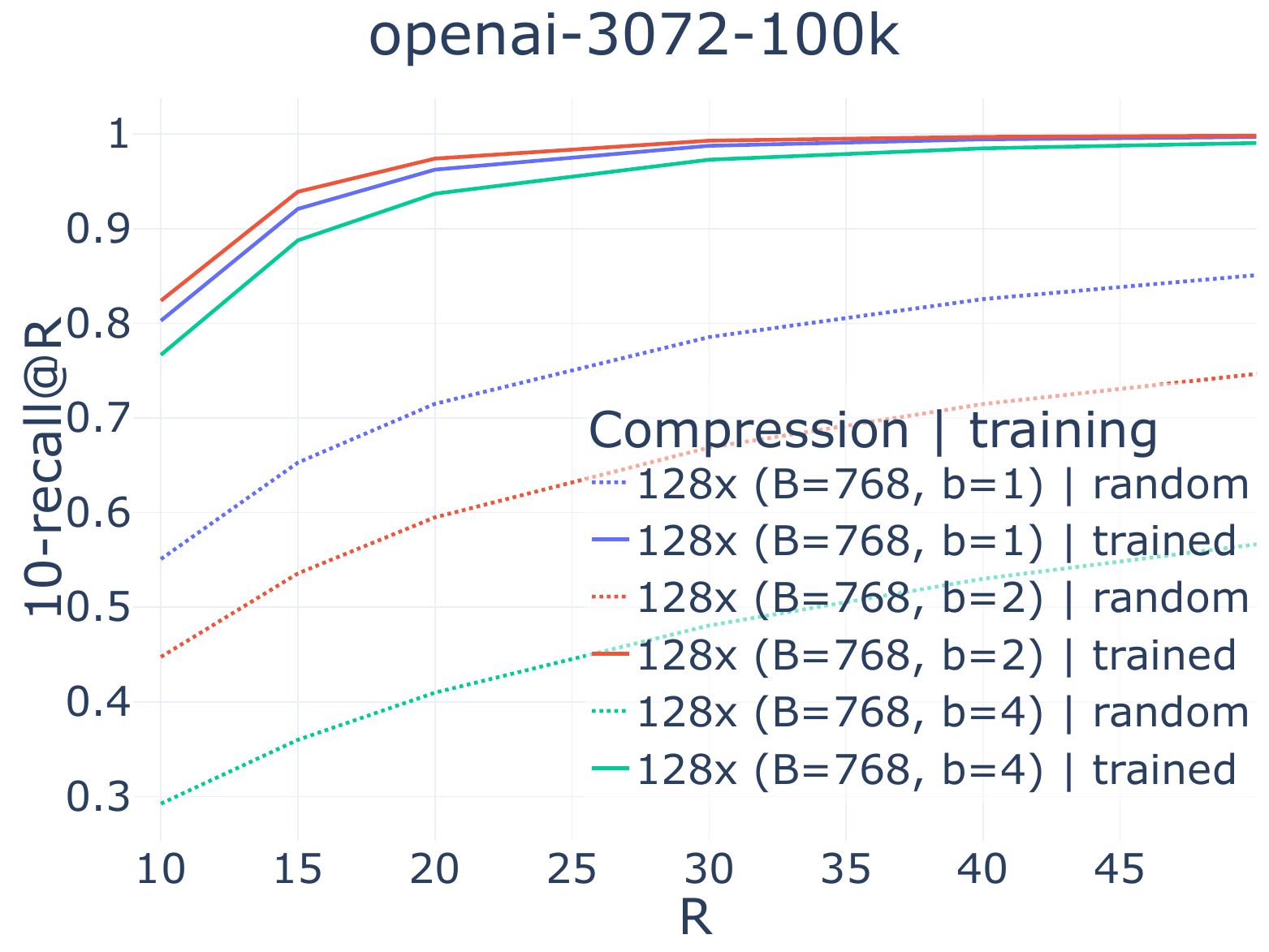}%
    \includegraphics[width=0.20\linewidth]{figures/ash_training_nv-qa-v4-100k_64x.pdf}%

    \caption{Learning the projection matrix $\mat{W}$ leads to significant improvements on search accuracy, measured as 10-recall@R  for $B=D$ (top) and $B=D/4$ (bottom). In ASH, increasing the bitrate $b$ with $B$ fixed means decreasing the target dimensionality $d$. When $D > d$, the advantage of the learned parameters becomes wider. 
    Notably, ASH with $b=2$ consistently beats $b=1$, meaning that reducing the dimensionality while increasing the bitrate pays off.}
    \label{fig:ash_training_continued}
\end{figure*}

\begin{figure*}
    \centering
    \includegraphics[width=0.25\linewidth]{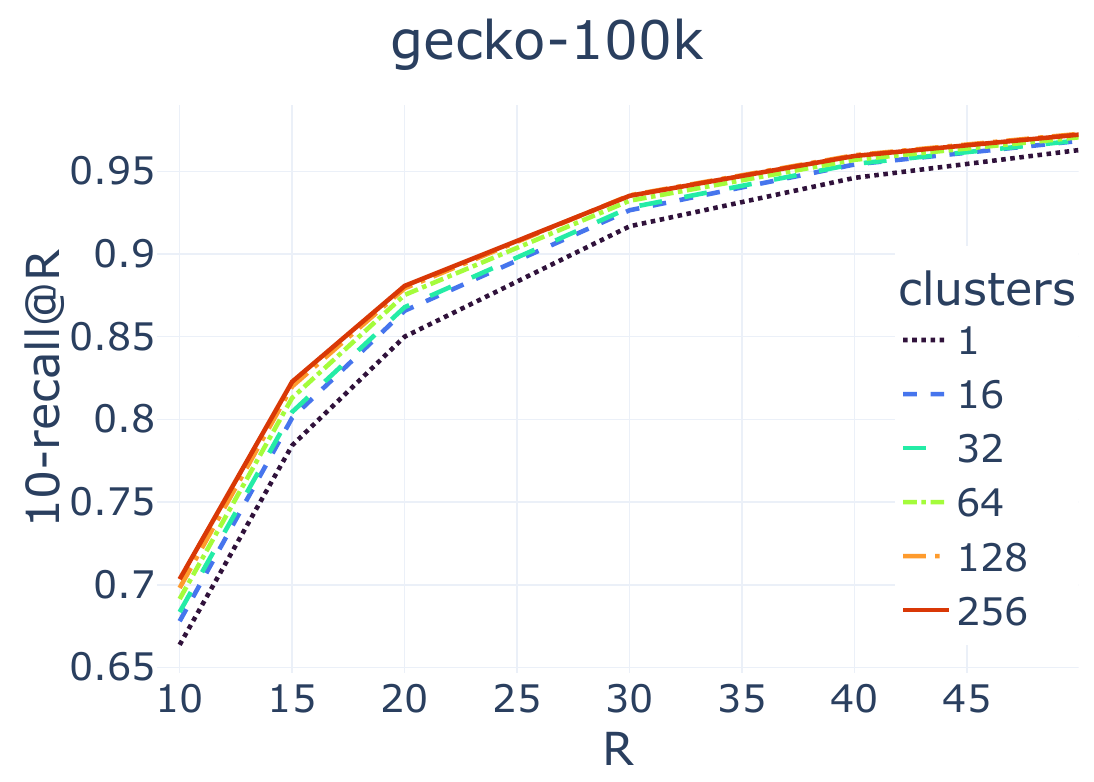}%
    \includegraphics[width=0.25\linewidth]{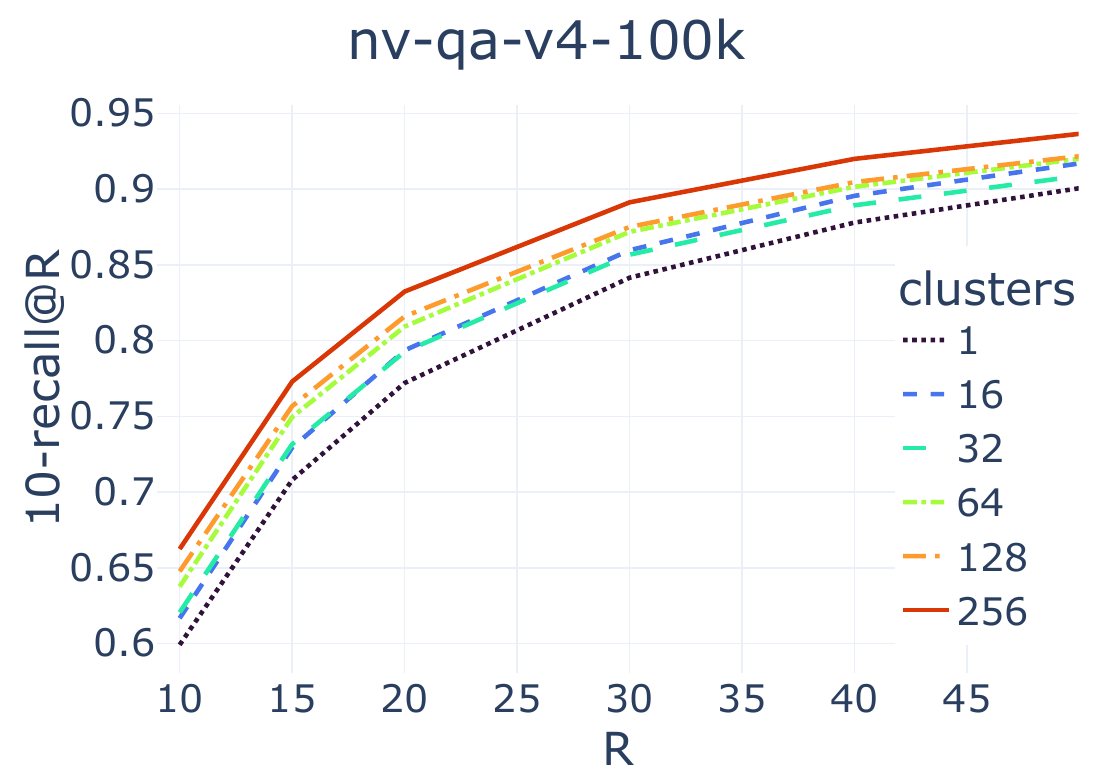}%
    \includegraphics[width=0.25\linewidth]{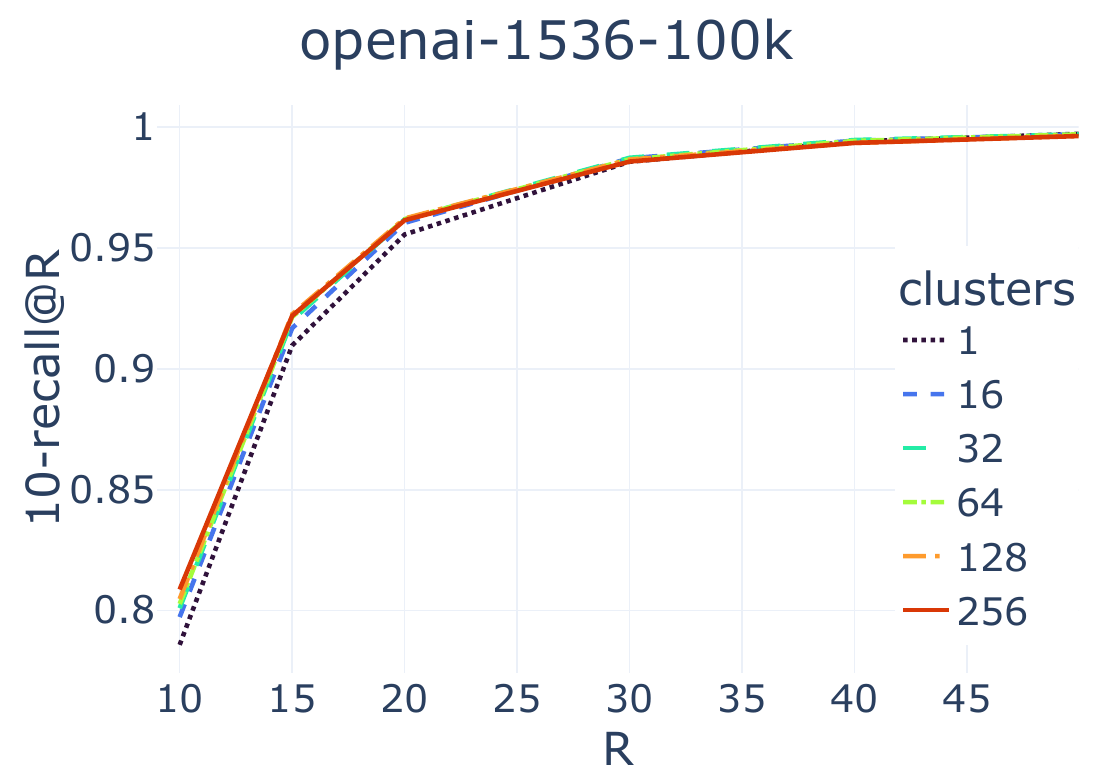}%
    \includegraphics[width=0.25\linewidth]{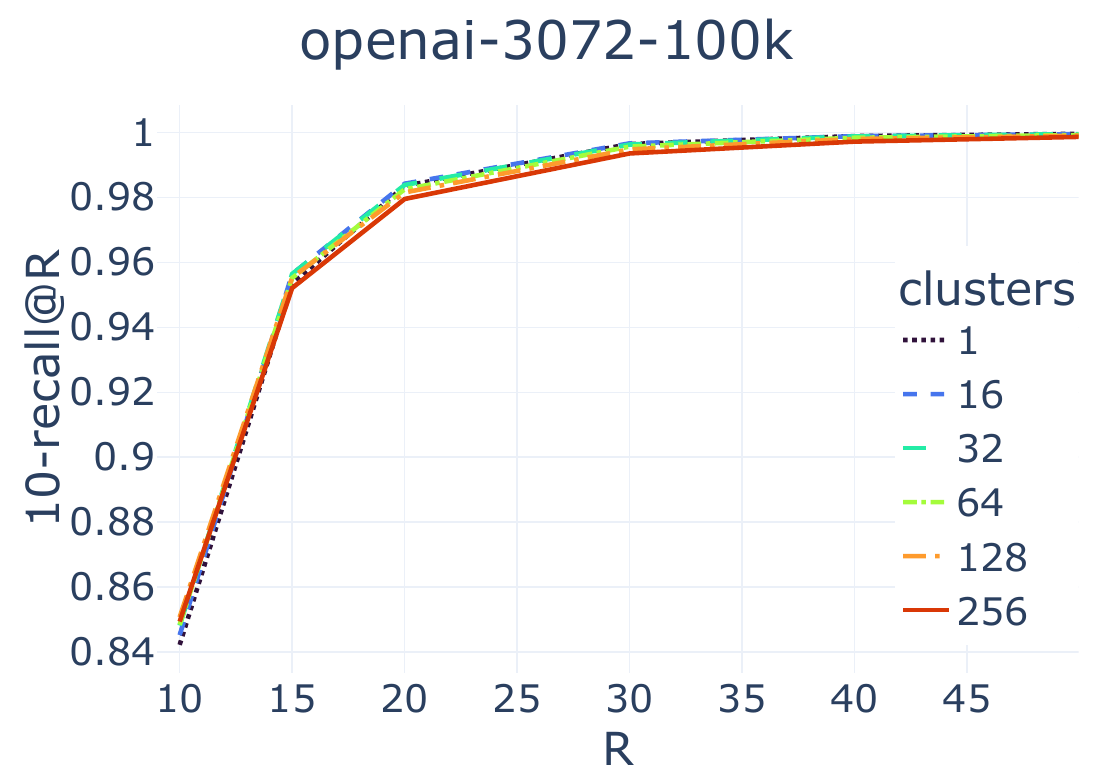}%
    
    \includegraphics[width=0.25\linewidth]{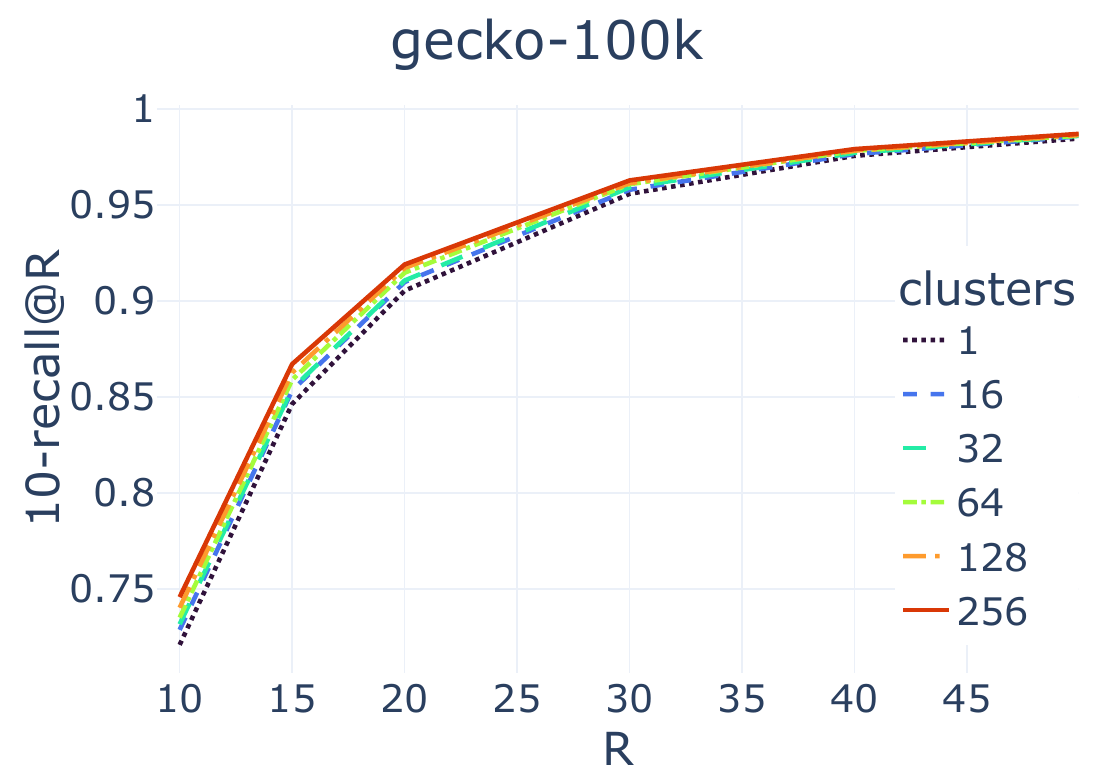}%
    \includegraphics[width=0.25\linewidth]{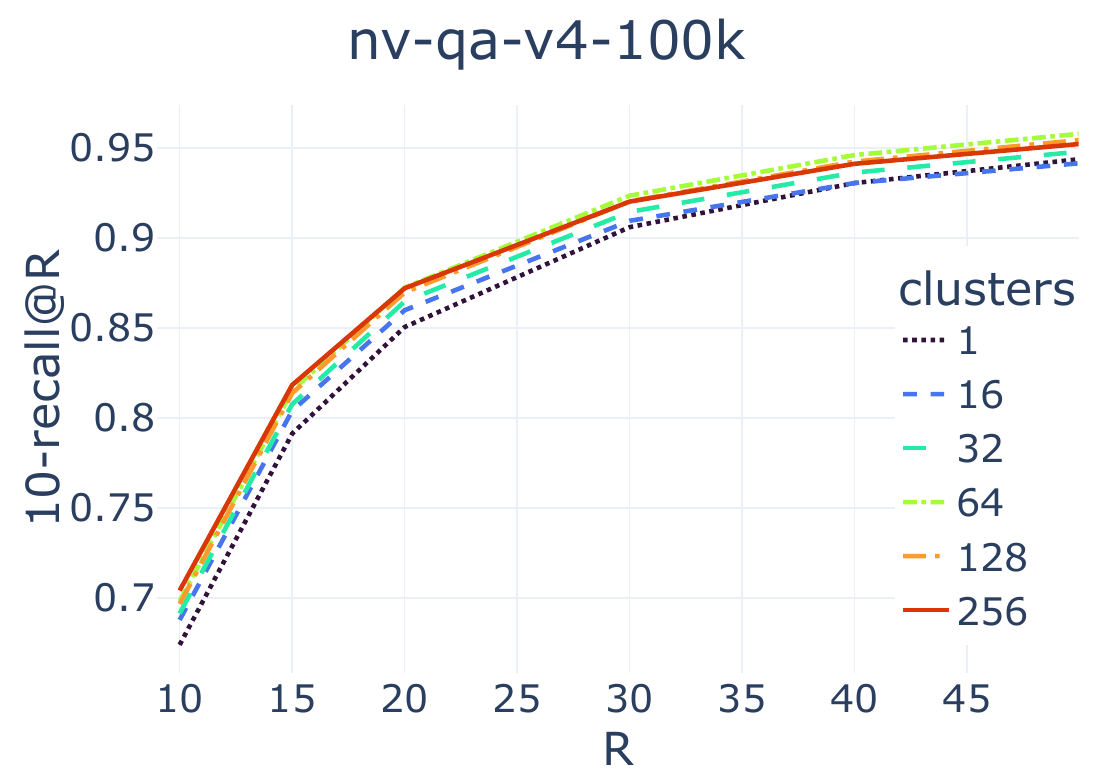}%
    \includegraphics[width=0.25\linewidth]{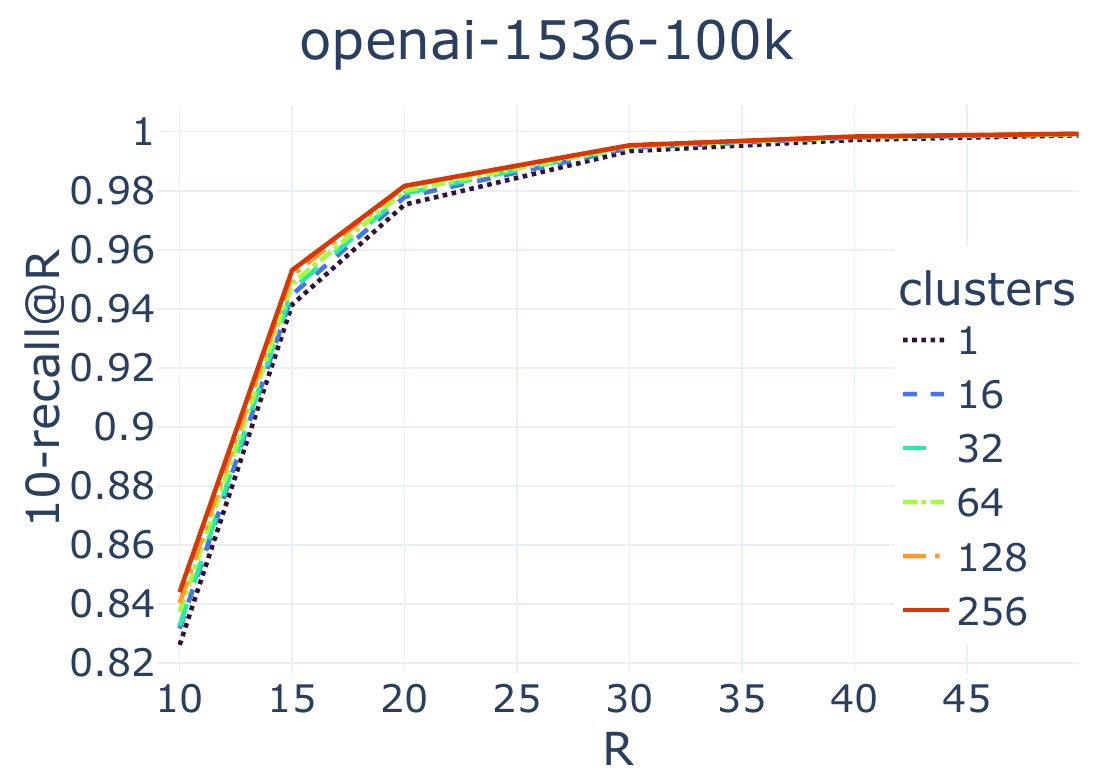}%
    \includegraphics[width=0.25\linewidth]{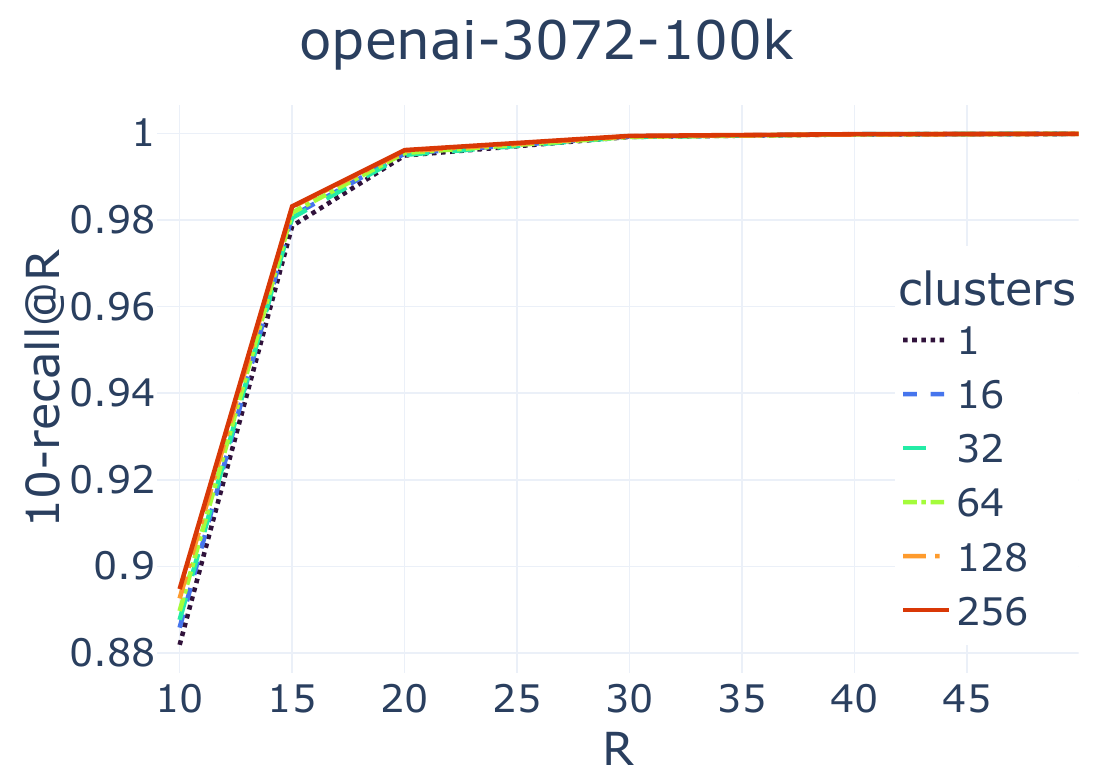}%
    
    \caption{The search accuracy increases with the number of ASH landmarks, defined in \zcref{eq:x_normalized}. We set $B = D / 2$, $b=1$ (top row) and $b=2$ (bottomr row, and use $C=1,16,64,128,256$.}
    \label{fig:ash_groups_continued}
\end{figure*}

\begin{figure*}
    \centering
    \includegraphics[width=0.2\linewidth]{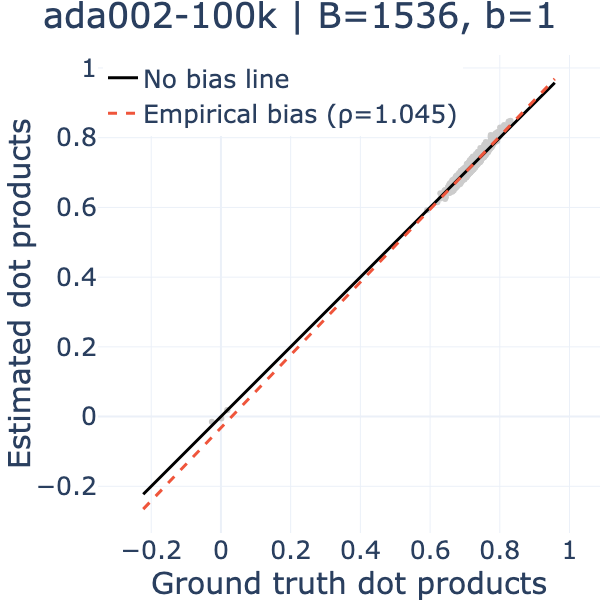}%
    \includegraphics[width=0.2\linewidth]{figures/ash_bias_openai-v3-large-1536-100k_B=1536_b=1.png}%
    \includegraphics[width=0.2\linewidth]{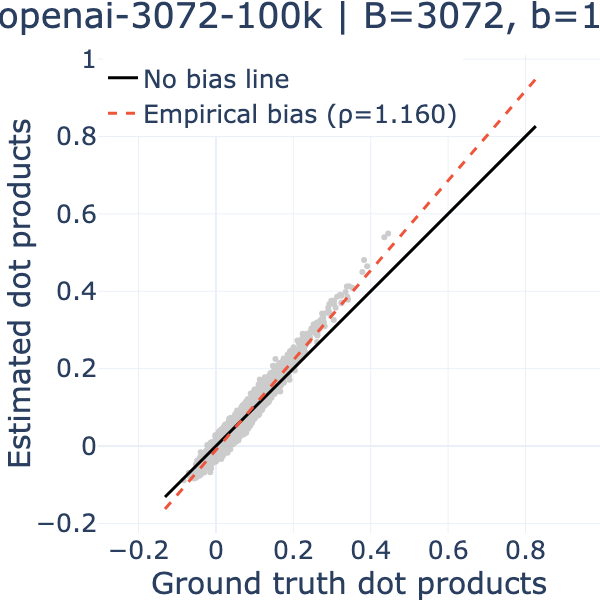}%
    \includegraphics[width=0.2\linewidth]{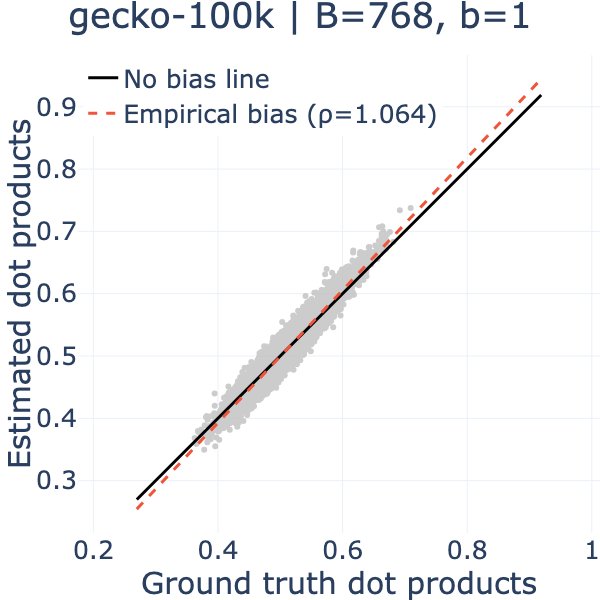}%
    \includegraphics[width=0.2\linewidth]{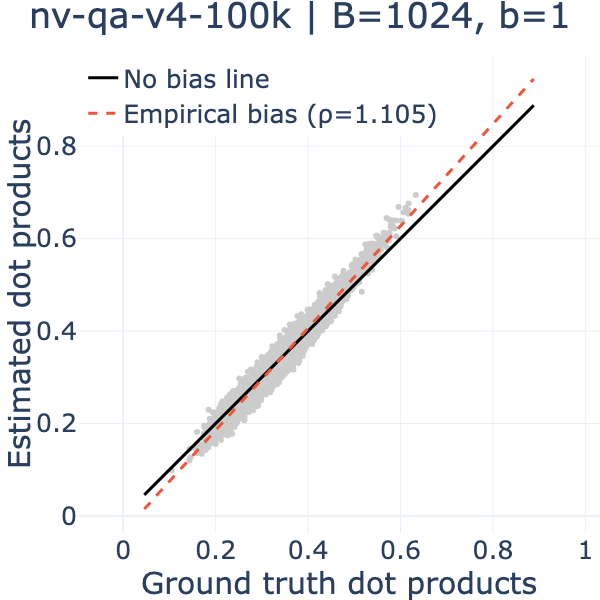}%

    \includegraphics[width=0.2\linewidth]{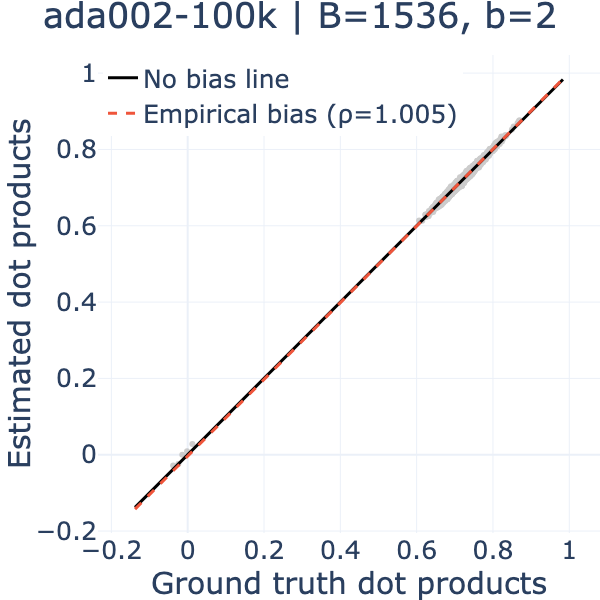}%
    \includegraphics[width=0.2\linewidth]{figures/ash_bias_openai-v3-large-1536-100k_B=1536_b=2.png}%
    \includegraphics[width=0.2\linewidth]{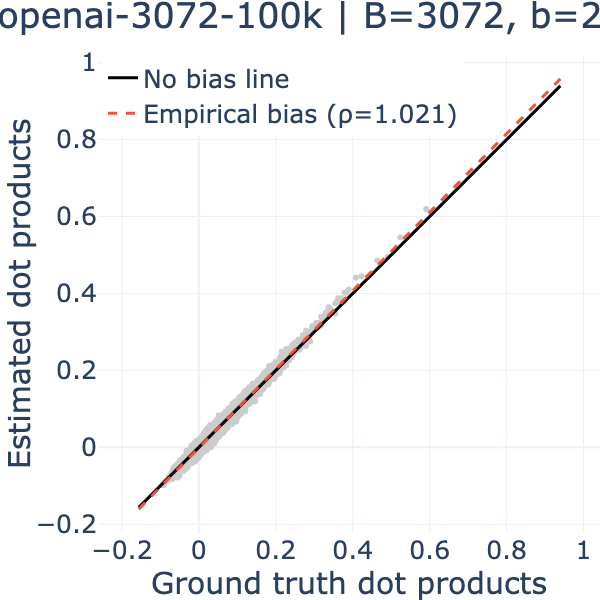}%
    \includegraphics[width=0.2\linewidth]{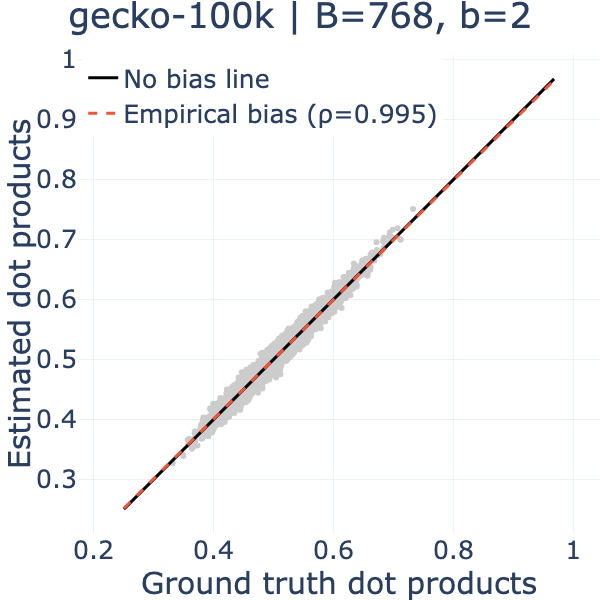}%
    \includegraphics[width=0.2\linewidth]{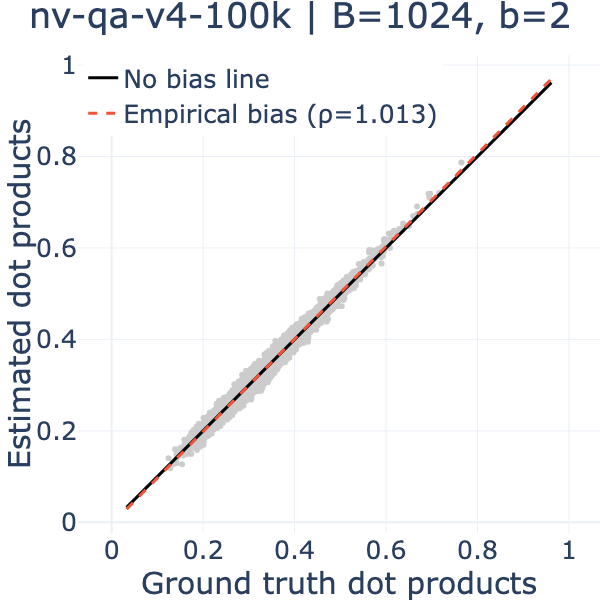}%

    \includegraphics[width=0.2\linewidth]{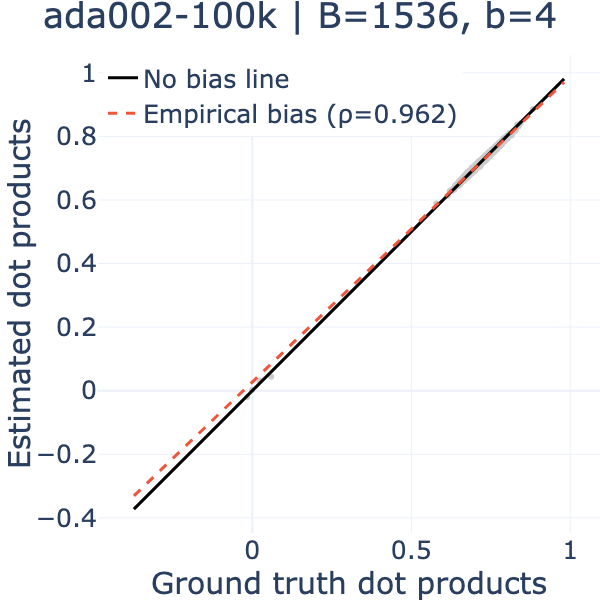}%
    \includegraphics[width=0.2\linewidth]{figures/ash_bias_openai-v3-large-1536-100k_B=1536_b=4.png}%
    \includegraphics[width=0.2\linewidth]{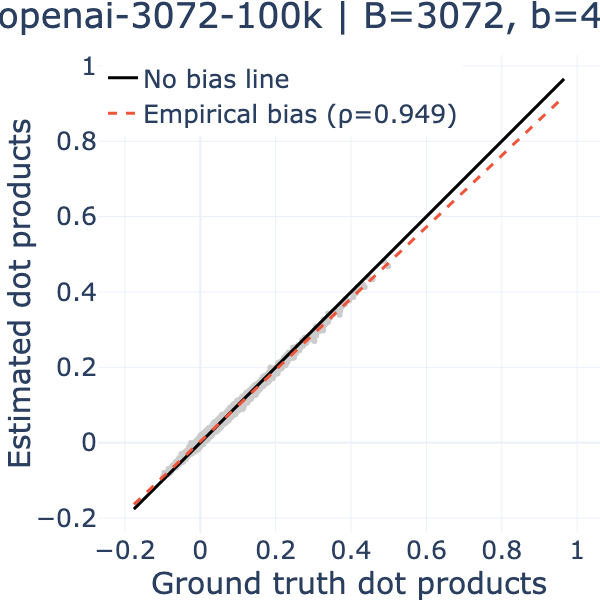}%
    \includegraphics[width=0.2\linewidth]{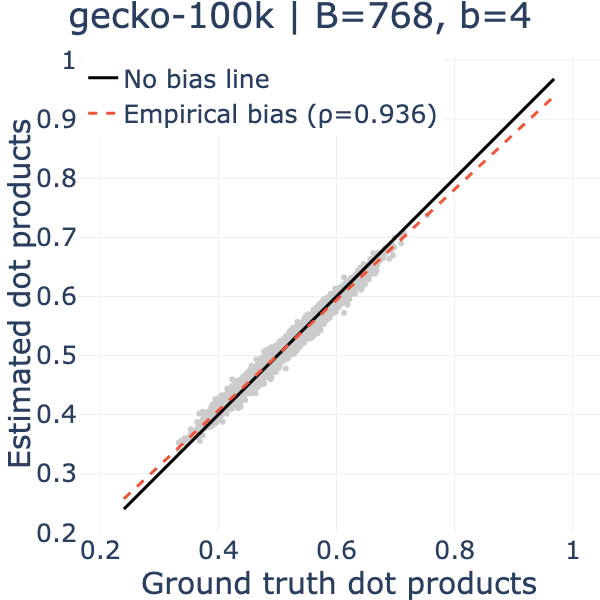}%
    \includegraphics[width=0.2\linewidth]{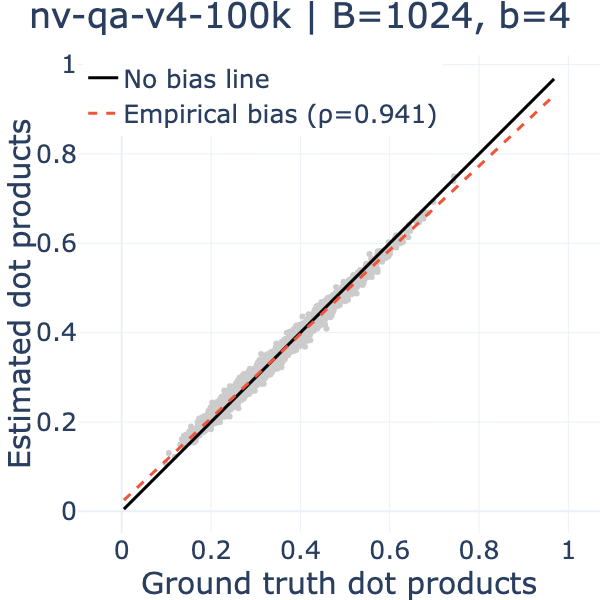}%
    
    \caption{The ASH estimator has a slight bias, see \zcref{eq:regression_coefficient_bias}, when comparing  the ground truth $\langle \vect{q}, \vect{x} \rangle$ with the estimated $\langle \vect{q}, \operatorname{quant}(\vect{x}) \rangle$ (each pair $(\langle \vect{q}, \vect{x} \rangle, \langle \vect{q}, \operatorname{quant}(\vect{x}) \rangle)$ corresponds to a circle in the plot) using the dataset queries. The bias slope $\rho$ depends number of bits $B$ and the bitrate $b$ (in these examples $d = \lfloor (B - 32) / b \rfloor$). In this case, $B = D$.}
    \label{fig:ash_bias_continued}
\end{figure*}

\begin{figure*}
    \centering
    \includegraphics[width=0.2\linewidth]{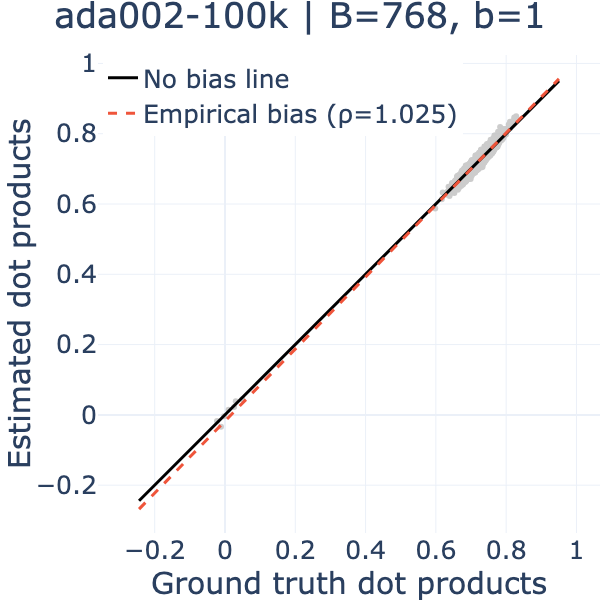}%
    \includegraphics[width=0.2\linewidth]{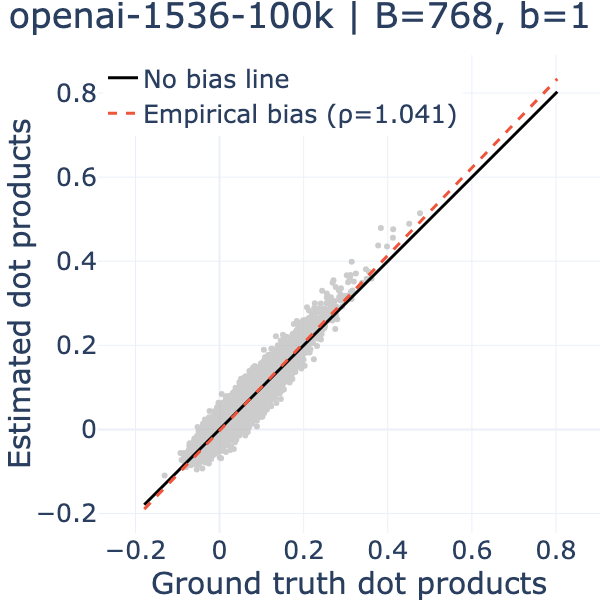}%
    \includegraphics[width=0.2\linewidth]{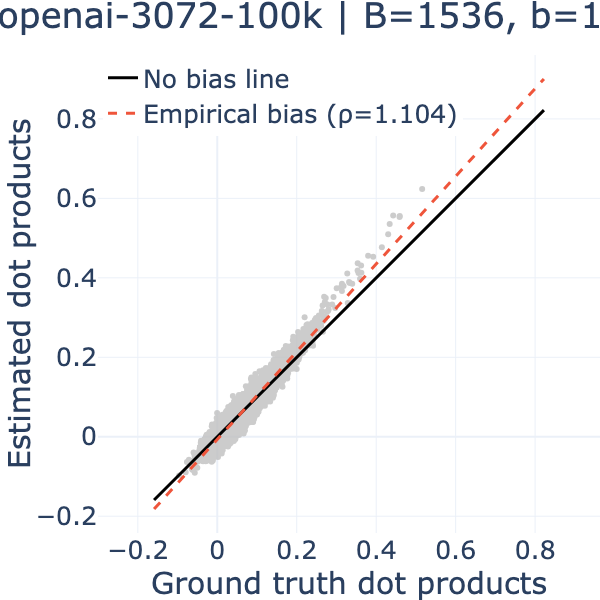}%
    \includegraphics[width=0.2\linewidth]{figures/ash_bias_gecko-768-100k_B=768_b=1.png}%
    \includegraphics[width=0.2\linewidth]{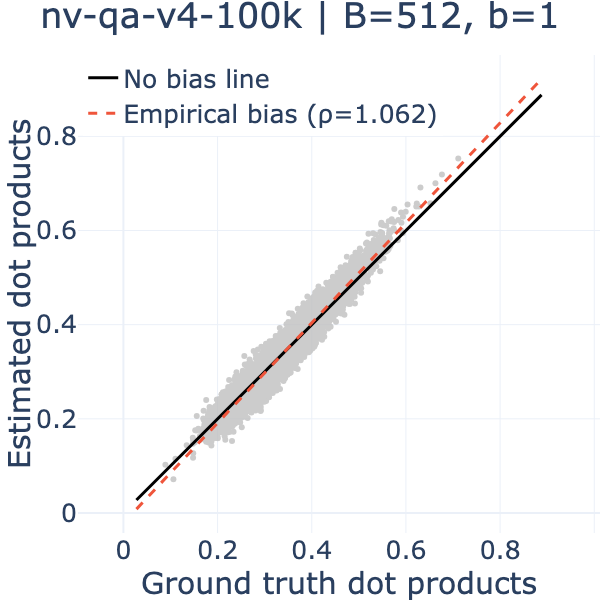}%

    \includegraphics[width=0.2\linewidth]{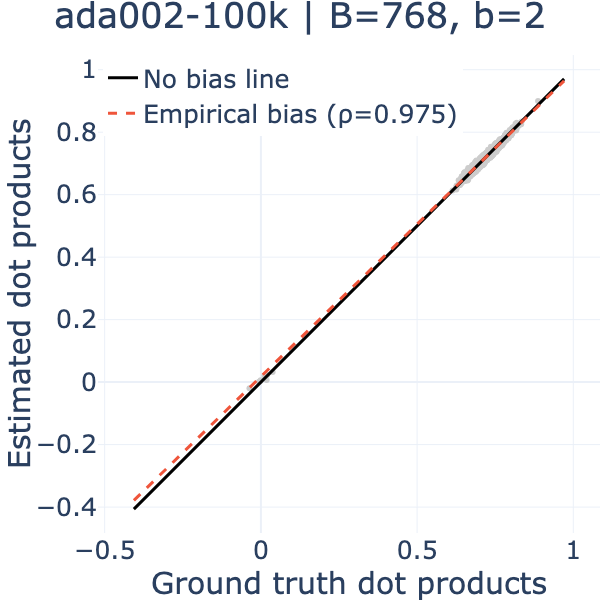}%
    \includegraphics[width=0.2\linewidth]{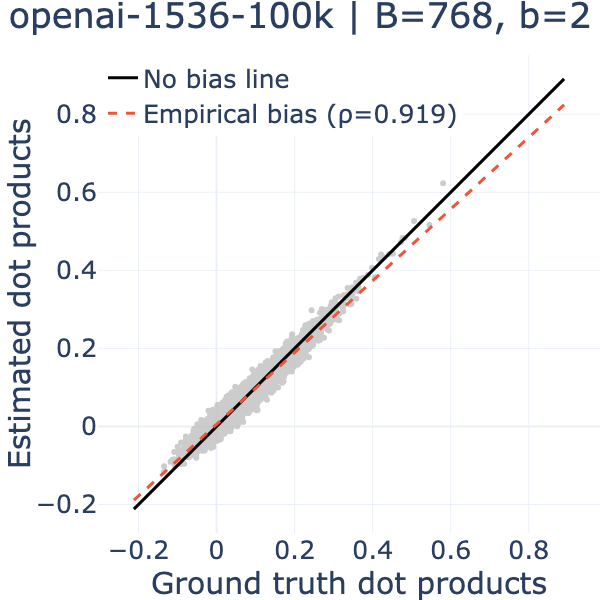}%
    \includegraphics[width=0.2\linewidth]{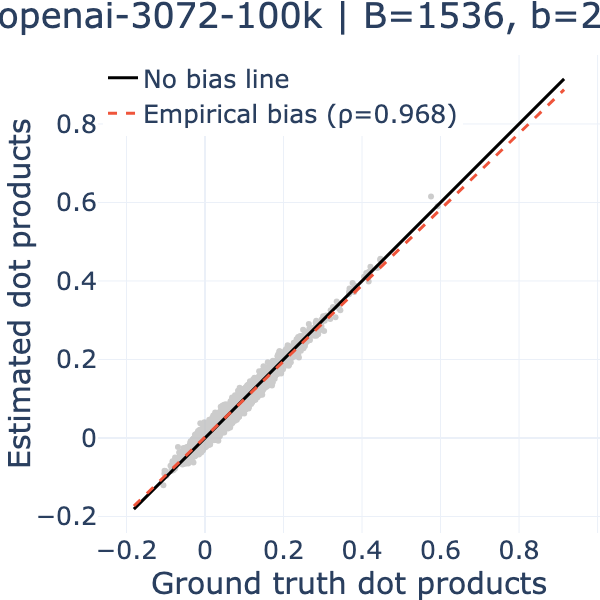}%
    \includegraphics[width=0.2\linewidth]{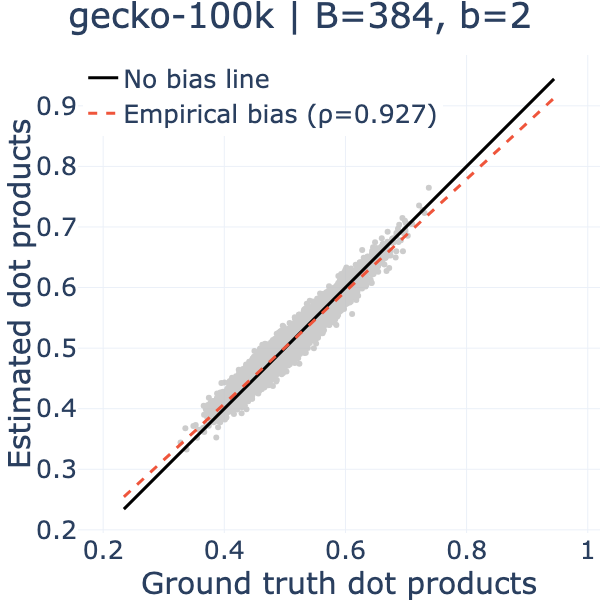}%
    \includegraphics[width=0.2\linewidth]{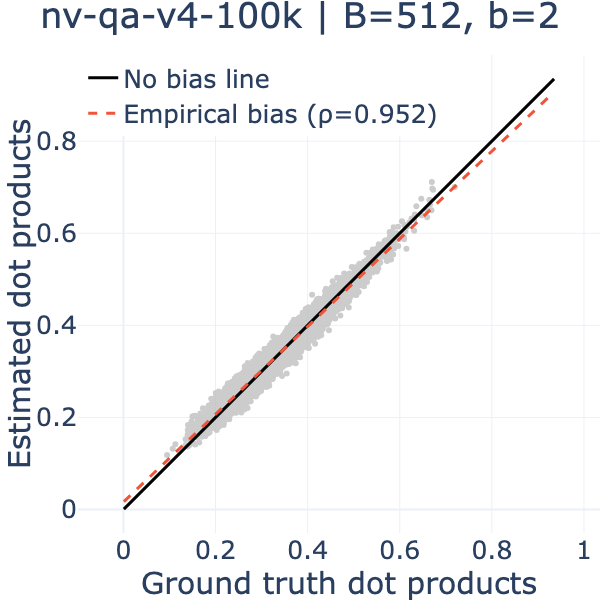}%

    \includegraphics[width=0.2\linewidth]{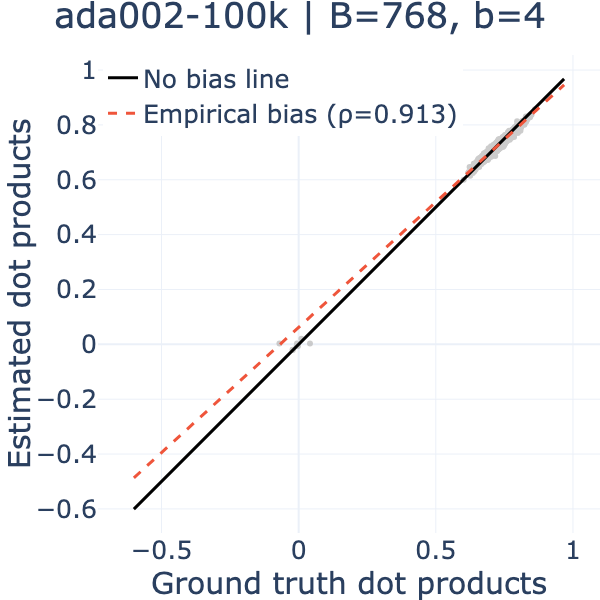}%
    \includegraphics[width=0.2\linewidth]{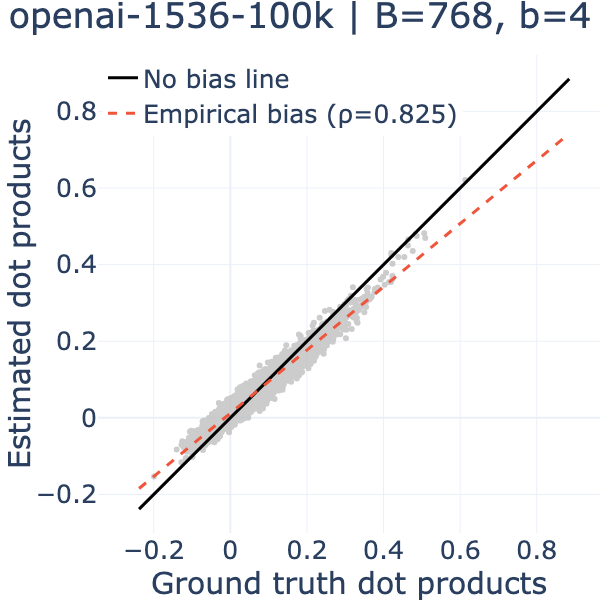}%
    \includegraphics[width=0.2\linewidth]{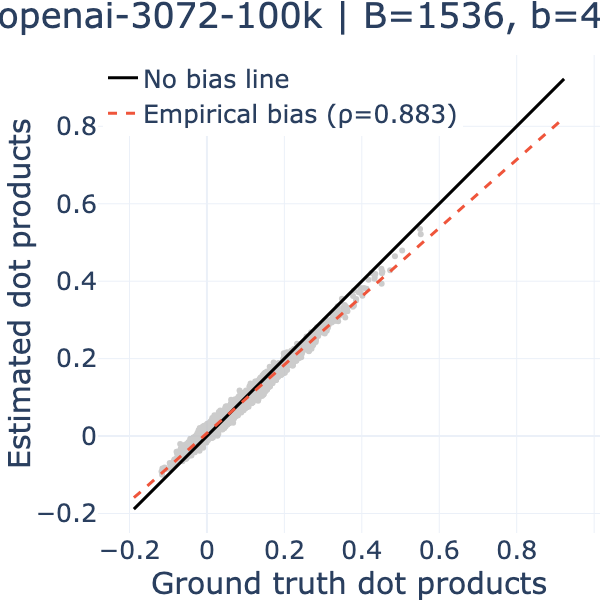}%
    \includegraphics[width=0.2\linewidth]{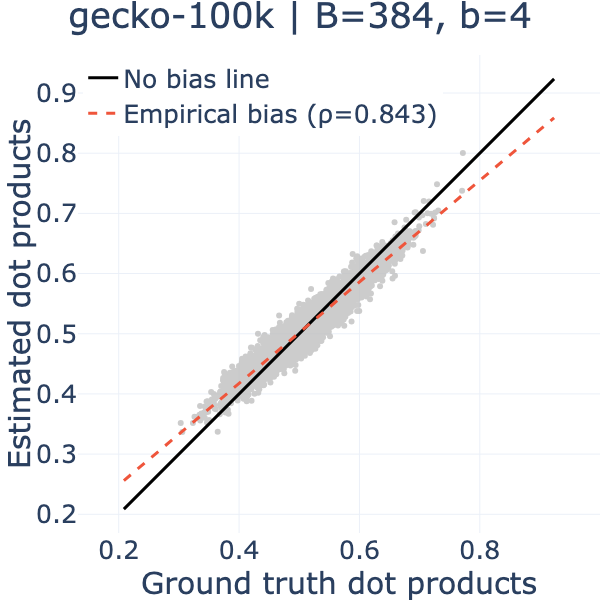}%
    \includegraphics[width=0.2\linewidth]{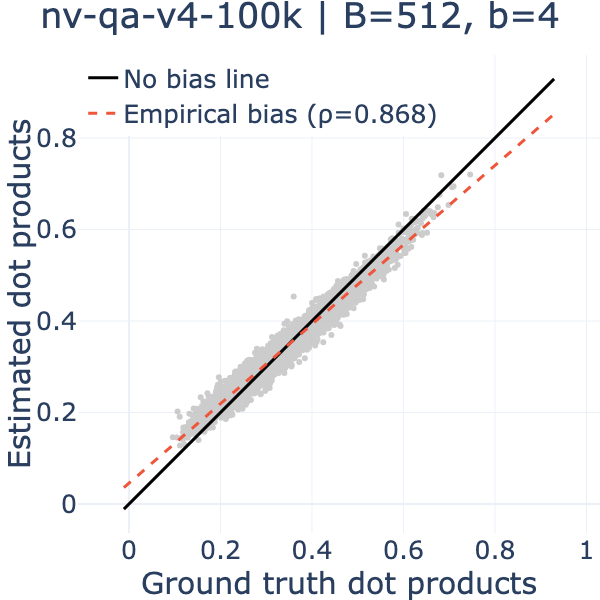}%
    
    \caption{The ASH estimator has a slight bias, see \zcref{eq:regression_coefficient_bias}, when comparing  the ground truth $\langle \vect{q}, \vect{x} \rangle$ with the estimated $\langle \vect{q}, \operatorname{quant}(\vect{x}) \rangle$ (each pair $(\langle \vect{q}, \vect{x} \rangle, \langle \vect{q}, \operatorname{quant}(\vect{x}) \rangle)$ corresponds to a circle in the plot) using the dataset queries. The bias slope $\rho$ depends number of bits $B$ and the bitrate $b$ ($d = \lfloor (B - 32) / b \rfloor$ in these examples). In this case, $B = D / 2$.}
    \label{fig:ash_bias_continued2}
\end{figure*}

\begin{figure*}
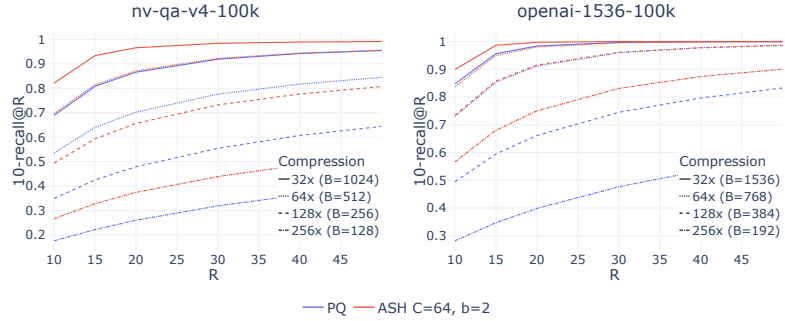

    \centering
    
    \includegraphics[width=0.3\linewidth]{figures/ash_vs_pq_nv-qa-v4-100k.pdf}%
    \hspace{1em}%
    \includegraphics[width=0.3\linewidth]{figures/ash_vs_pq_openai-v3-large-1536-100k.pdf}%

    \includegraphics[width=0.15\linewidth]{figures/ash_vs_pq_legend.pdf}
    
    \caption{ASH outperforms PQ in search accuracy, measured using 10-recall@R for different values of R. ASH is even competitive at several compression levels with PQ configurations that use twice the space.}
    \label{fig:ash_vs_pq_continued}
\end{figure*}

\begin{figure*}
    \centering
    \includegraphics[width=0.2\linewidth]{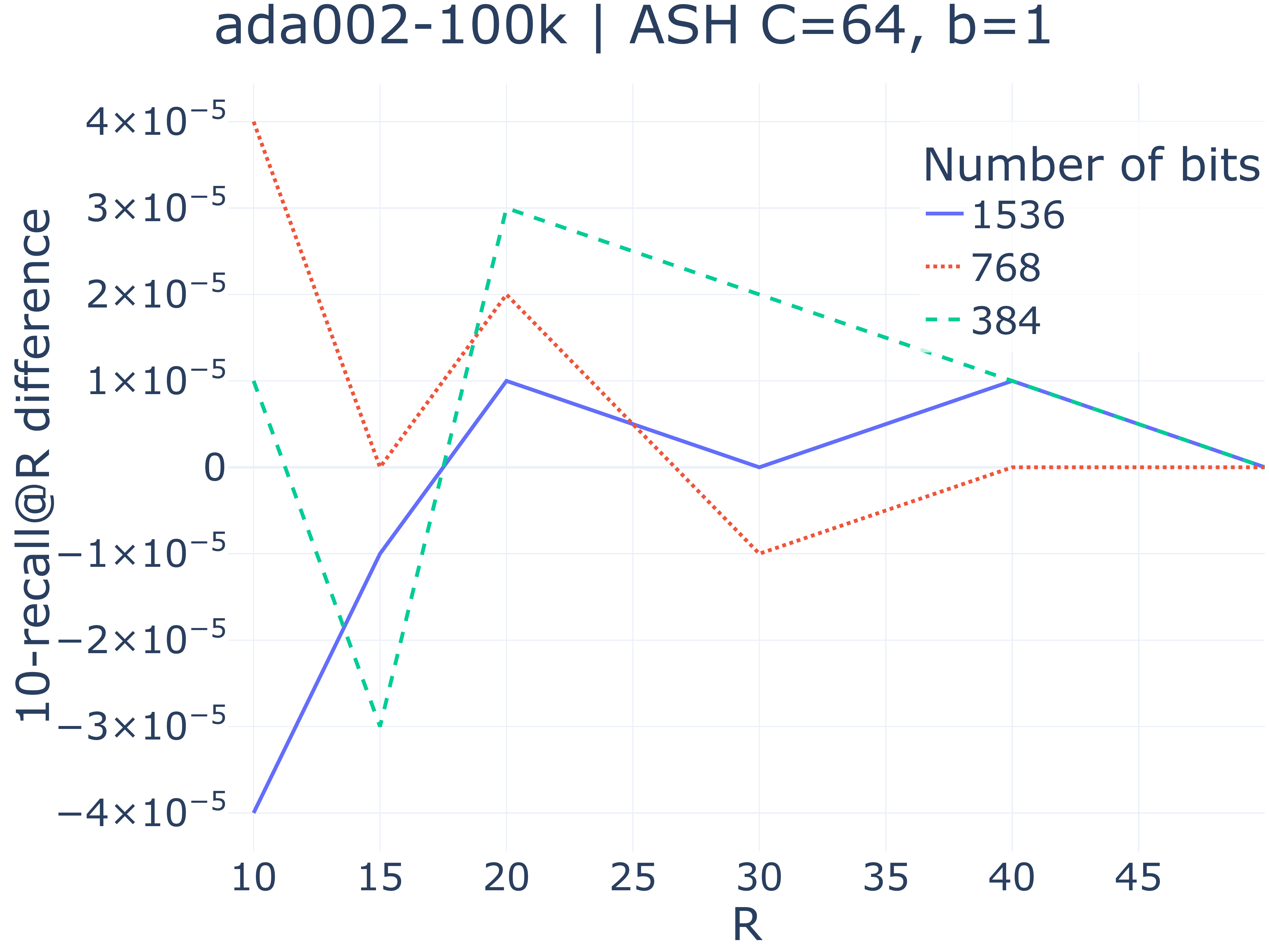}%
    \includegraphics[width=0.2\linewidth]{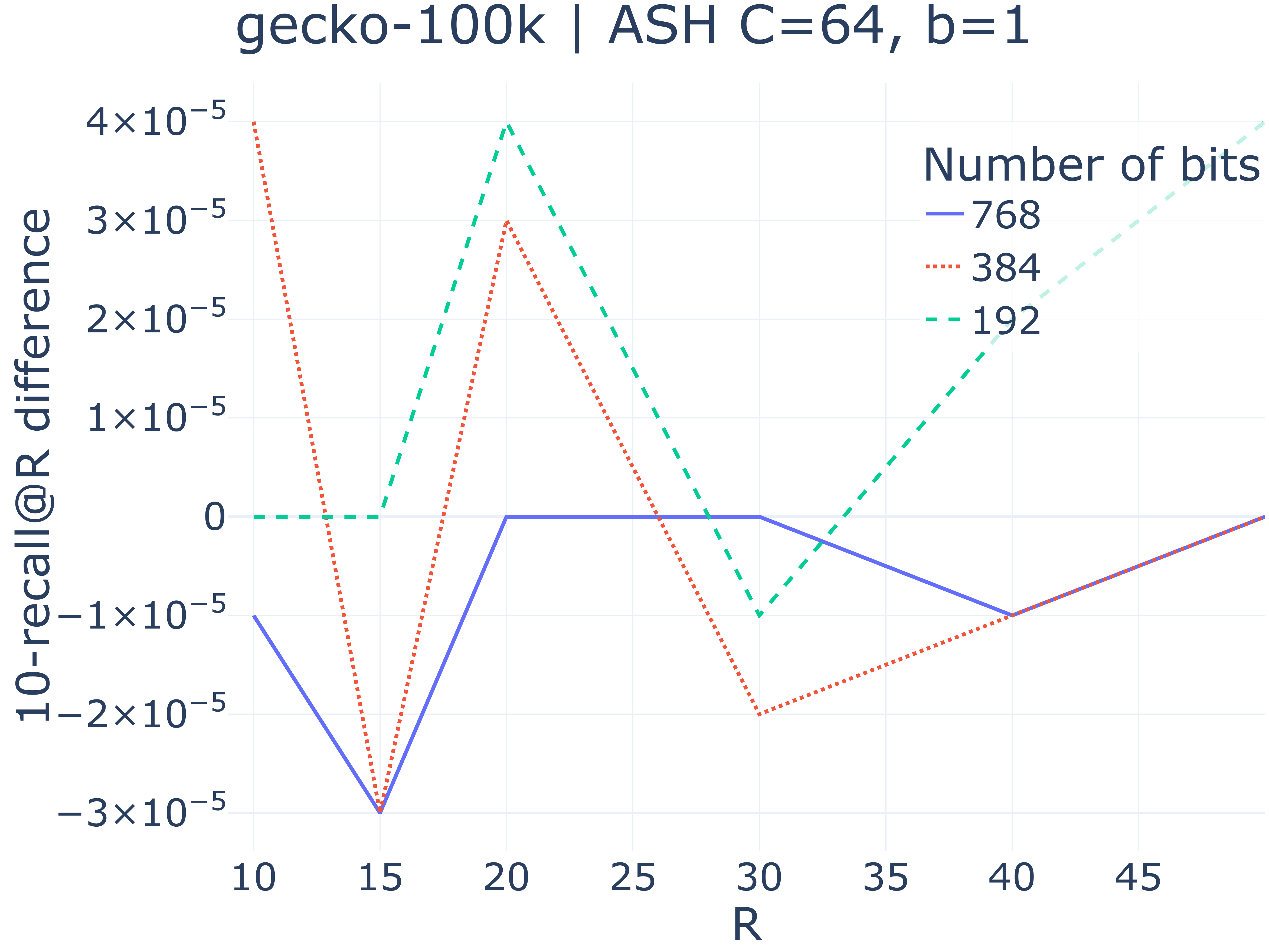}%
    \includegraphics[width=0.2\linewidth]{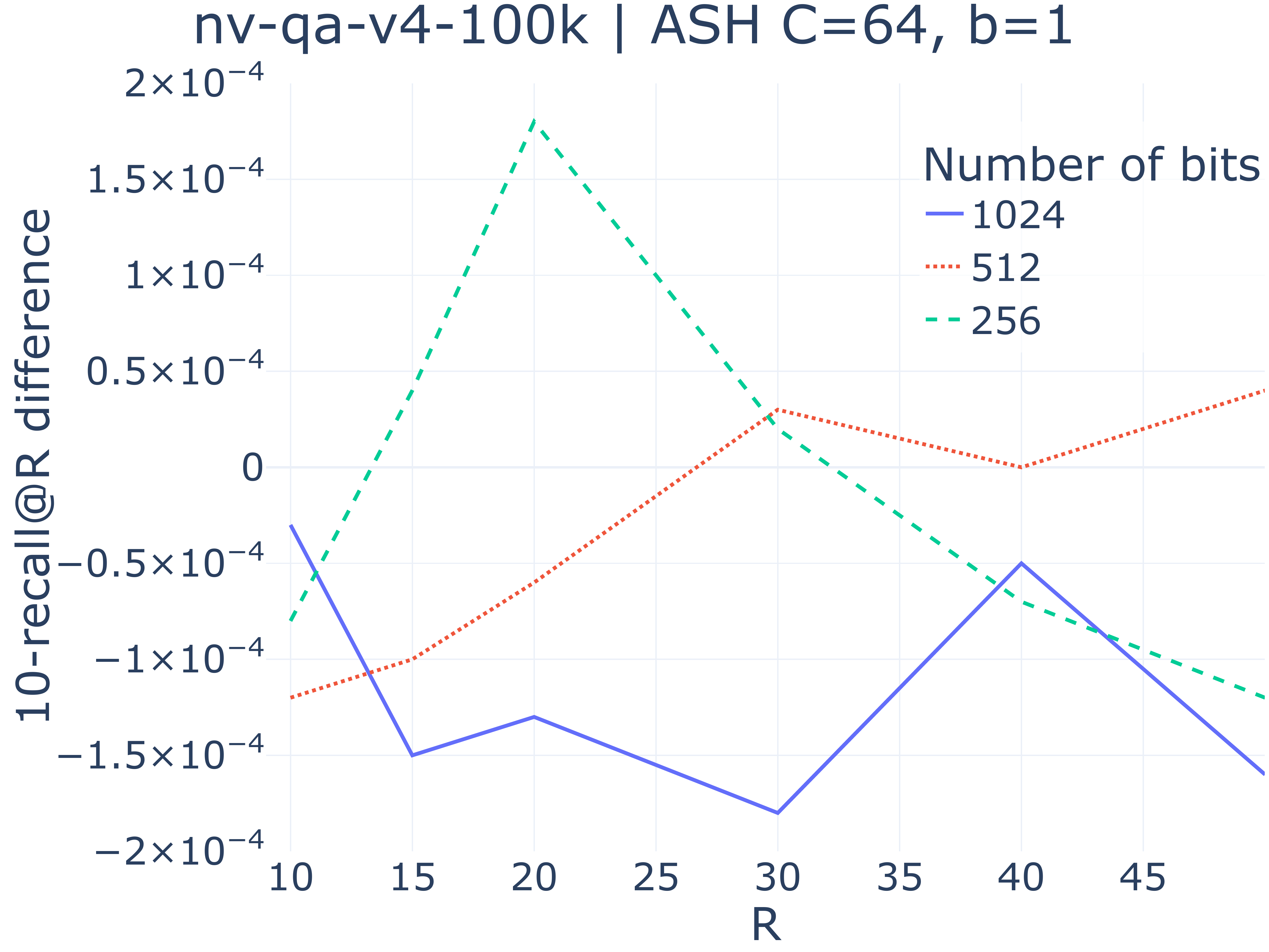}%
    \includegraphics[width=0.2\linewidth]{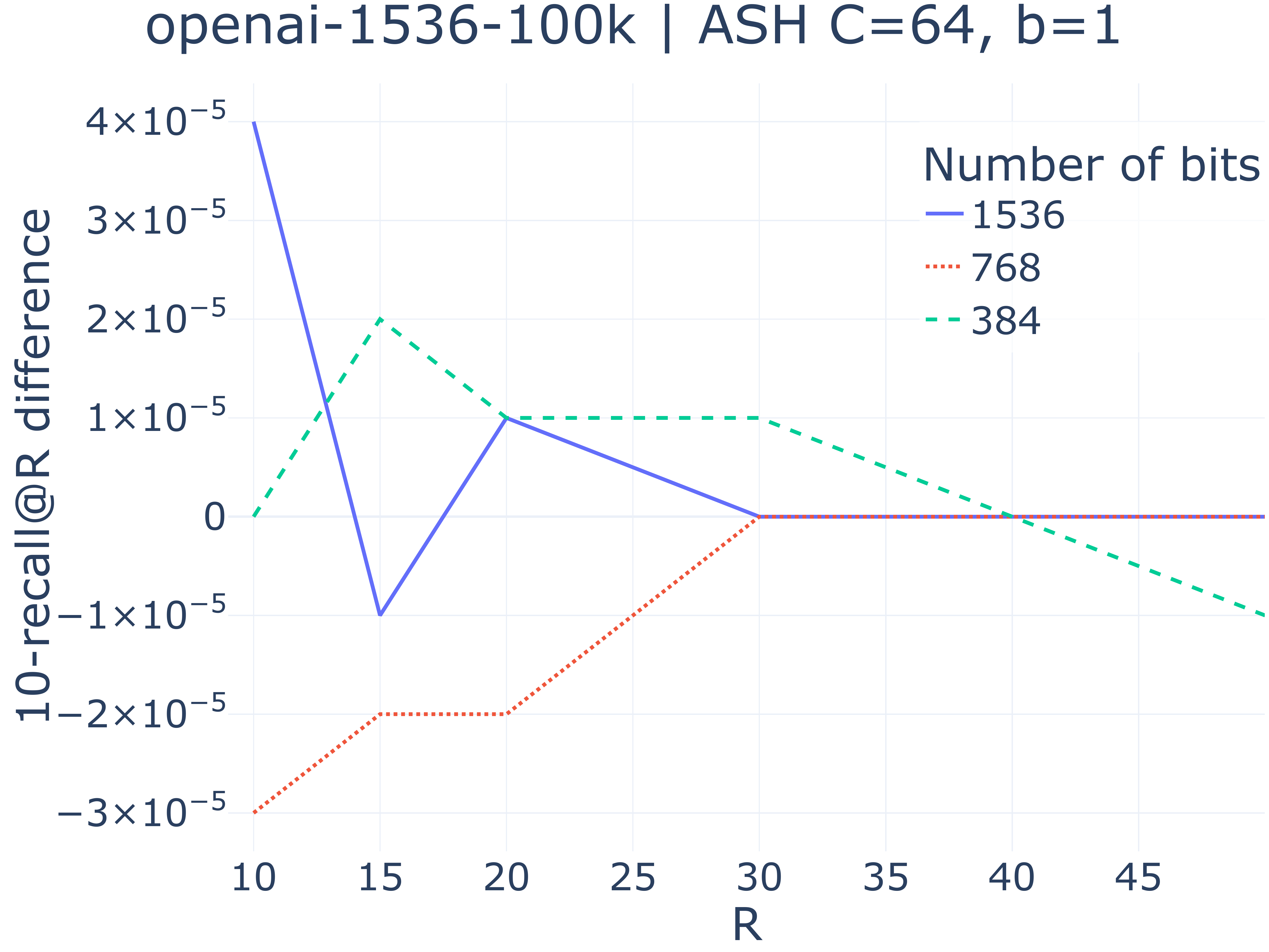}%
    \includegraphics[width=0.2\linewidth]{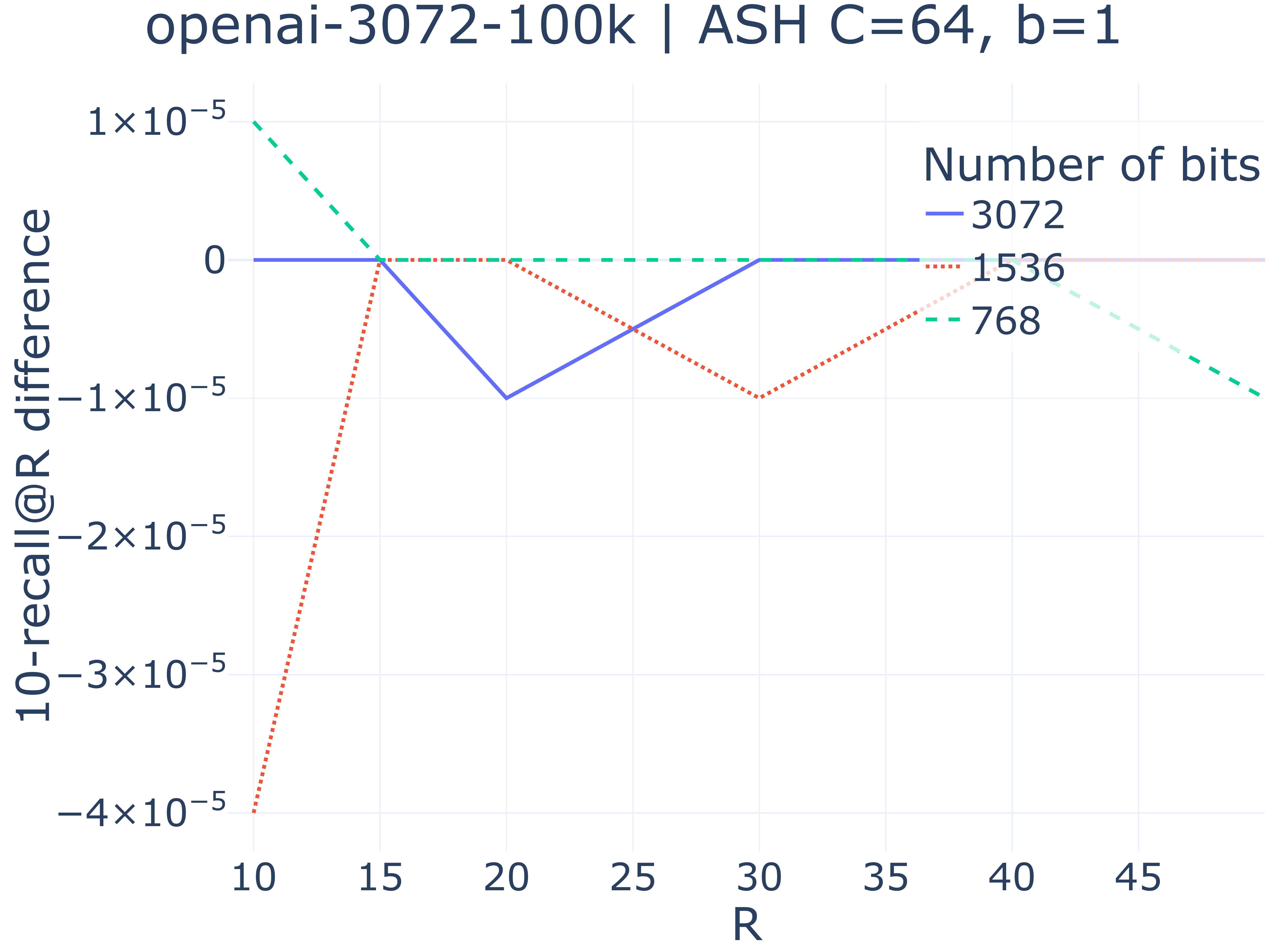}%
        
    \includegraphics[width=0.2\linewidth]{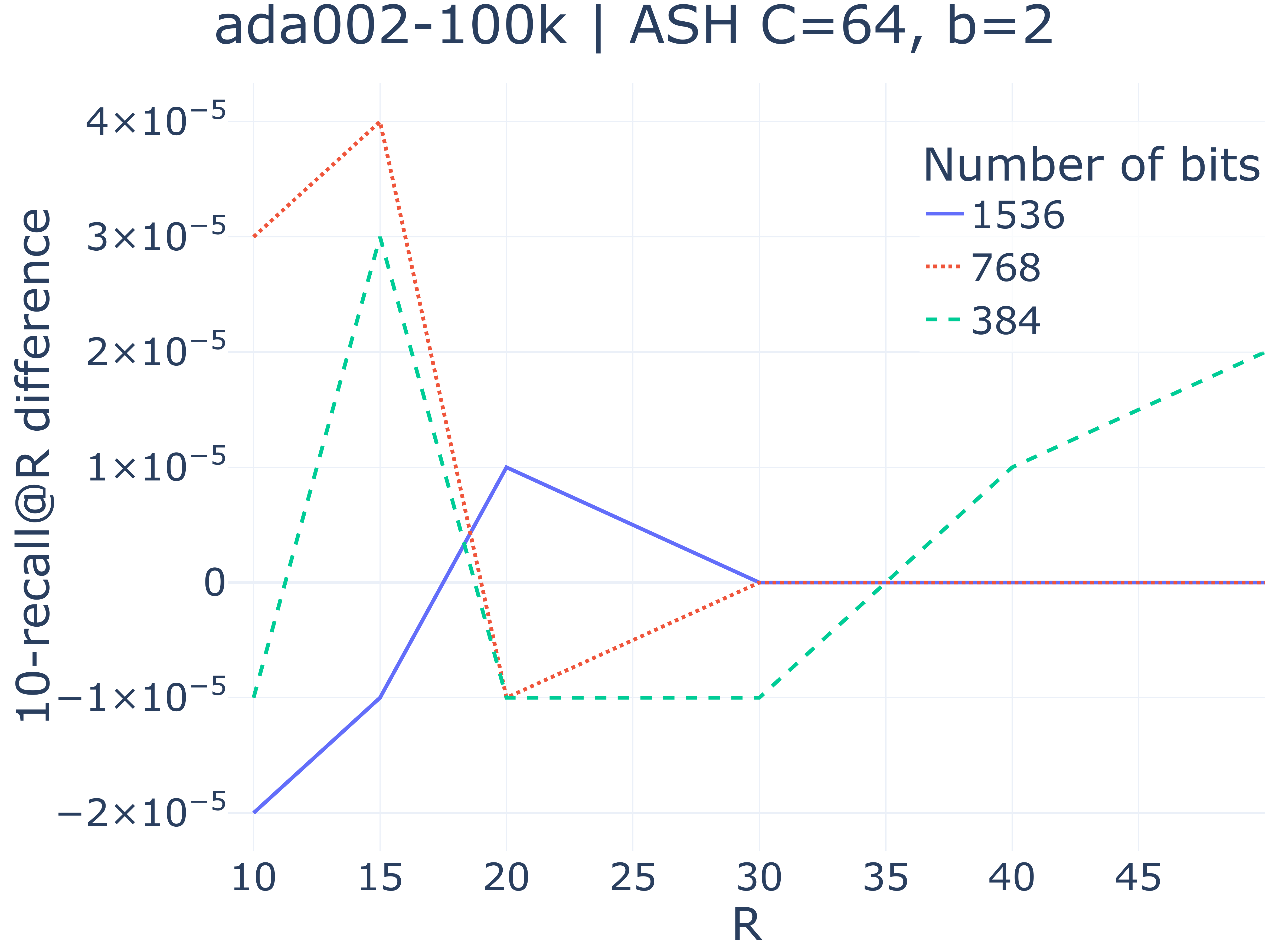}%
    \includegraphics[width=0.2\linewidth]{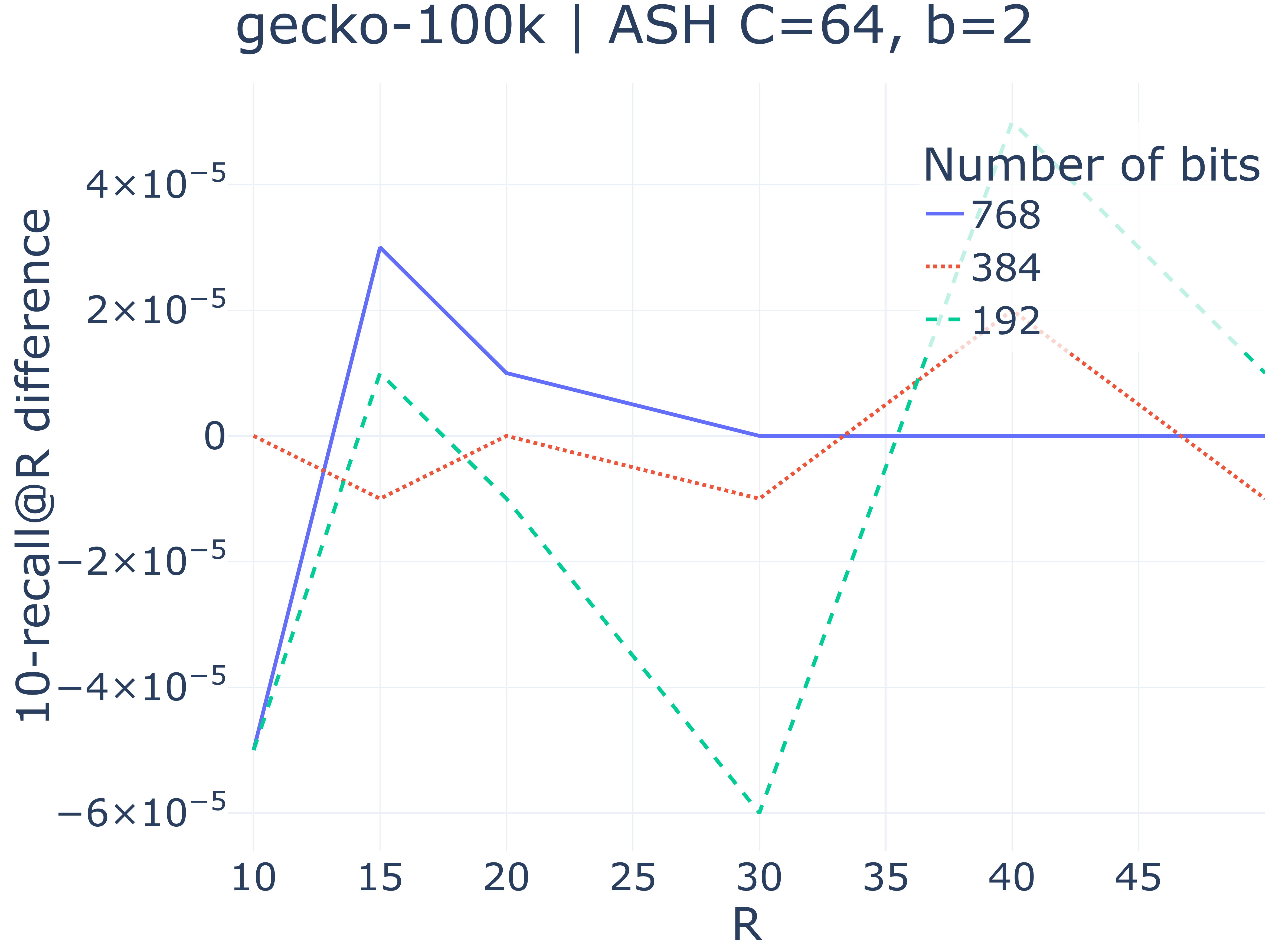}%
    \includegraphics[width=0.2\linewidth]{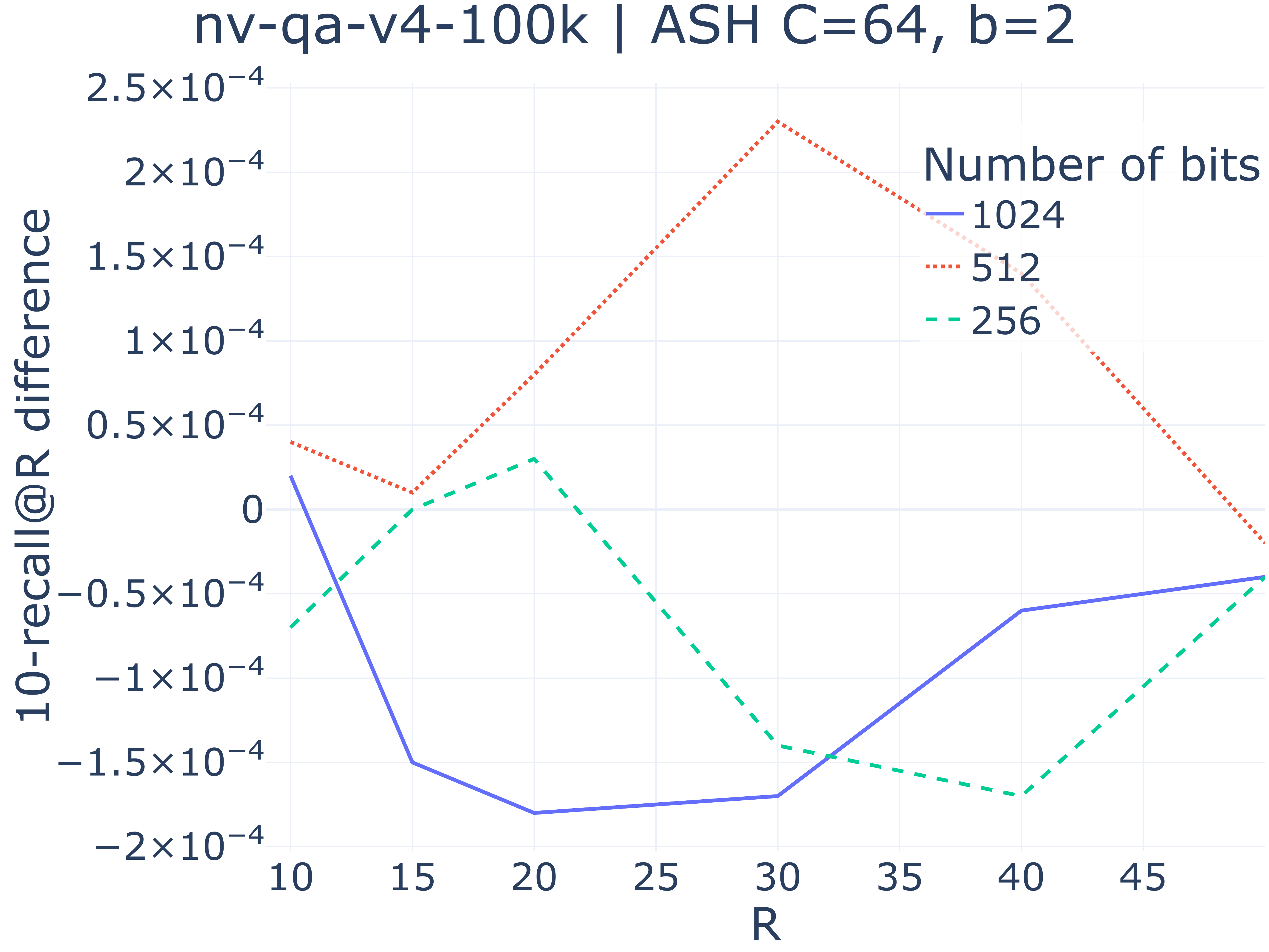}%
    \includegraphics[width=0.2\linewidth]{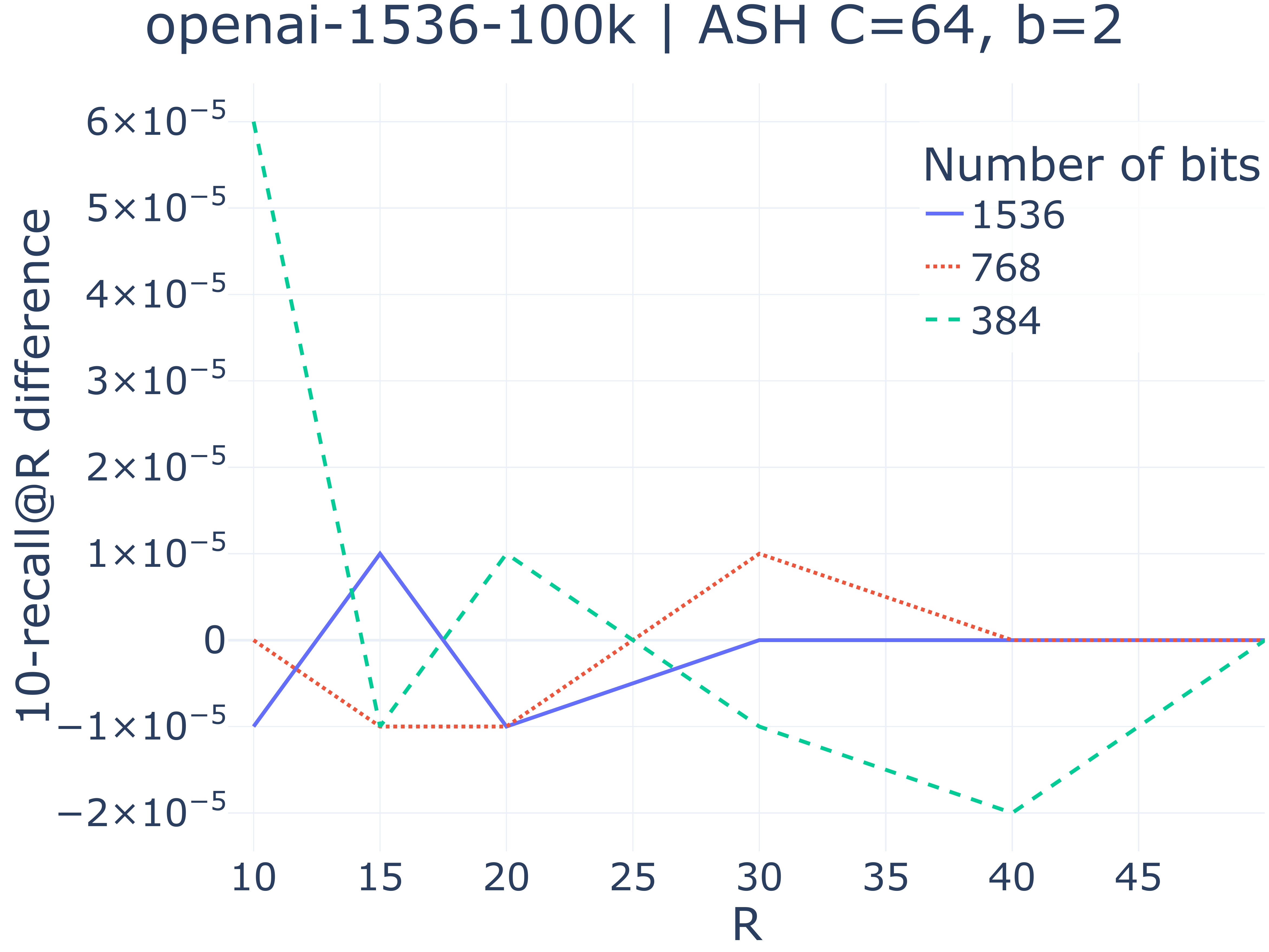}%
    \includegraphics[width=0.2\linewidth]{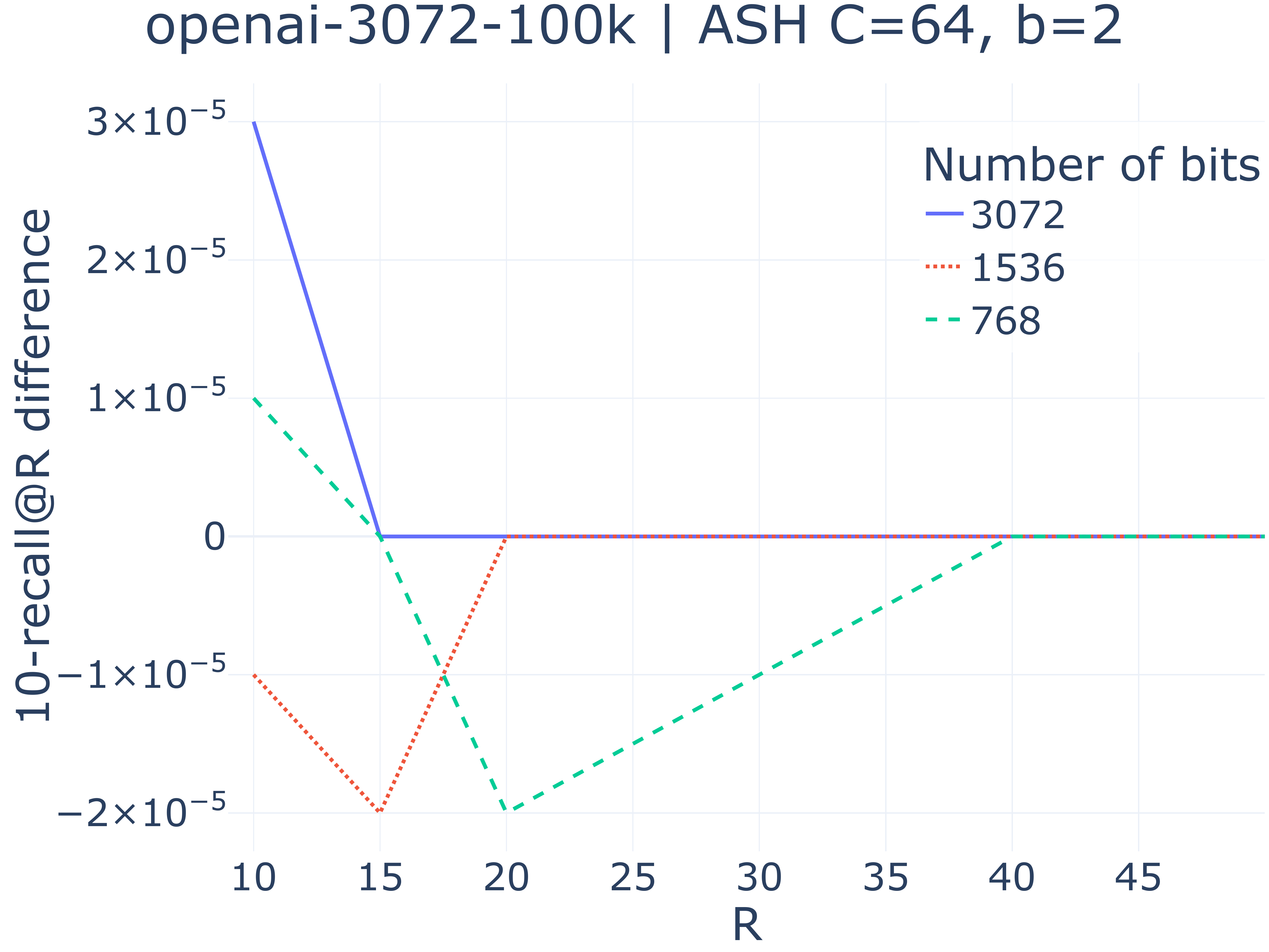}%
        
    \caption{The 10-recall@R difference between queries encoded as float32 or as float16 for different values of R. In the float16 case, the values of the query vector are simply downcasted.}
    \label{fig:ash_query_fp16}
\end{figure*}

\begin{figure*}
    \centering
    \includegraphics[width=0.2\linewidth]{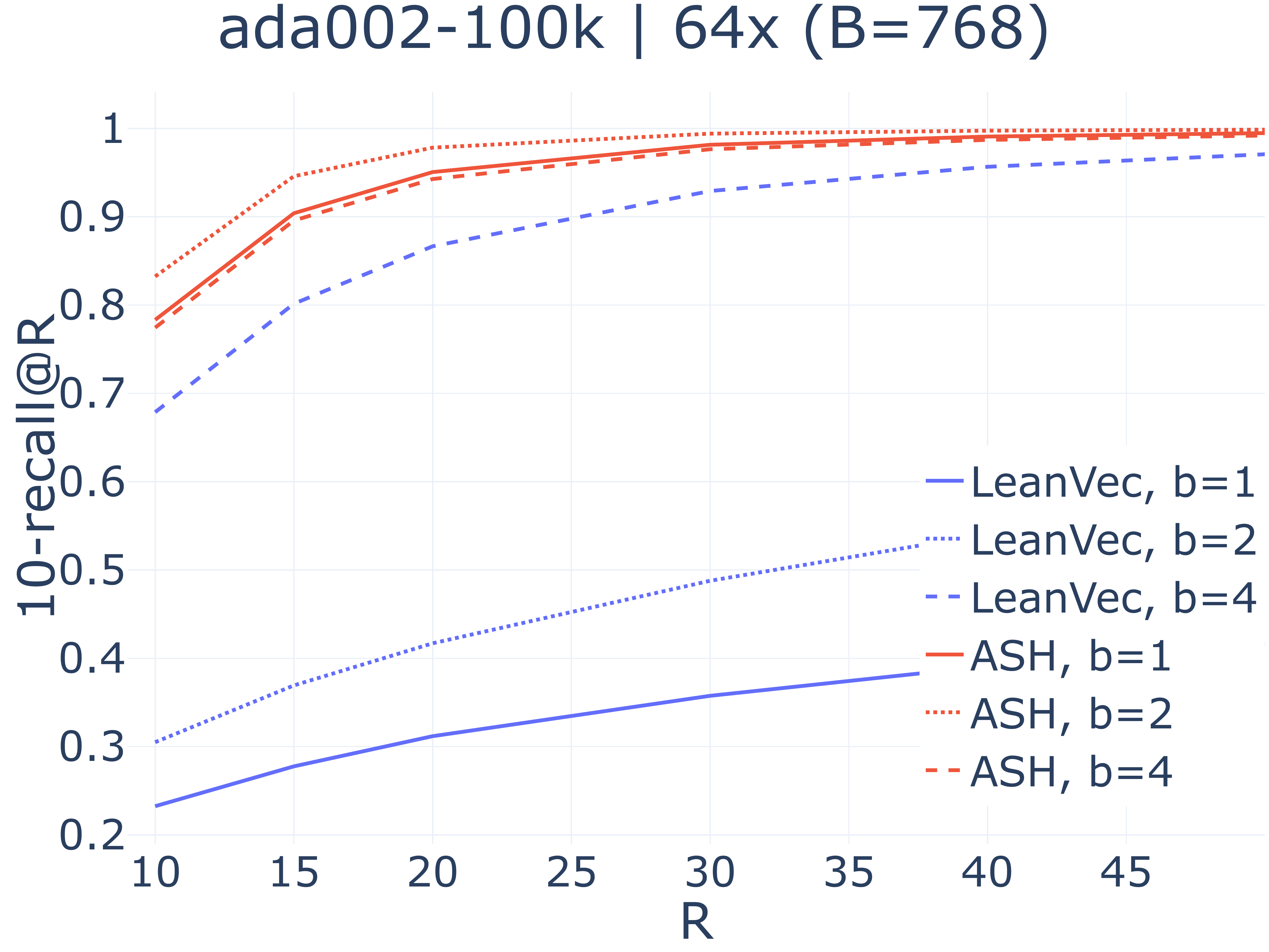}%
    \hfill%
    \includegraphics[width=0.2\linewidth]{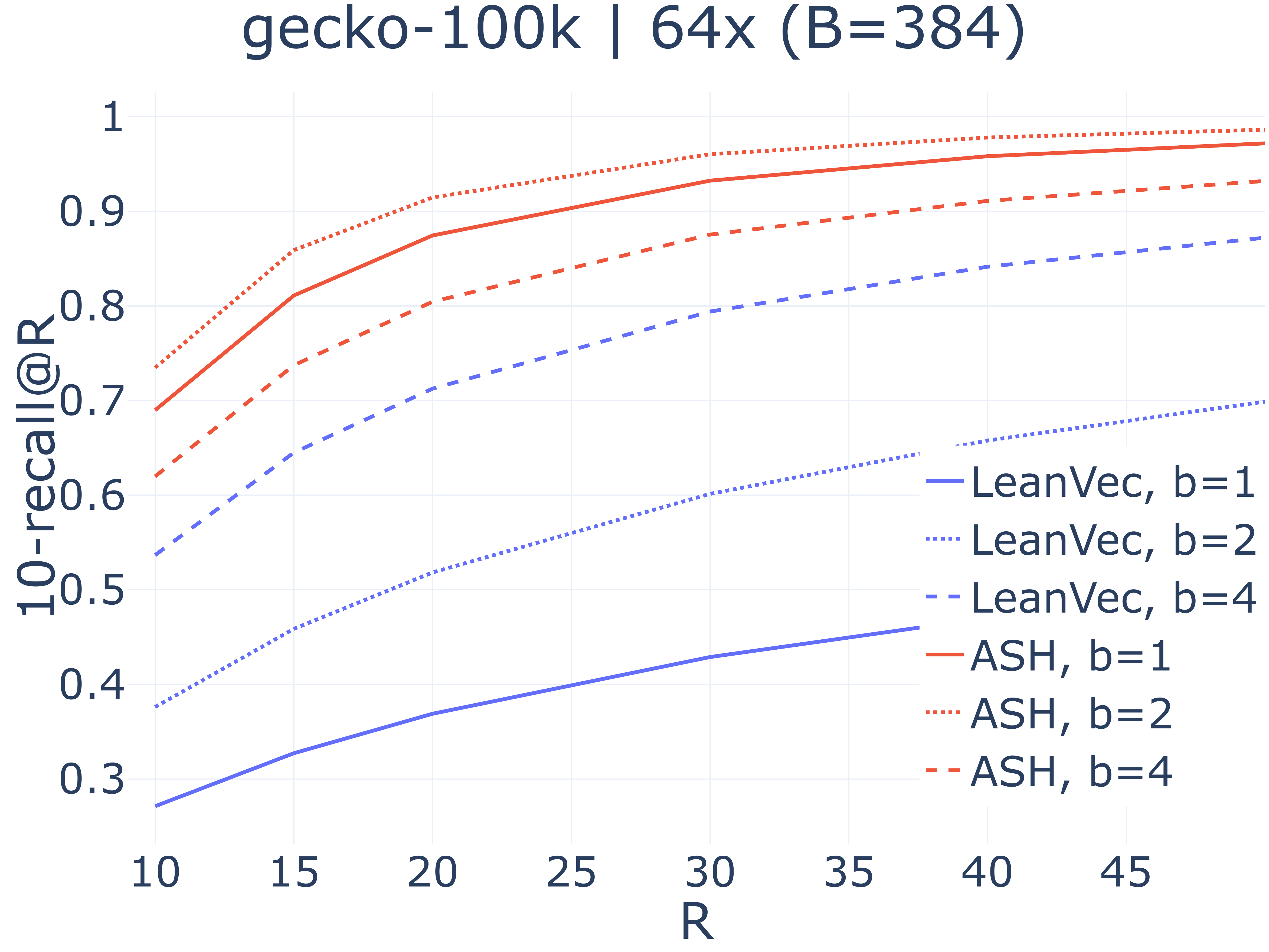}%
    \hfill%
    \includegraphics[width=0.2\linewidth]{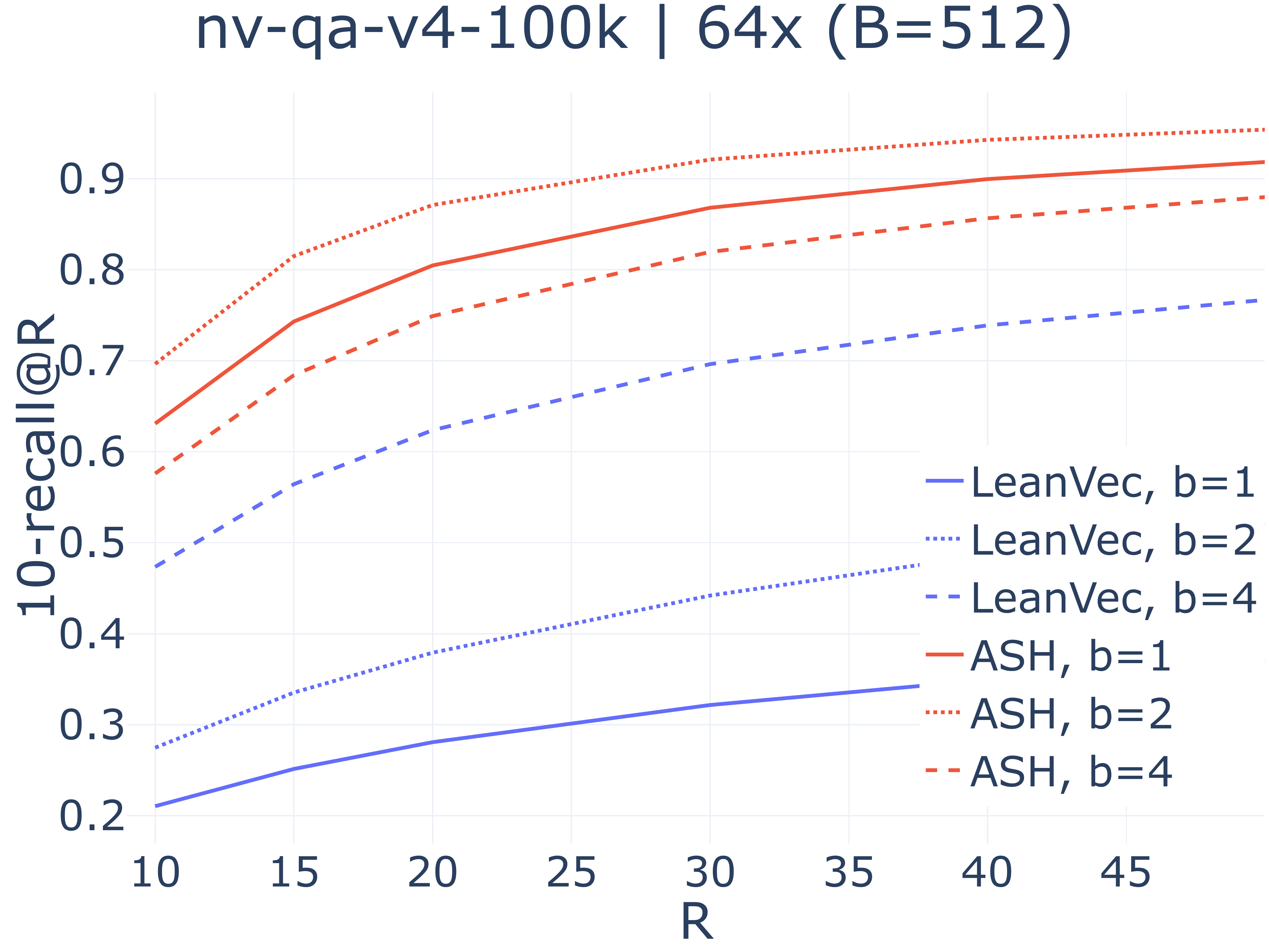}%
    \hfill%
    \includegraphics[width=0.2\linewidth]{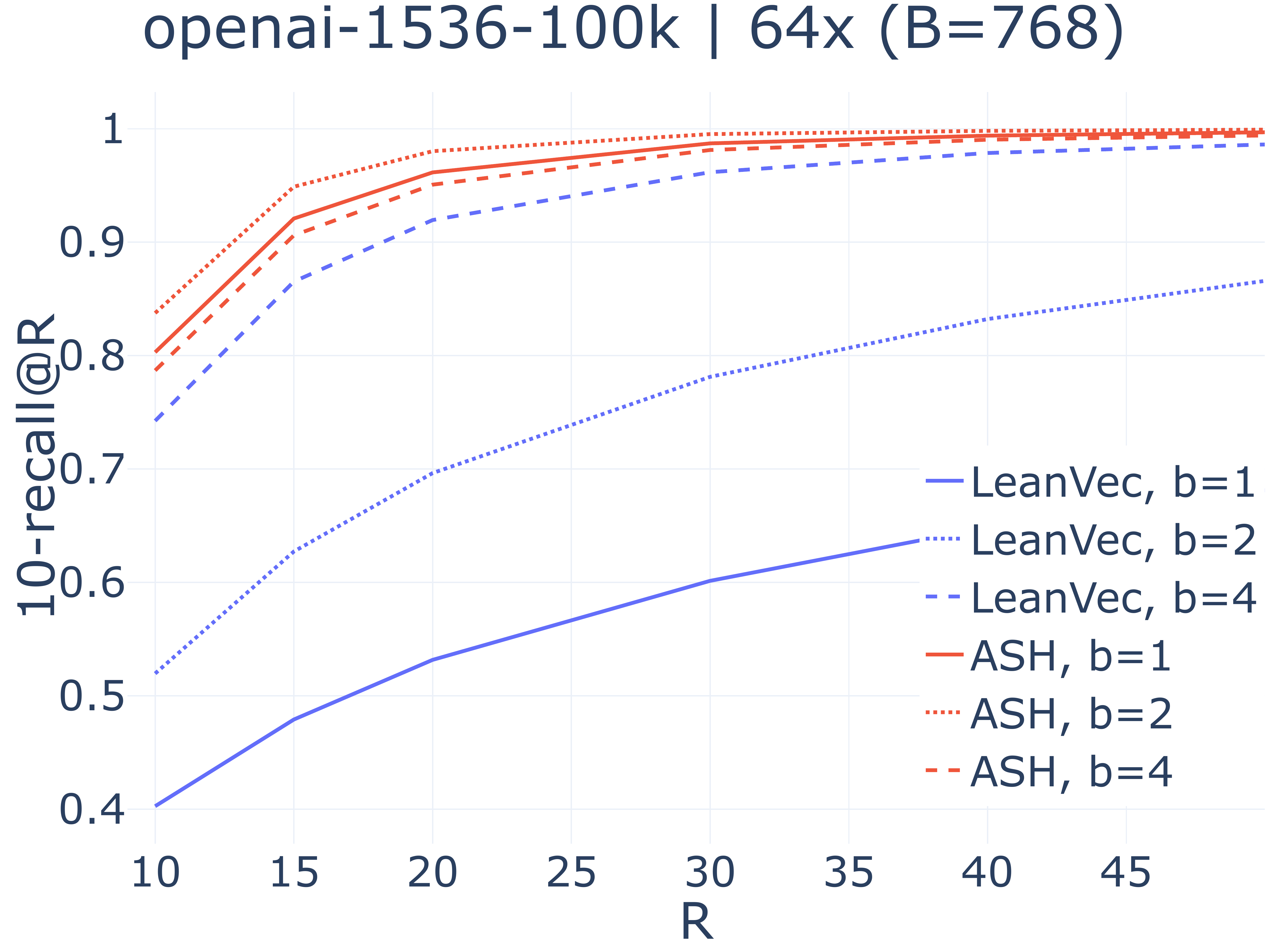}%
    \hfill%
    \includegraphics[width=0.2\linewidth]{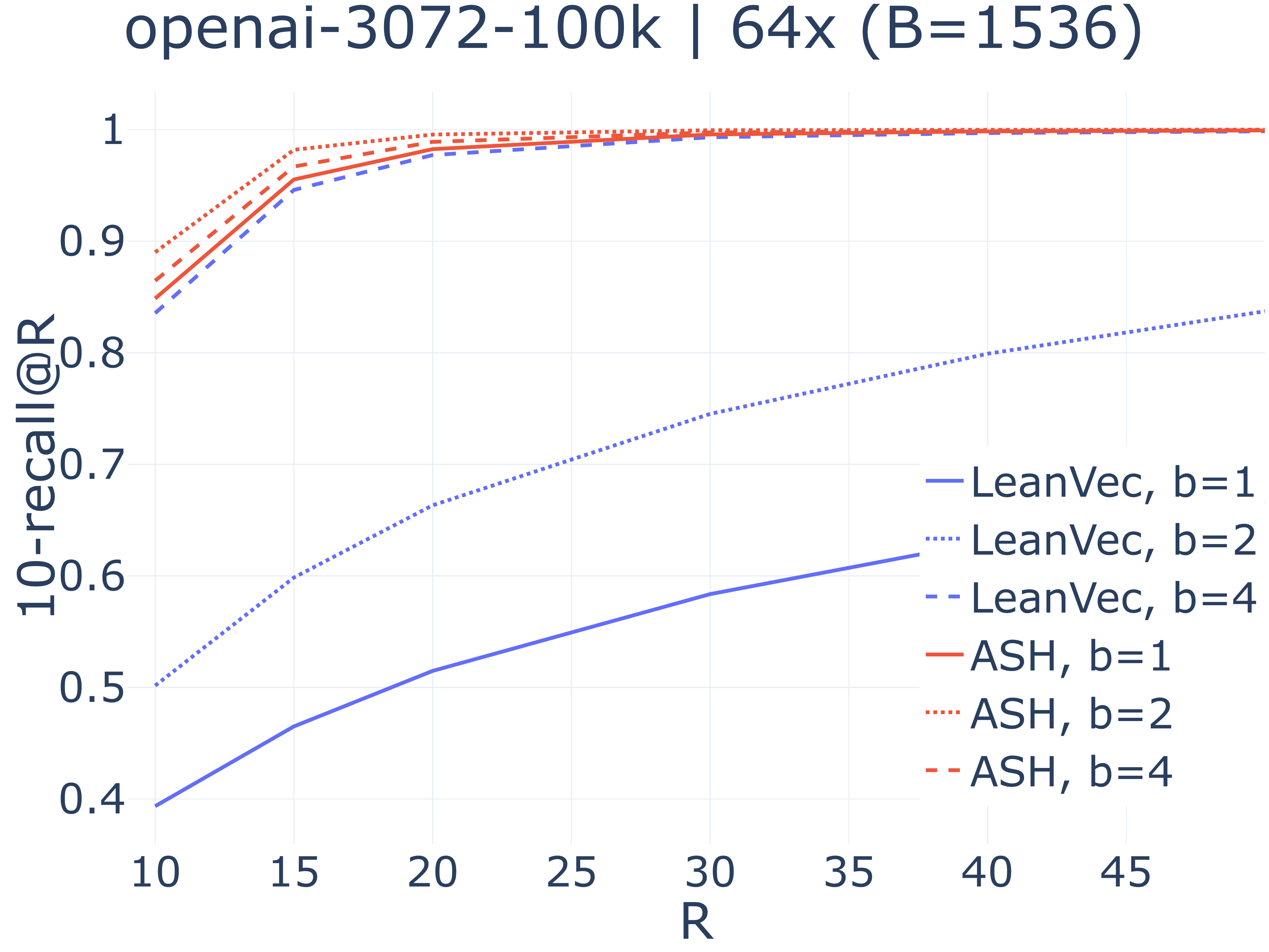}%

    \includegraphics[width=0.2\linewidth]{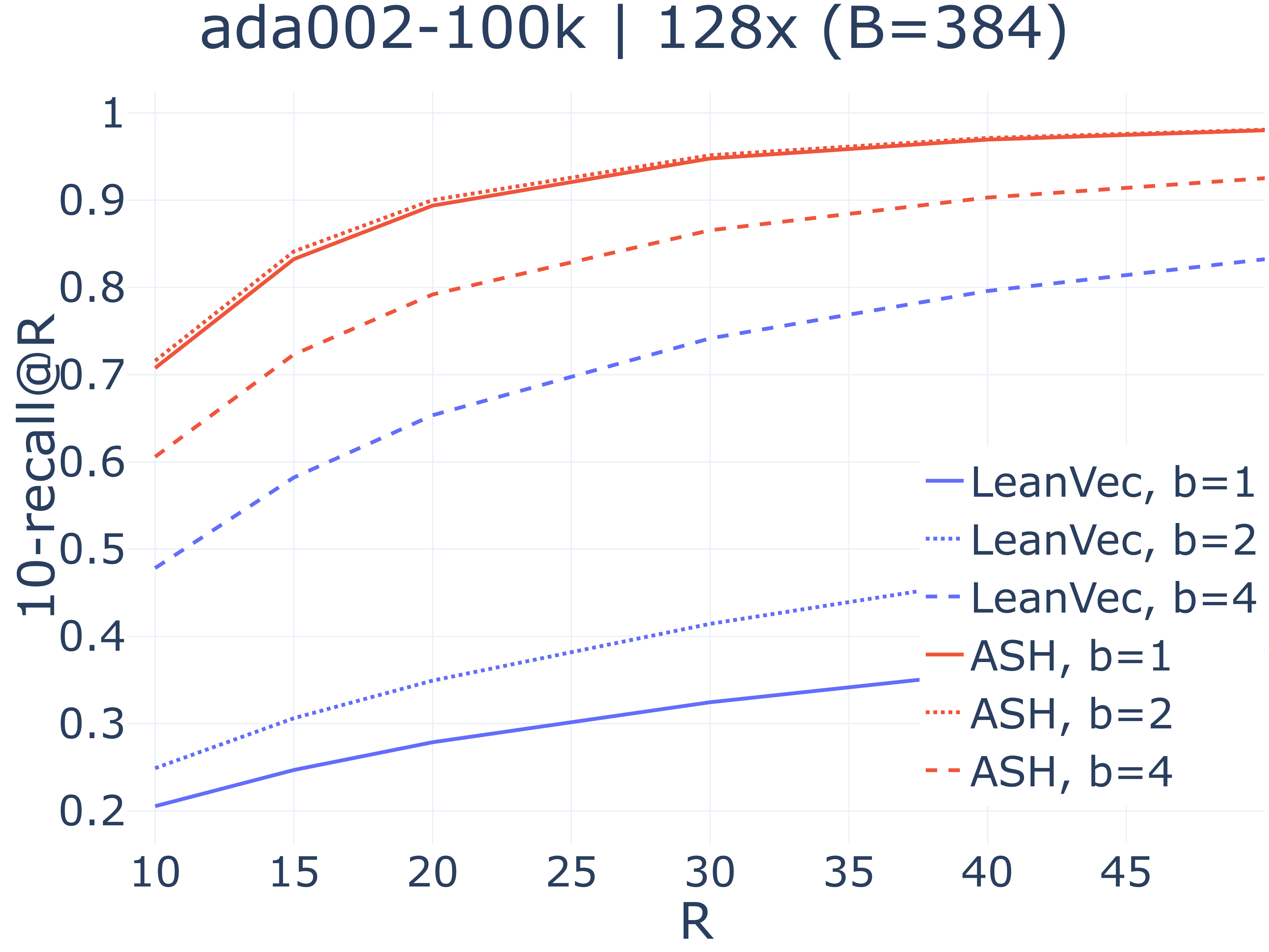}%
    \hfill%
    \includegraphics[width=0.2\linewidth]{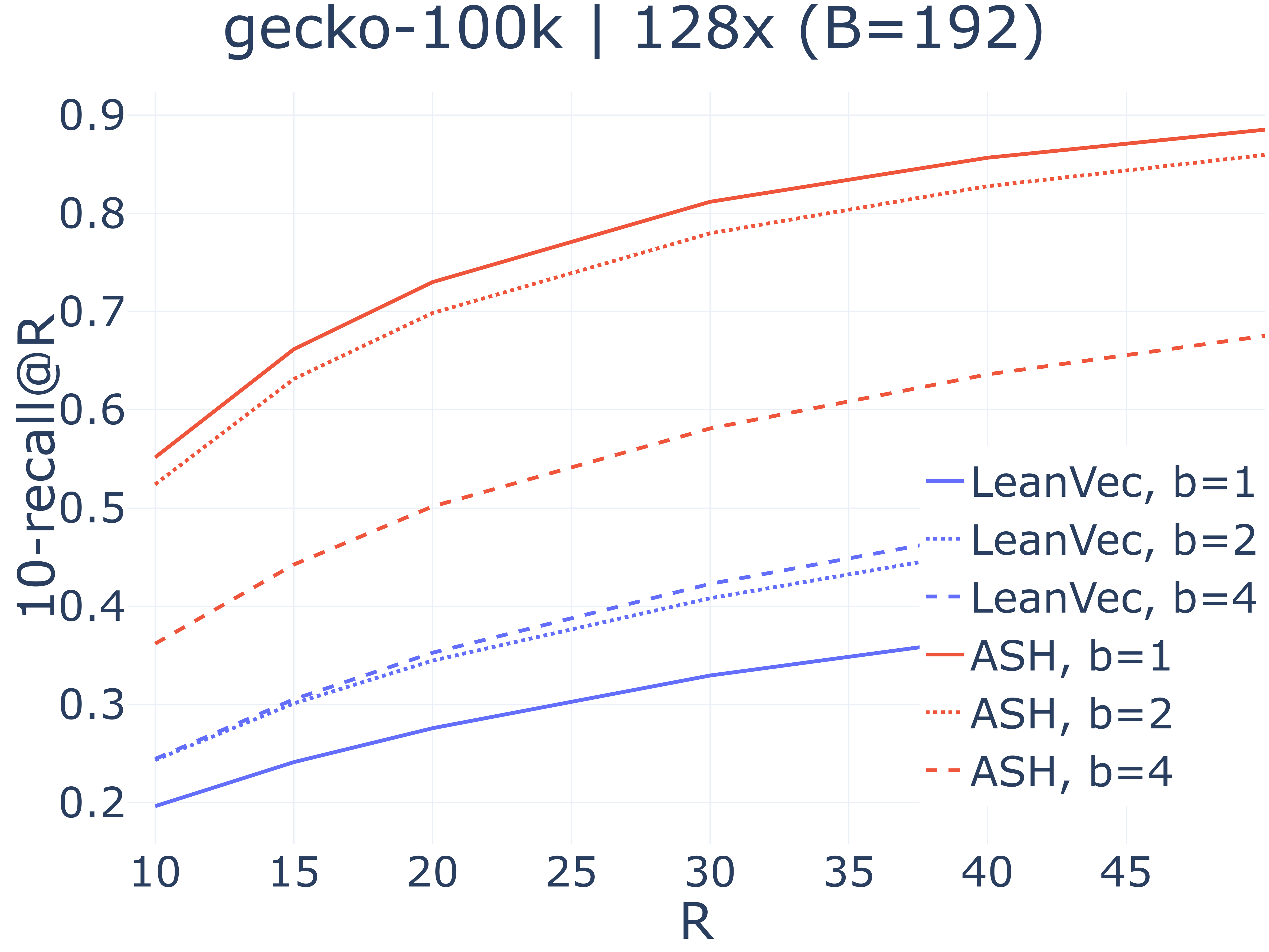}%
    \hfill%
    \includegraphics[width=0.2\linewidth]{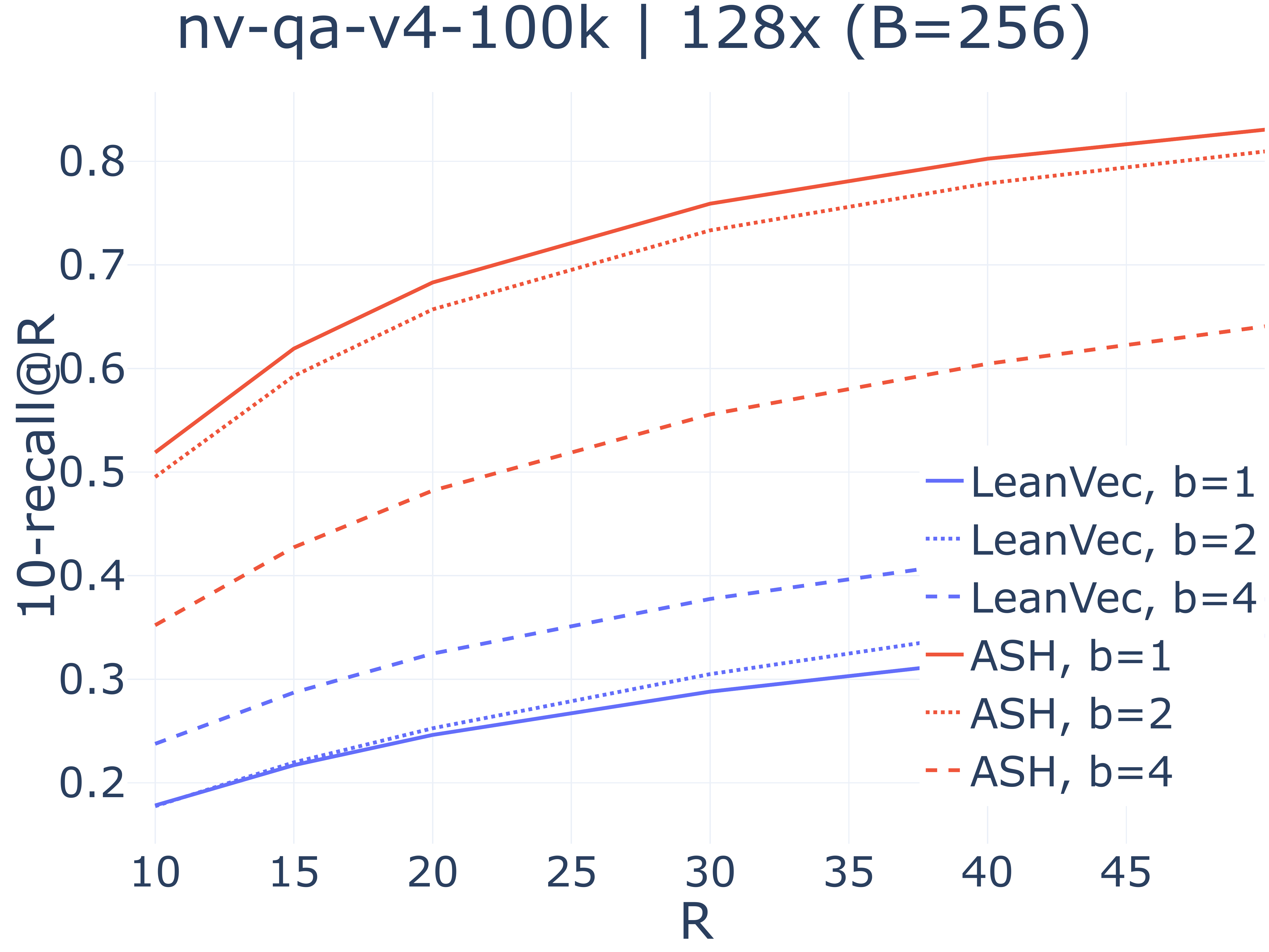}%
    \hfill%
    \includegraphics[width=0.2\linewidth]{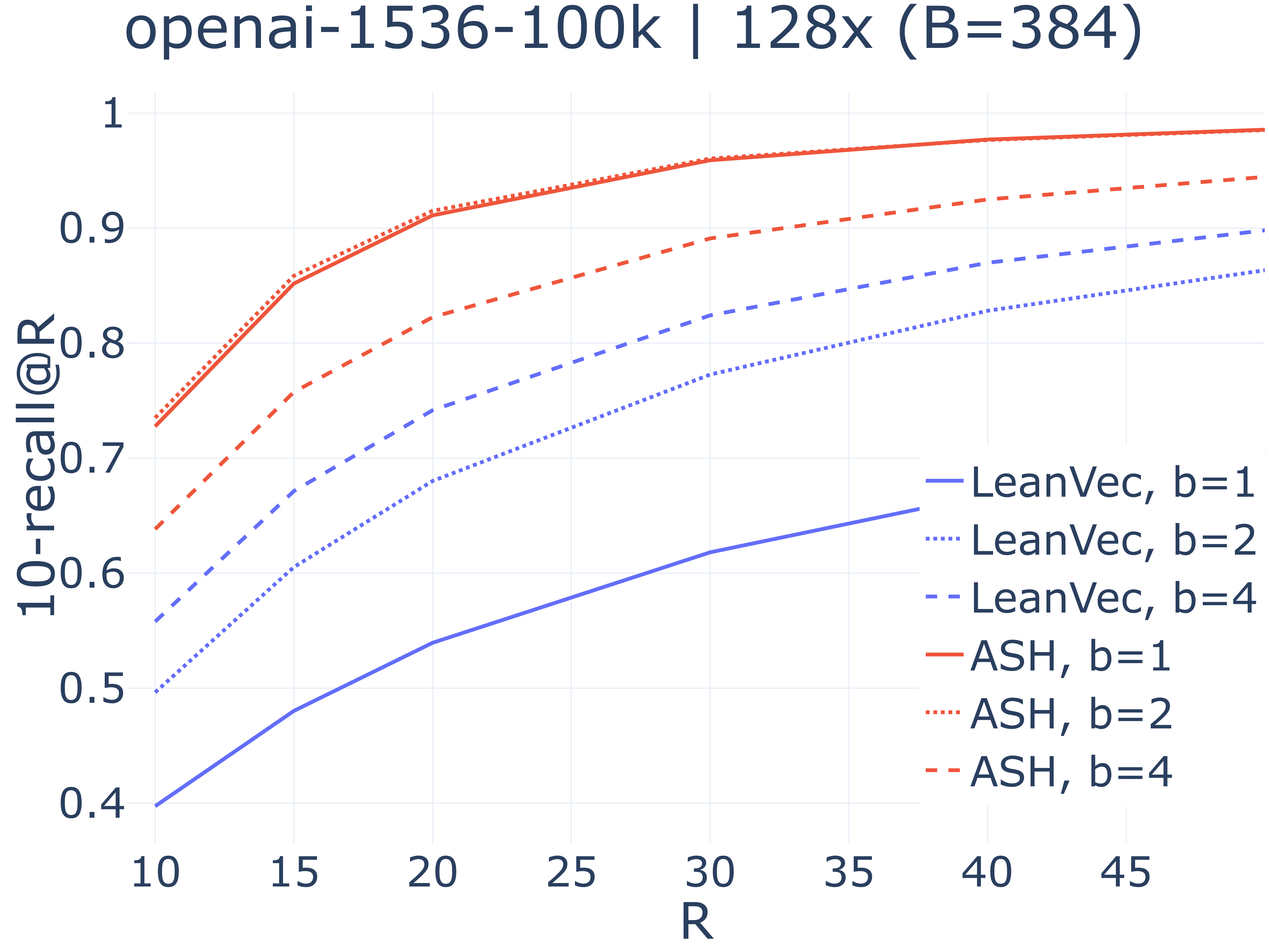}%
    \hfill%
    \includegraphics[width=0.2\linewidth]{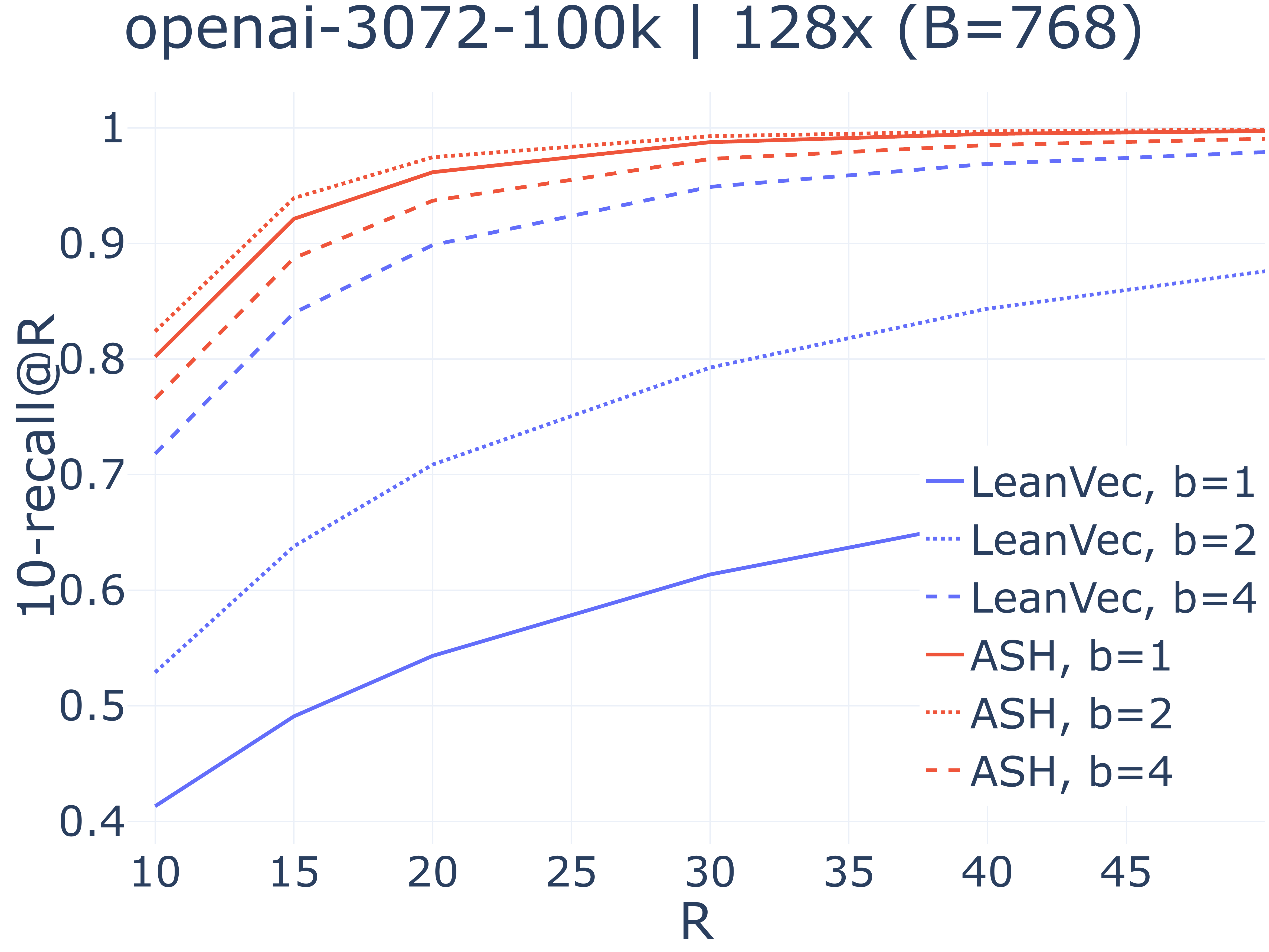}%
    
    \caption{ASH outperforms LeanVec \cite{tepper_leanvec_2024} (that also combines dimensionality reduction and scalar hashing) in search accuracy. We measure the search accuracy using 10-recall@R for different values of R.}
    \label{fig:ash_vs_leanvec_continued}
\end{figure*}

\begin{figure*}
    \centering
    \includegraphics[width=0.3\linewidth]{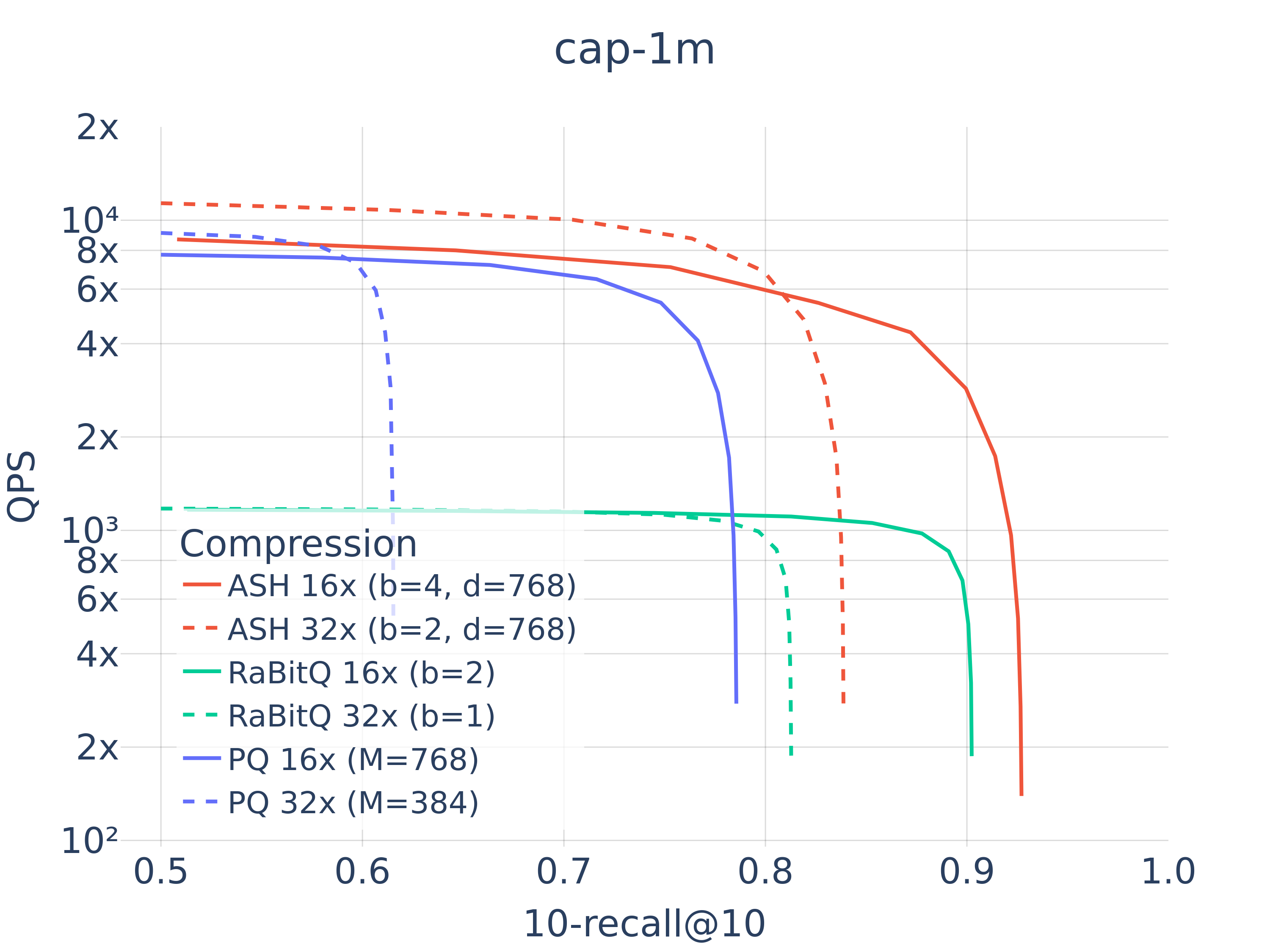}%
    \hspace{1em}%
    \includegraphics[width=0.3\linewidth]{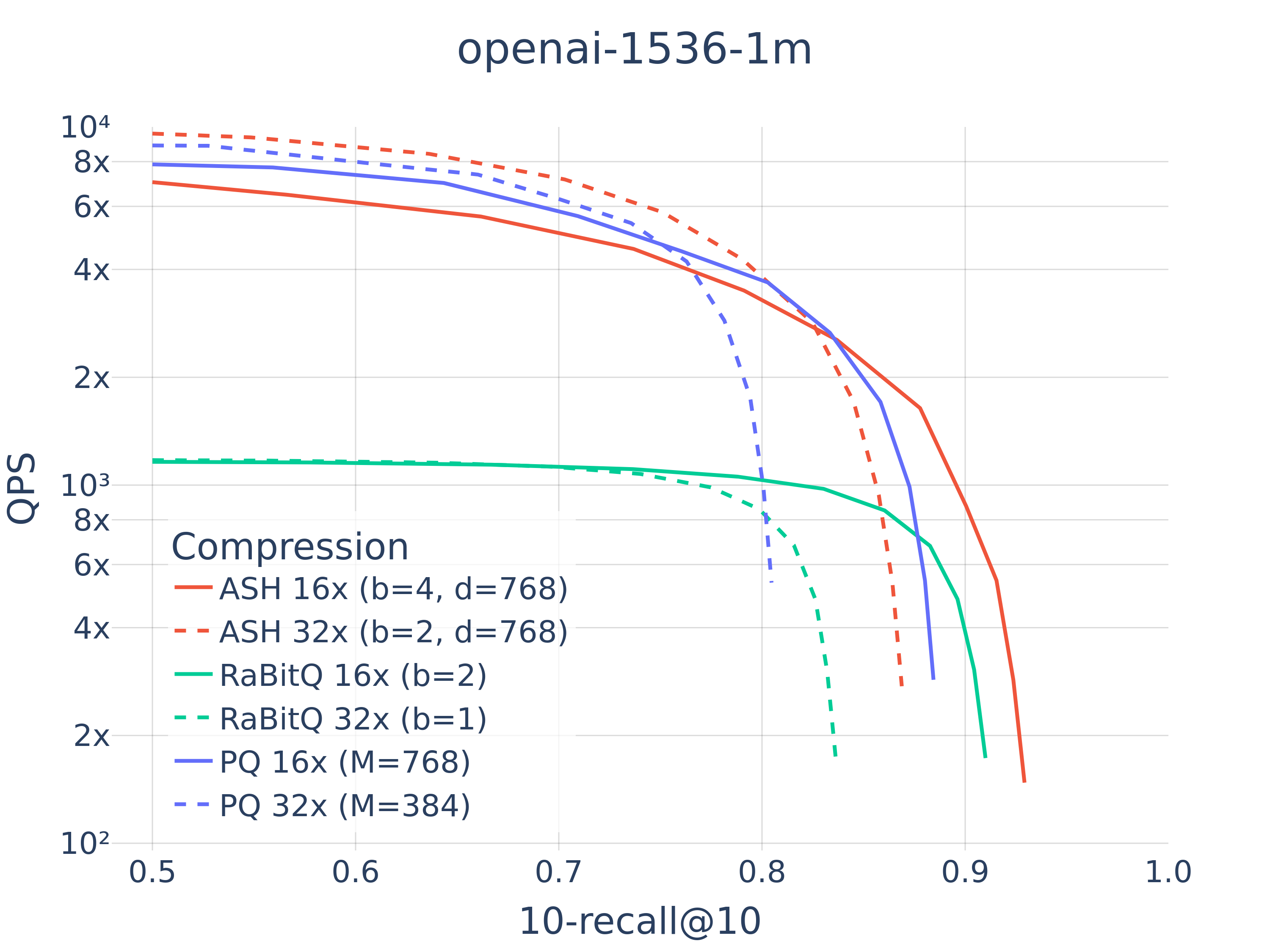}%

    \caption{ASH outperforms RaBitQ \cite{gao_rabitq_2024,gao_practical_2025} and PQ \cite{jegou_product_2011} in search accuracy (10-recall@R) and throughput.}
    \label{fig:qps_recall_continued}
\end{figure*}

We include additional experimental results in \zcref{fig:rabitq_expected_loss,fig:ash_training_continued,fig:ash_groups_continued,fig:ash_bias_continued,fig:ash_bias_continued2,fig:ash_vs_pq_continued,fig:ash_query_fp16,fig:ash_vs_leanvec_continued,fig:qps_recall_continued}. These figures extend the results in the main body of the manuscript.

\begin{figure*}
    \centering

    \includegraphics[width=0.3\linewidth]{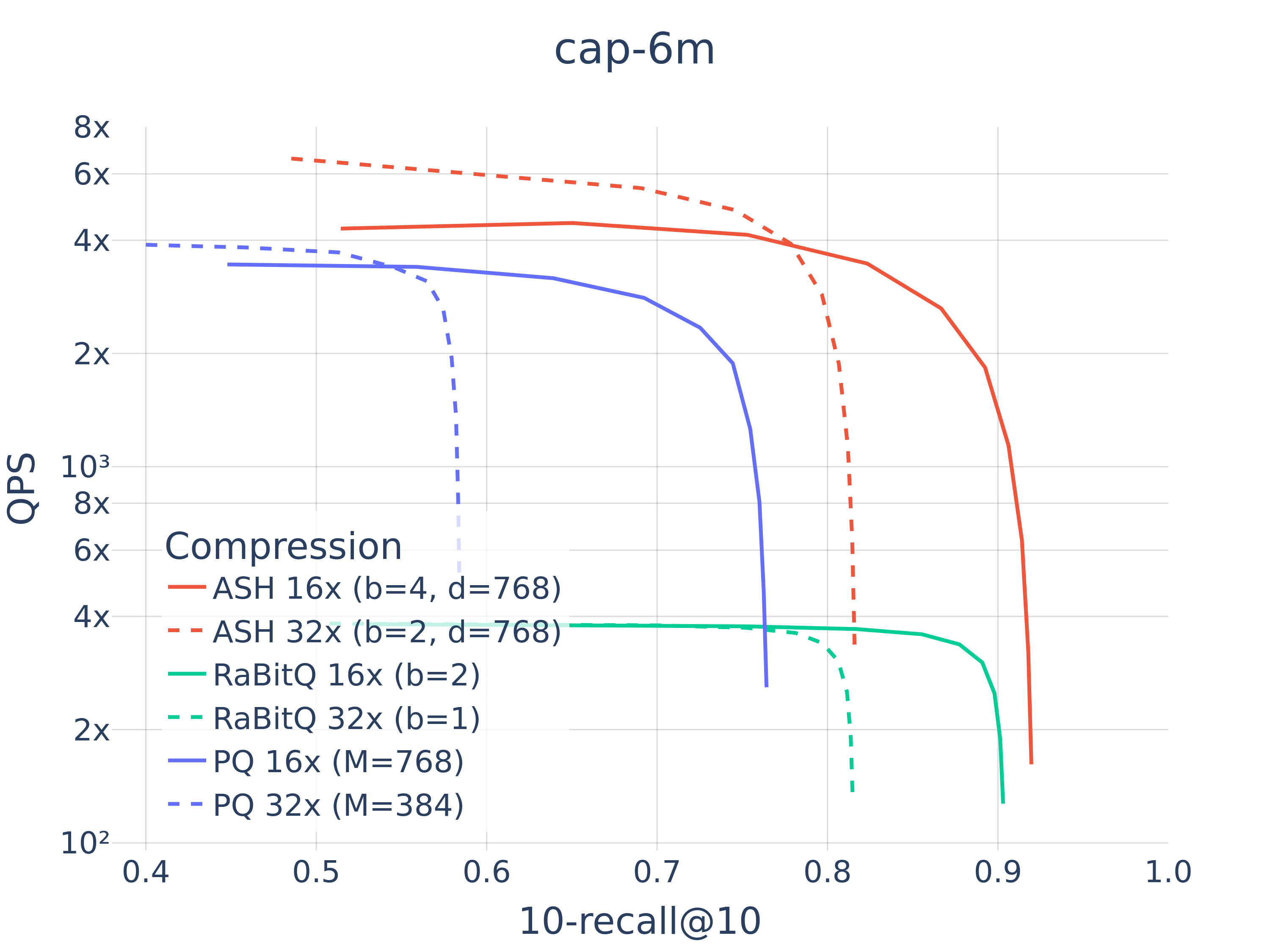}%
    \hspace{1em}%
    \includegraphics[width=0.3\linewidth]{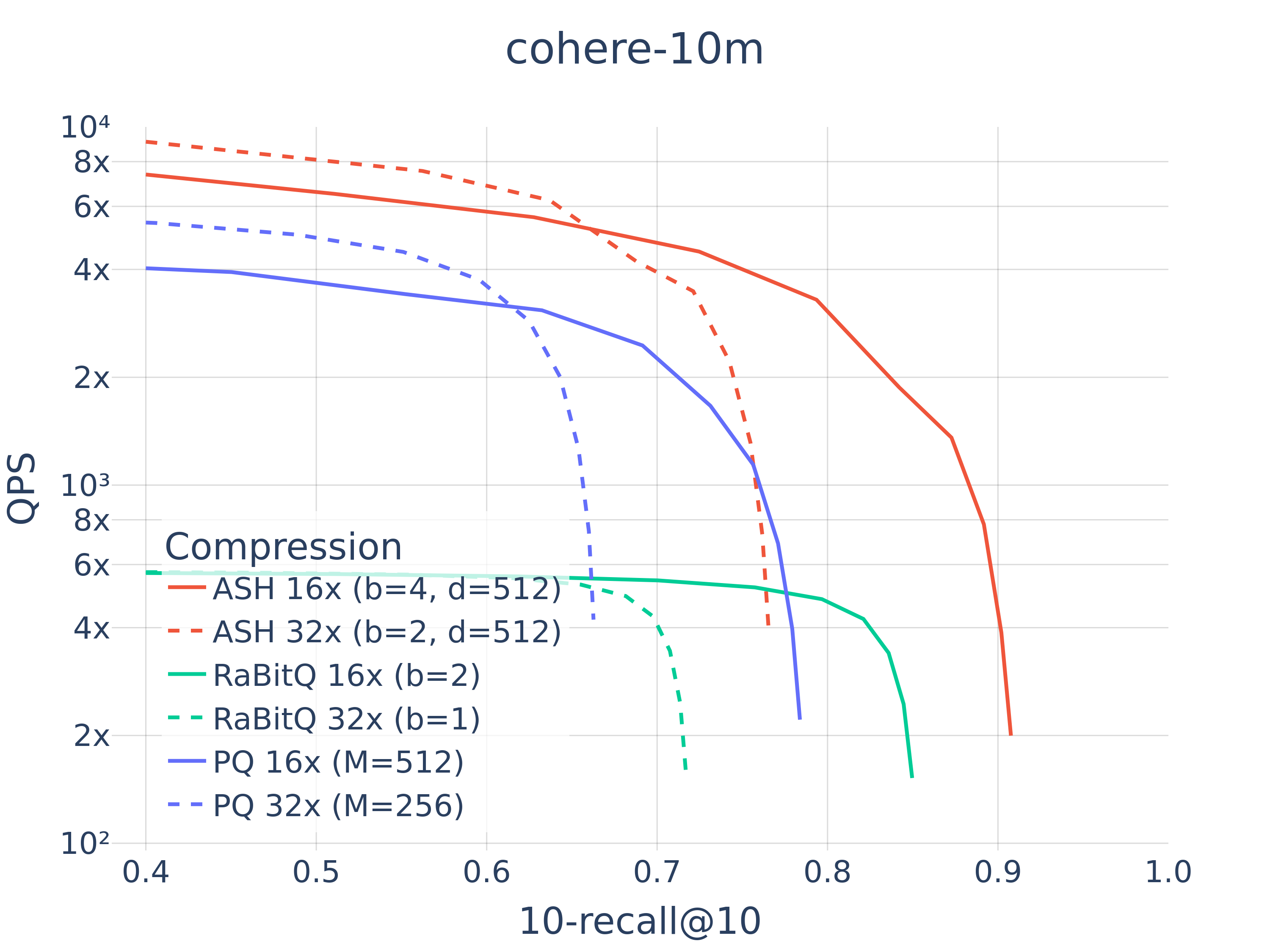}%
    \includegraphics[width=0.3\linewidth]{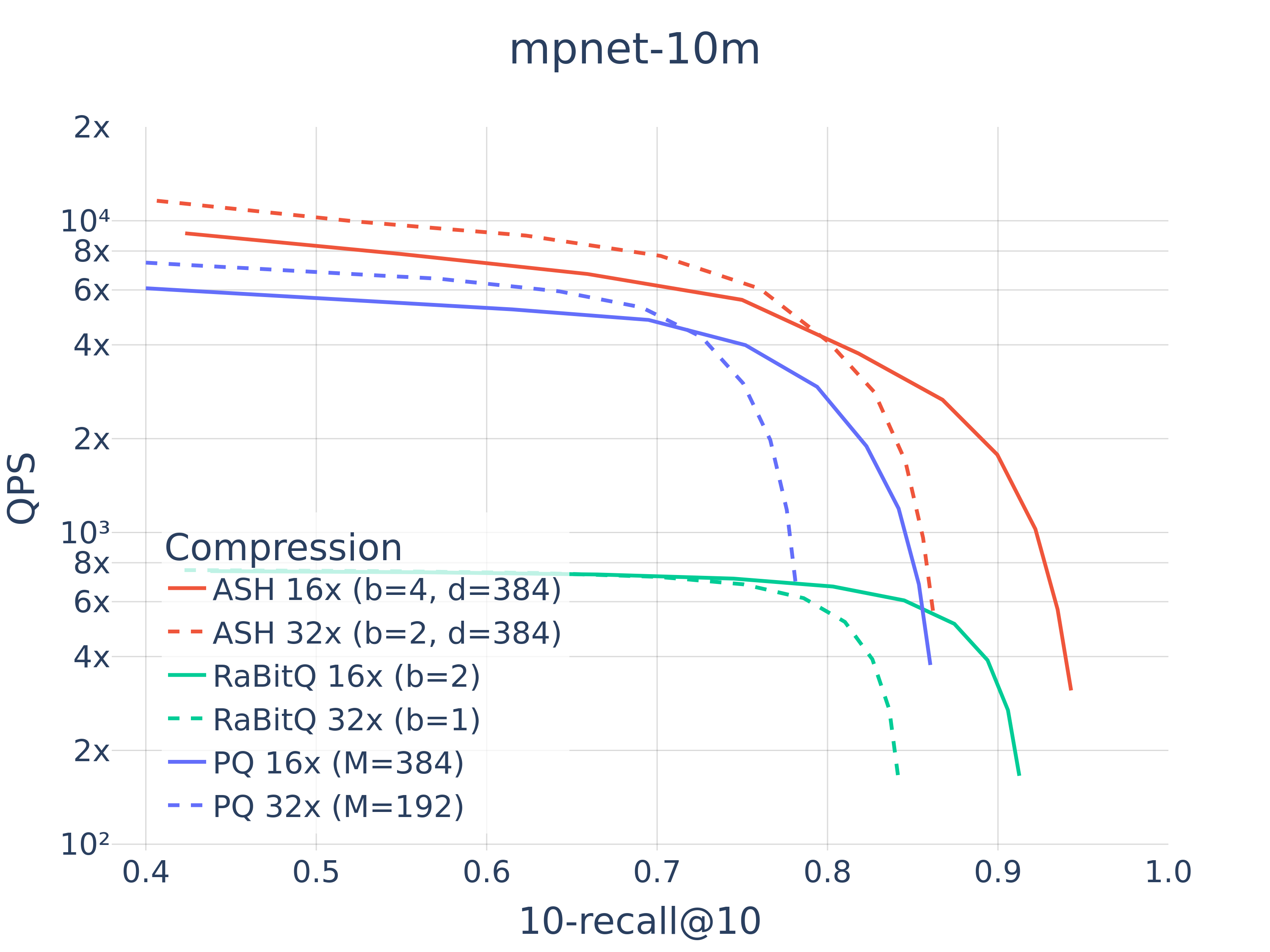}%

    \caption{Selected studies on larger datasets. ASH continues to outperform RaBitQ \cite{gao_rabitq_2024,gao_practical_2025} and PQ \cite{jegou_product_2011} in search accuracy (10-recall@R) and throughput. ASH uses $C=1$ in the comparison.}
    \label{fig:qps_recall_10M}
\end{figure*}

\end{document}